\documentclass[iop]{emulateapj}
\usepackage{amsmath}

\slugcomment{Accepted to the Astrophysical Journal}

\shorttitle{HAT-P-2b Phase Variations}
\shortauthors{Lewis et al.}

\pdfoutput=1

\begin{document}

\title{Orbital Phase Variations of the Eccentric Giant Planet HAT-P-2b}

\author{
Nikole K. Lewis\altaffilmark{1,2,3}, Heather A. Knutson\altaffilmark{4}, Adam P. Showman\altaffilmark{1}, 
Nicolas B. Cowan\altaffilmark{5},  Gregory Laughlin\altaffilmark{6}, Adam Burrows\altaffilmark{7},
Drake Deming\altaffilmark{8}, Justin R. Crepp\altaffilmark{9}, Kenneth J. Mighell\altaffilmark{10}, 
Eric Agol\altaffilmark{11}, G\'asp\'ar \'A. Bakos\altaffilmark{7}, David Charbonneau\altaffilmark{12}, 
Jean-Michel D\' esert\altaffilmark{4,3}, Debra A.\ Fischer\altaffilmark{13}, Jonathan J. Fortney\altaffilmark{6}, 
Joel D.\ Hartman\altaffilmark{7}, Sasha Hinkley\altaffilmark{14}, Andrew W.\ Howard\altaffilmark{15},
John Asher Johnson\altaffilmark{14},  Melodie Kao\altaffilmark{4}, Jonathan Langton\altaffilmark{16}, 
Geoffrey W.\ Marcy\altaffilmark{17}, Joshua N. Winn\altaffilmark{18}
}

\altaffiltext{1}{Department of Planetary Sciences and Lunar and Planetary Laboratory, The University of Arizona, Tucson, AZ 85721, USA}
\altaffiltext{2}{Department of Earth, Atmospheric and Planetary Sciences, Massachusetts Institute of Technology, Cambridge, MA 02139, USA;
\email{nklewis@mit.edu}}
\altaffiltext{3}{Sagan Postdoctoral Fellow}
\altaffiltext{4}{Division of Geological and Planetary Sciences, California Institute of Technology, Pasadena, CA 91125, USA} 
\altaffiltext{5}{Center for Interdisciplinary Exploration and Research in Astrophysics and Department of Physics and Astronomy, 
Northwestern University, 2131 Tech Drive, Evanston, IL 60208, USA}
\altaffiltext{6}{Department of Astronomy \& Astrophysics, University of California, Santa Cruz, CA 95064, USA}
\altaffiltext{7}{Dept. of Astrophysical Sciences, Princeton University, Princeton, NJ 08544, USA}
\altaffiltext{8}{Dept. of Astronomy, University of Maryland, College Park, MD, 20742, USA}
\altaffiltext{9}{Department of Physics, University of Notre Dame, Notre Dame, IN 46556, USA}
\altaffiltext{10}{National Optical Astronomy Observatories, Tucson, AZ 85726, USA}
\altaffiltext{11}{Dept. of Astronomy, University of Washington, Seattle, WA 98195, USA}
\altaffiltext{12}{Harvard-Smithsonian Center for Astrophysics, 60 Garden St., Cambridge, MA 02138, USA}
\altaffiltext{13}{Department of Astronomy, Yale University, New Haven, CT 06511, USA}
\altaffiltext{14}{Department of Astrophysics, California Institute of Technology, MC 249-17, Pasadena, CA 91125, USA}
\altaffiltext{15}{Institute for Astronomy, University of Hawaii, 2680 Woodlawn Drive, Honolulu, HI 96822, USA}
\altaffiltext{16}{Department of Physics, Principia College, 1 Maybeck Place, Elsah, IL 62028, USA}
\altaffiltext{17}{Department of Astronomy, University of California, Berkeley, CA 94720, USA}
\altaffiltext{18}{Department of Physics, and Kavli Institute for Astrophysics and Space Research, Massachusetts Institute of Technology, Cambridge, MA 02139}

\begin{abstract}
We present the first secondary eclipse and phase curve observations for the highly eccentric hot Jupiter HAT-P-2b 
in the 3.6, 4.5, 5.8, and 8.0~$\mu$m bands of the {\it Spitzer Space Telescope}.  The 3.6 and 4.5~$\mu$m data 
sets span an entire orbital period of HAT-P-2b ($P=5.6334729$ d), making them the longest continuous phase curve observations 
obtained to date and the first full-orbit observations of a planet with an eccentricity exceeding 0.2.  We present an improved
non-parametric method for removing the intrapixel sensitivity variations in {\it Spitzer} data 
at 3.6 and 4.5~$\mu$m that robustly maps position-dependent flux variations.   We find 
that the peak in planetary flux occurs at 4.39$\pm$0.28, 5.84$\pm$0.39, and 4.68$\pm$0.37 hours 
after periapse passage with corresponding maxima in the planet/star flux ratio of  
$0.1138\%\pm0.0089\%$, $0.1162\%\pm0.0080\%$, and $0.1888\%\pm0.0072\%$
in the 3.6, 4.5, and 8.0~$\mu$m bands respectively.  
Our measured secondary eclipse depths of $0.0996\%\pm0.0072\%$, $0.1031\%\pm0.0061\%$, $0.071\%^{+0.029\%}_{-0.013\%}$, and 
$0.1392\%\pm0.0095\%$ in the 3.6, 4.5, 5.8, and 8.0~$\mu$m bands respectively indicate that the planet cools significantly from 
its peak temperature before we measure the dayside flux during secondary eclipse.  
We compare our measured secondary eclipse depths to the predictions from a one-dimensional radiative transfer model, which suggests 
the possible presence of a transient day side inversion in HAT-P-2b's atmosphere near periapse.
We also derive improved estimates for the system parameters, including its mass, radius, and orbital ephemeris.  
Our simultaneous fit to the transit, secondary eclipse, and radial velocity data allows us to determine
the eccentricity ($e=0.50910\pm0.00048$) and argument of periapse ($\omega=188.09^{\circ}\pm0.39^{\circ}$)
of HAT-P-2b's orbit with a greater precision than has been achieved for any other eccentric extrasolar planet.   We also find 
evidence for a long-term linear trend in the radial velocity data.  This trend suggests the presence of another substellar 
companion in the HAT-P-2 system, which could have caused HAT-P-2b to migrate inward to its present-day orbit 
via the Kozai mechanism.

\end{abstract}

\keywords{planets and satellites: general, planets and satellites: individual: HAT-P-2b, 
techniques: photometric, methods:  numerical, atmospheric effects}

\section{Introduction}

The supermassive ($M_p=9~M_J$) Jupiter sized ($R_p=1~R_J$) planet HAT-P-2b (aka HD~147506b) was 
first discovered from transit observations by \citet{bak07a} using the HATNet \citep{bak02,bak04}
network of ground-based telescopes.  Follow-up radial velocity measurements of the HAT-P-2 system 
revealed that the orbit of HAT-P-2b is highly eccentric \citep[$e=0.5$][]{bak07a,loe08}.  
Only a handful of transiting exoplanets have been shown to posses eccentricities in excess of that of Mercury ($e=0.2$), 
which has made HAT-P-2b an interesting target for many theoretical studies concerning the evolution of the HAT-P-2 system \citep{jac08, fab08, mat08, bar08}.
Because of its mass, HAT-P-2b represents a class of exoplanets that provides 
an important link between extrasolar giant planets and low-mass brown dwarfs \citep{bar08}.  
Observational constraints on the basic atmospheric properties of HAT-P-2b will provide an important probe into the structure and evolution 
of not only HAT-P-2b, but an entire class of massive extrasolar planets.

Atmospheric circulation models for planets on eccentric orbits show significant 
variations in atmospheric temperature and wind speeds that provide an important probe 
into atmospheric radiative and dynamical timescales \citep{lan08,lew10,cow11,kat12}. 
The incident flux on HAT-P-2b from its stellar host at periapse is ten times that at apoapse, which should cause 
large variations in atmospheric temperature, wind speeds, and chemistry.  
Heating and cooling rates for HAT-P-2b can be constrained by measuring planetary brightness as a function of time.
Such observations are analogous to previous phase curve observations of HD~189733b using the {\it Spitzer Space Telescope}, 
which provided the first clear observational evidence for atmospheric circulation in an exoplanet atmosphere \citep{knu07, knu09b, knu12}.

HAT-P-2b's large orbital eccentricity makes it an ideal target for investigating hot Jupiter migration 
mechanisms.  Gas giant planets such as HAT-P-2b are expected to form beyond the ice line
in their protoplanetary disk far from their stellar hosts.  HAT-P-2b currently resides 
at a semi-major axis of 0.07~AU from its host star, indicating that it must have migrated inward via a process such as gas 
disk migration \citep{lin96}, planet-planet scattering \citep{ras96}, secular interactions \citep{wu11}, 
or Kozai migration \citep{wei96, nao11}.  HAT-P-2b's close-in and highly eccentric orbit favors one of the latter three 
mechanisms since disk migration tends to damp out orbital eccentricities.  For Kozai 
migration, the presence of a third body with at least as much mass as HAT-P-2b is needed.  
In this study we present six years of radial velocity measurements for this system, which allow us 
to search for the presence of a massive third body in the HAT-P-2 system.  

Here we present our analysis of the 3.6 and 4.5~$\mu$m full-orbit phase curves of the HAT-P-2 system, which 
include two secondary eclipses and one transit at each wavelength.  These full-orbit phase curves 
represent the longest continuous phase observations ever obtained by the {\it Spitzer Space Telescope}.  The orbital 
period of HAT-P-2b (5.6334729 d) is more than 2.5 times that of other exoplanets with published full-orbit phase curve observations:  
WASP-12b \citep{cow12a} and HD~189733b \citep{knu12}.
Additionally, we present an analysis of previous
partial orbit phase curve and secondary eclipse {\it Spitzer} observations at 8.0 and 5.8~$\mu$m respectively.  
We use these observations to characterize the changes in the planet's emission spectrum as a function of orbital
phase and to probe the atmospheric chemistry and circulation regime of HAT-P-2b.
The following sections describe our observations and data reduction methods (\S\ref{hat2_obs})
and the results of our analysis (\S\ref{hat2_results}).  Section~\ref{hat2_discussion} discusses the results 
from our analysis of the {\it Spitzer} data and compares them to predictions from atmospheric models for HAT-P-2b.  
Additionally, we discuss trends in our radial velocity data that indicate the presence of an additional body in HAT-P-2 system 
in Section~\ref{hat2_discussion}.  Section~\ref{hat2_conclusions} overviews the 
main conclusions from our analysis and presents ideas for future work.

\section{Observations}\label{hat2_obs}
We analyze nearly 300 hours of observation of HAT-P-2 at 3.6~$\mu$m and 4.5~$\mu$m
obtained during the post-cryogenic mission of the {\it Spitzer Space Telescope} \citep{wer04} 
using the IRAC instrument \citep{faz04} in subarray mode.  The observation periods were
UT 2010 March 28 to UT 2010 April 03 and UT 2011 July 09 to UT 2011 July 15 for the 3.6 and 4.5~$\mu$m 
bandpasses respectively.  Both observations cover a period just over 149 hours with 
two approximately 2-hour breaks for data downlink, corresponding to $\sim$1.2 million images in each bandpass.  
Each observation begins a few hours before the secondary eclipse of the planet, continues through planetary transit, and ends a few hours after 
the subsequent planetary secondary eclipse.

We also analyze observations of the HAT-P-2 system at 5.8~$\mu$m and 8.0~$\mu$m obtained during 
 the cryogenic phase of the {\it Spitzer Space Telescope} mission.  The 5.8~$\mu$m observations cover 
 a single secondary eclipse of the planet that occurs on UT 2009 March 16.  The 8.0~$\mu$m 
 observations cover a portion of the planet's orbit that includes transit, periapse passage, and secondary eclipse
 on UT 2007 September 10-11.  The 5.8~$\mu$m observations were obtained using 
 subarray mode, while the 8.0~$\mu$m observations were obtained using the full IRAC array. 
 We obtain our 3.6, 4.5, 5.8, and 8.0~$\mu$m photometry from the Basic Calibrated (BCD) 
 files produced by version 18.18.0 of the {\it Spitzer} analysis pipeline.  
 
In subarray mode, 32$\times$32 pixel images are stored in sets of 64 as a single FITS file with a single header.  
We calculate the \mbox{BJD\_UTC} at mid-exposure for each image from time stamp stored in the \mbox{MBJD\_OBS} keyword 
of the FITS header, which corresponds to the start of the first image in each set of 64.  We assume
uniform spacing of the images over the time period defined by the \mbox{AINTBEG} 
and \mbox{ATIMMEEND} header keywords.  The image spacing is roughly equal to the 0.4 s exposure time
selected for the 3.6, 4.5, and 5.8~$\mu$m observations of HAT-P-2 ($K_{mag}=7.60$).  For the 8.0~$\mu$m full array
observations, the \mbox{MBJD\_OBS}, \mbox{AINTBEG}, and \mbox{ATIMMEEND} header keywords are used to calculate 
the \mbox{BJD\_UTC} at mid-exposure for each image, which are spaced by roughly the 12.0 s exposure time.
In order to convert from UTC to TT timing standards as suggested by \citet{eas10}, 66.184 s would be added to 
our \mbox{BJD\_UTC} values.  We report all of our timing measurements using \mbox{BJD\_UTC} for consistency with other 
studies. 

\subsection{3.6 and 4.5~$\mu$m Photometry}\label{phot12}
We determine the background level in each image from the region outside of a ten pixel radius 
from the central pixel. This minimizes contributions from the wings of the star's 
point spread function while still retaining a substantial statistical sample.  We iteratively 
trim 3$\sigma$ outliers from the background pixels
then fit a Gaussian to a histogram of the remaining pixel values to estimate a sky value.
The background flux is 0.6\%  and 0.2\% of the total flux in the science aperture for 3.6 and 
4.5~$\mu$m observations respectively.  We correct for transient hot pixels by flagging pixels more than 
4.5$\sigma$ away from the median flux at a given pixel position across each set of 64 images then replacing 
flagged pixels by their corresponding position median value. 

Several different methods exist to determine the stellar centroid on the {\it Spitzer} array such as 
flux-weighted centroiding \citep[e.g.][]{knu08}, Gaussian centroiding, and least asymmetry methods \citep[e.g.][]{ste10, agol10}.  
We compare photometry calculated using both Gaussian and flux-weighted centroiding estimates and  
find that for the HAT-P-2 data flux-weighted centroiding produces the smallest scatter in the final light 
curve solutions.  This is in contrast with the work by \citet{ste10} and \citet{agol10}, which advocate 
Gaussian fits and least asymmetry methods to determine the stellar centroid for observations lasting less than 10~hours.   We find that 
flux-weighted centroids give more stable position estimates over long periods, especially if the stellar 
centroid crosses a pixel boundary during the duration of the observation.
For these data, we calculate the flux-weighted centroid for each 
background subtracted image using a range of aperture sizes from 2.0 to 5.0 pixels in 0.5 pixel increments.  We find that the stellar centroid 
aperture sizes that best reduce the scatter in the final time series are 4.5 and 3.5 pixels for the 3.6~$\mu$m and 4.5~$\mu$m 
observations respectively.

We estimate the stellar flux from each background subtracted image using circular aperture photometry.  A range of 
aperture sizes from 2.0 to 5.0 pixels in 0.25 pixel increments were tested to determine the optimal aperture size for 
each observation.  Additionally, we tested time-varying aperture sizes based on the noise pixel parameter described in the Appendix 
We find that we obtain the lowest standard deviation of the residuals from our best-fit solution at 3.6 $\mu$m using a variable aperture size 
given by the square-root of the measured noise pixel value for each image (median aperture size of 2.4 pixels). 
For the 4.5 $\mu$m observations we find that a fixed 2.25 pixel aperture gives the lowest scatter in the final solution. 
 We remove outliers from our final photometric data sets by discarding 
points more than 4.5$\sigma$ away from a moving boxcar median 16 points wide.  We find that using a narrower median filter or 
larger value for the outlier cutoff will miss some of the significant outliers.  We also find that using a wider median filter or smaller 
value for the outlier cutoff will often selectively trim the points at the top and bottom of the `saw-tooth' pattern seen in the resulting 
photometry for the 3.6 and 4.5~$\mu$m observations (Figures~\ref{ch1_raw} and \ref{ch2_raw}), 
which is the result of intrapixel sensitivity variations discussed in the Appendix. 

\begin{figure}
\centering
  \includegraphics[width=0.5\textwidth]{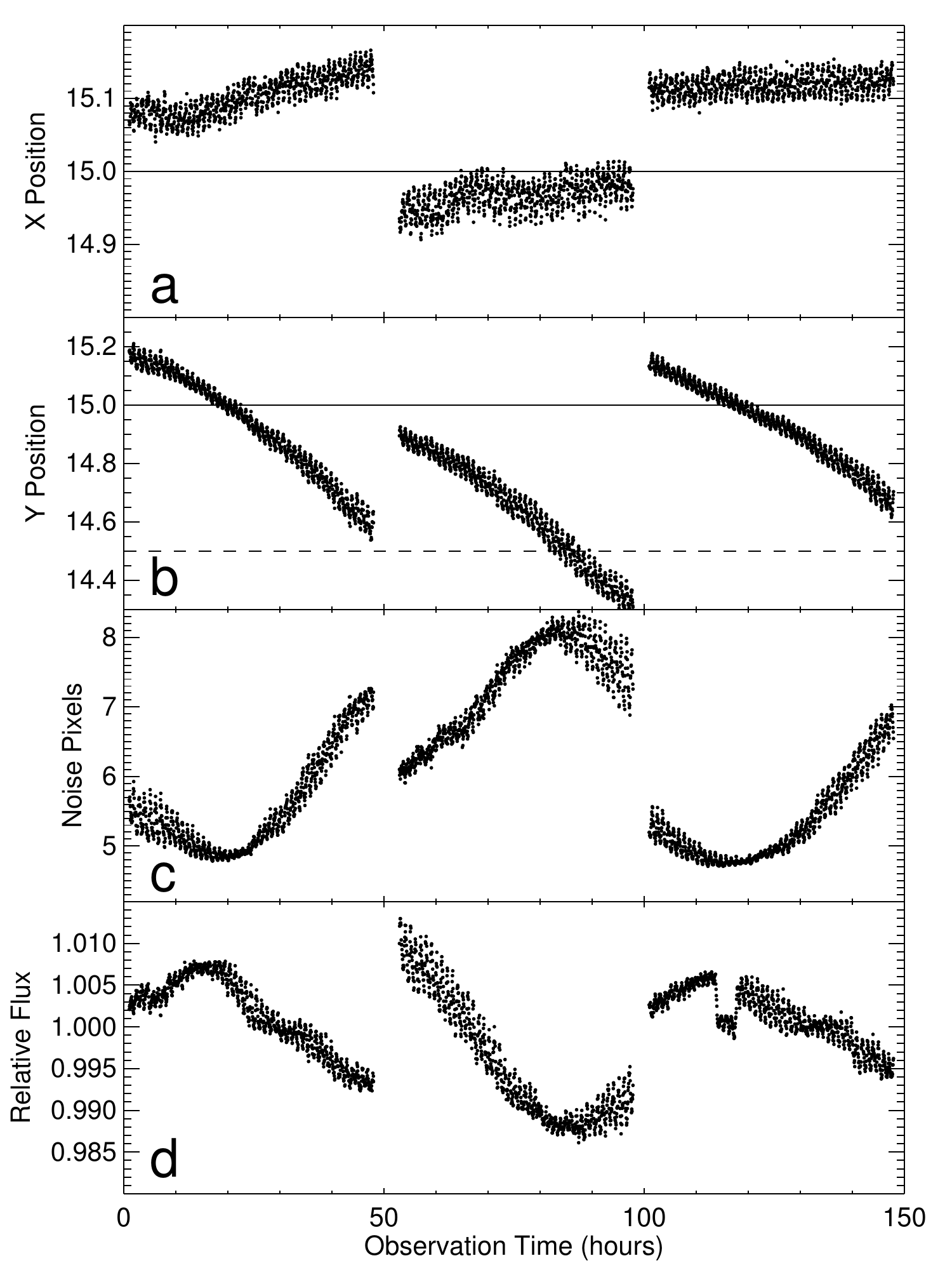}
  \caption{Measured $x$ (a), $y$ (b), noise pixels (c), and raw photometry (d) 
  as a function of time from the start of observation for the 3.6 $\mu$m phase 
  curve observations.  Data have been binned into three minute intervals.  
  In panels a and b solid horizontal lines indicate the pixel center while dashed 
  horizontal lines indicate a pixel edge.  Gaps
  in the data are due to spacecraft downlinks.  Jumps in position, noise pixels, and 
  relative flux after each downlink period are the result of stellar reacquisition.  
  The pointing oscillations and long-term drift in the $y$ direction are due to well-known
  instrumental effects, and are present in all {\it Spitzer} observations.}\label{ch1_raw}
\end{figure}

\begin{figure}
\centering
  \includegraphics[width=0.5\textwidth]{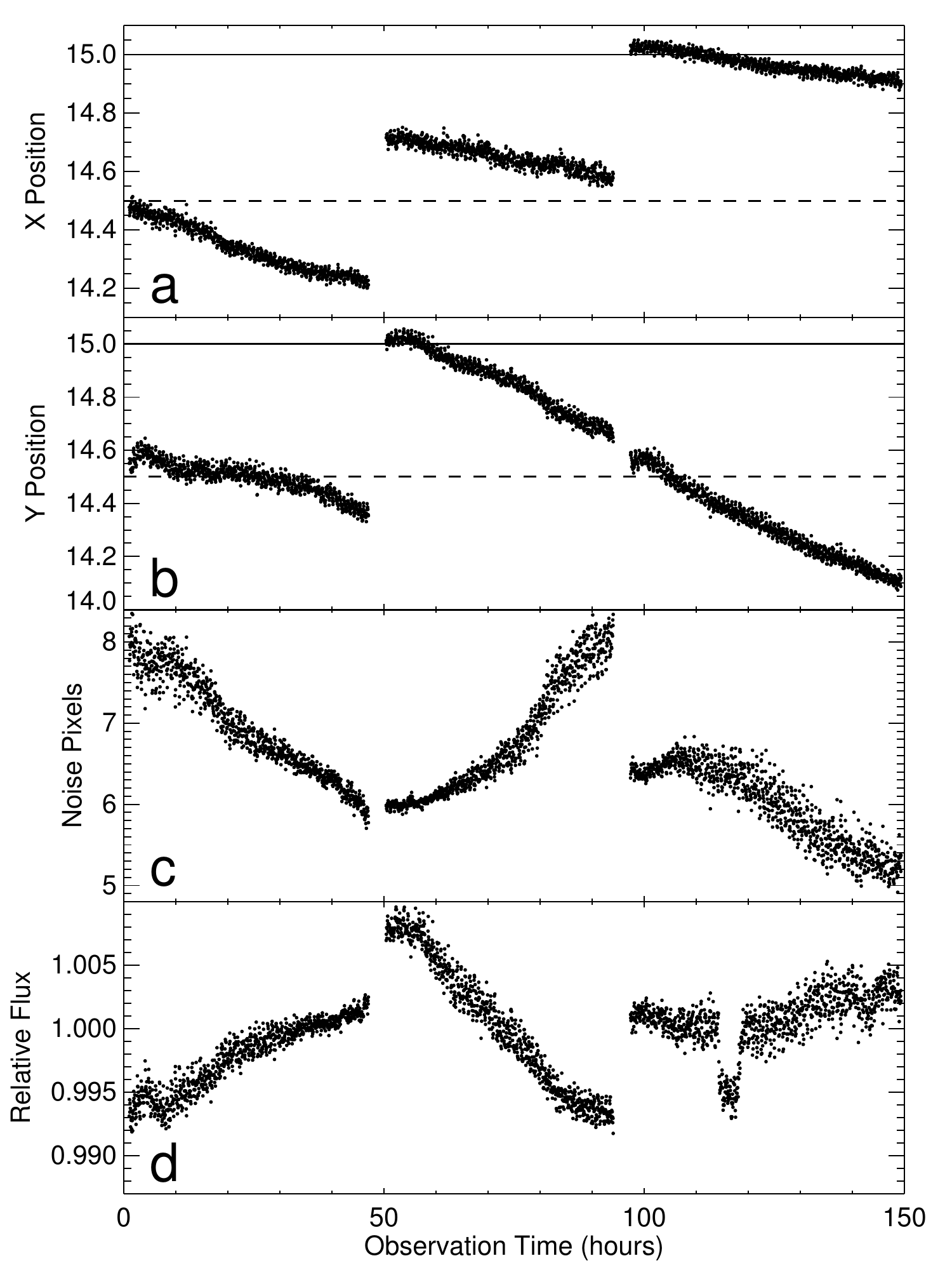}
  \caption{Measured $x$ (a), $y$ (b), noise pixels (c), and raw photometry (d) 
  as a function of time from the start of observation for the 4.5 $\mu$m phase 
  curve observations.  Data have been binned into three minute intervals.  
  In panels a and b solid horizontal lines indicate the pixel center while dashed 
  horizontal lines indicate a pixel edge;  
  see Figure \ref{ch1_raw} for a complete description.}\label{ch2_raw}
\end{figure}

\subsection{5.8 $\mu$m Photometry}

For our 5.8~$\mu$m data set, we determine the background level in each image using the same methodology as 
presented for the 3.6 and 4.5~$\mu$m data sets (\S\ref{phot12}).  The background flux is 0.6\% of the total flux in the science aperture.  
We tested both flux-weighted and Gaussian fits to the image methods of determining the stellar centroid.  For the Gaussian 
fits to the image we first fit the image allowing both the x and y width of the Gaussian to vary.  We then refit the same data set using a symmetric 
fixed width for the Gaussian that is the mean of the previous x and y widths.   We find that 
Gaussian fits to the image give a lower standard deviation of the residuals in the final fits compared with flux-weighted centroids.  This 
preference for the Gaussian centroiding method is likely the result of the short-time scale of these observations compared with the 
3.6 and 4.5~$\mu$m observations and the less than 0.1 pixel change in the stellar centroid position over the course of the observation.  

\begin{figure}
\centering
  \includegraphics[width=0.5\textwidth]{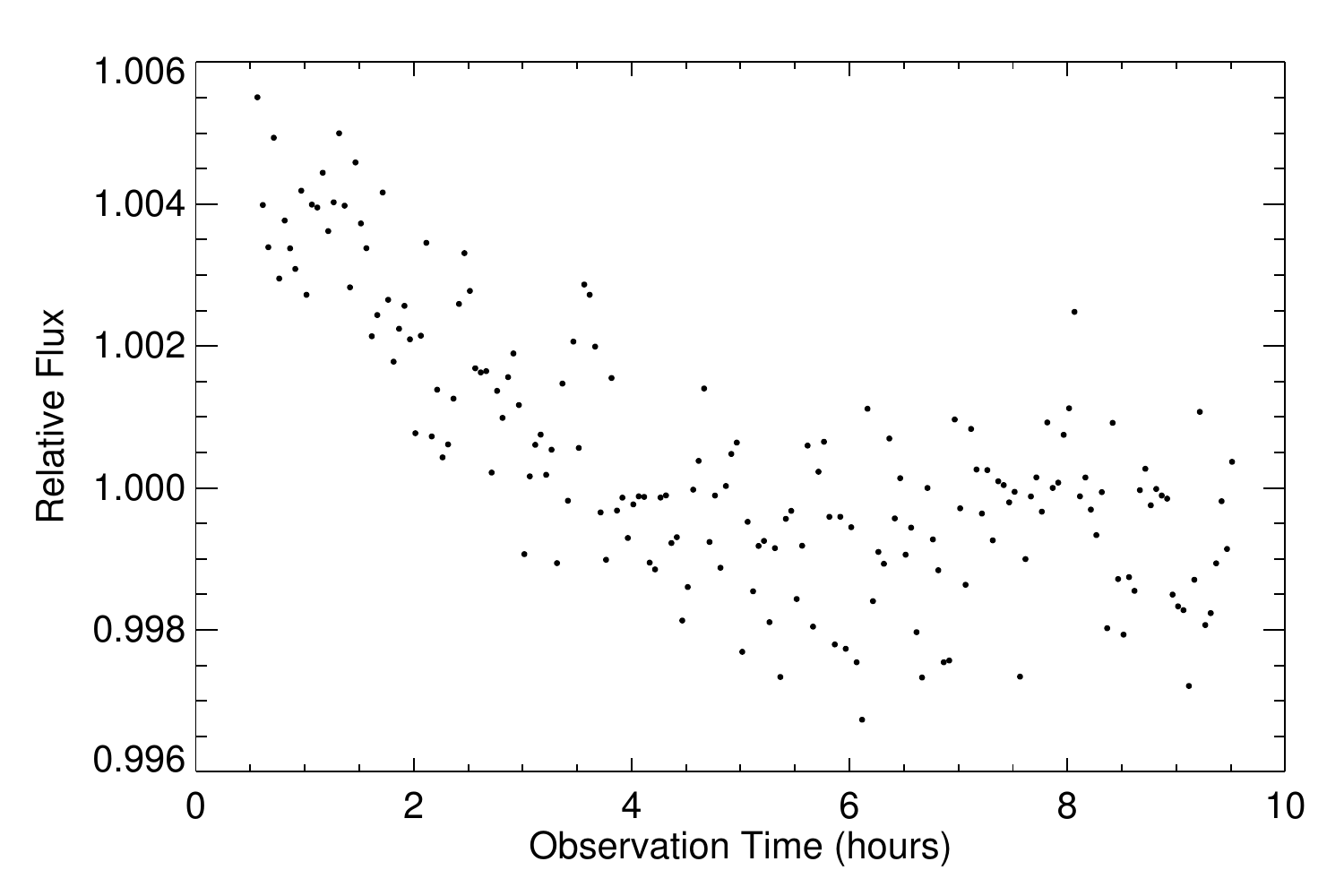}
  \caption{Raw photometry as a function of time from the start of observation for the 5.8 $\mu$m 
  secondary eclipse observation.  Data have been binned into three minute intervals.  Note 
  the downward slope in the relative flux values as a function of time.}\label{ch3_raw}
\end{figure}

We estimate the stellar flux from each background subtracted image using circular aperture photometry for a range 
of apertures sizes from 2.0 to 5.0 pixels in 0.25 pixel increments.  We find that an aperture size of 2.5 pixels gives the 
lowest standard deviation of the residuals in the final fits.  We remove outliers in our data set more than 3.0$\sigma$ away from a moving 
boxcar median 50 points wide.  Figure~\ref{ch3_raw} show our resulting photometry for the 5.8~$\mu$m observations.  
We test for possible intrapixel sensitivity variations using the methodology discussed in the Appendix, 
but find that including intrapixel sensitivity corrections actually degraded the precision of the final solution.

\subsection{8.0 $\mu$m Photometry}

Unlike the 3.6, 4.5, and 5.8~$\mu$m data, the 8.0~$\mu$m data 
were observed in the full-array (256$\times$256 pixels) mode of the IRAC instrument.  
Our photometry determines the flux from the star in a circular aperture 
of variable radius, as well as the background thermal emission surrounding the 
star.  We calculate two values of the background, using different spatial regions of the frame.  First, we 
calculate the background that applies to the entire frame.  Second, we isolate an annulus between 
6 and 35 pixels from the star.  In each region (entire frame, or annulus) we calculate the histogram of intensity 
values from the pixels within the region, and we fit a Gaussian to that histogram.  We use the centroid of that 
Gaussian as the adopted background value.  This method has the advantage of being 
insensitive to `hot' or otherwise discrepant values.  We find that 
using the background values calculated from the region near the star, 
as opposed to the entire frame, produces a lower scatter in the 
residuals in our final solution.

\begin{figure}
\centering
  \includegraphics[width=0.5\textwidth]{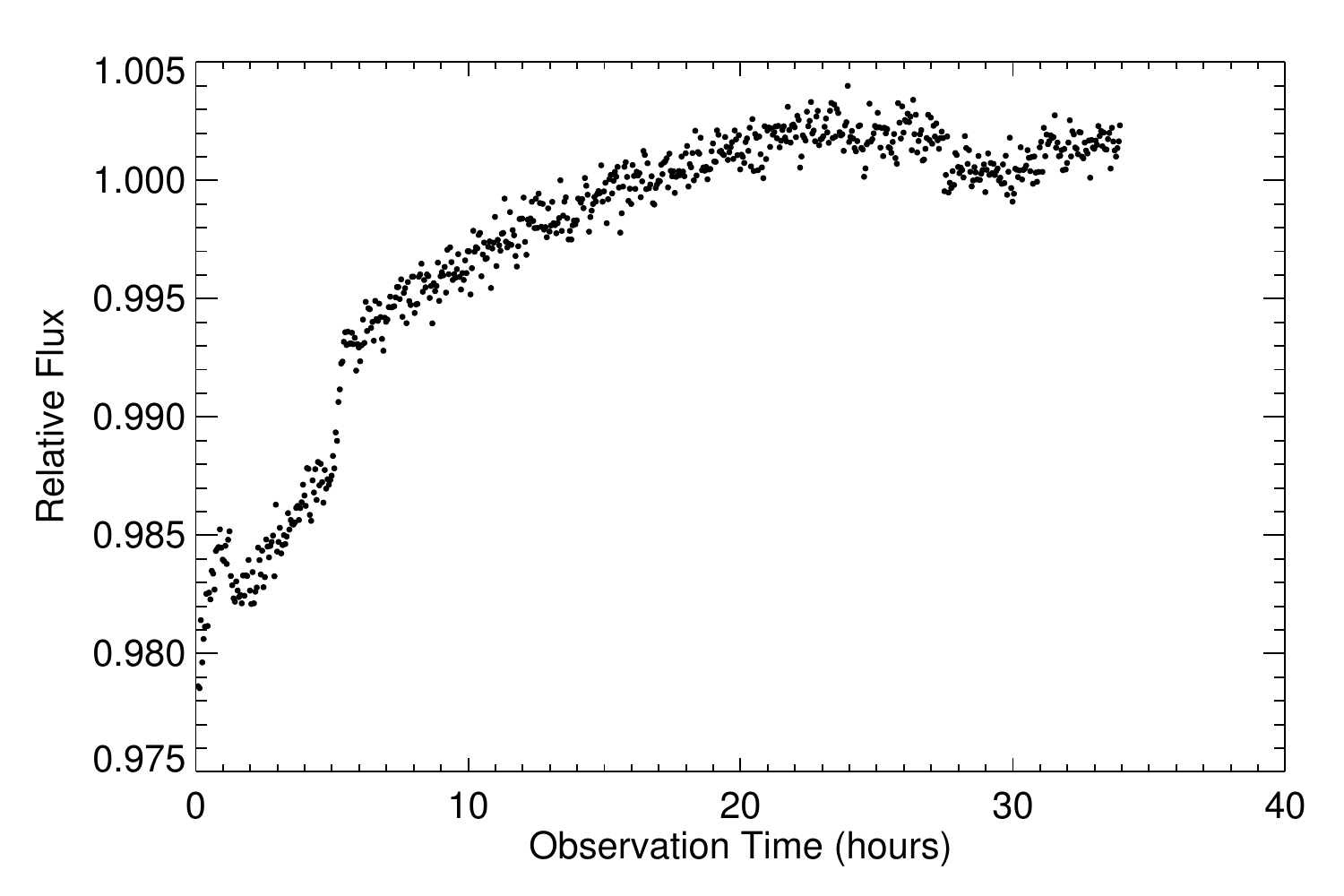}
  \caption{Raw photometry as a function of time from the start of observation for the 8.0 $\mu$m 
  partial orbit phase curve observation.  Data have been binned into three minute intervals.  Note 
  the strong exponential growth of the relative flux values as a function of time.}\label{ch4_raw}
\end{figure}

We locate the center of the star by fitting a 2-D Gaussian to a 3x3-pixel median filtered 
version of the frame; we find that this method produces slightly better photometric precision 
than using a center-of-light algorithm.
We estimate the stellar flux from each background subtracted image using circular aperture photometry for a range 
of apertures sizes from 2.0 to 5.0 pixels in 0.25 pixel increments.  We find that an aperture size of 3.5 pixels gives the 
lowest standard deviation of the residuals in the final fits.  We remove outliers in our data set more than 3.0$\sigma$ away from a moving 
boxcar median 50 points wide.  Figure~\ref{ch4_raw} show our resulting photometry for the 8.0~$\mu$m observations. 
We test for possible intrapixel sensitivity variations using the methodology discussed in the Appendix, 
but find that including intrapixel sensitivity corrections actually degraded the precision of the final solution.

\subsection{Flux Ramp Correction}\label{ramp_fit}

Our data exhibit a ramp-like increase, or decrease, in the flux values at the start of each observation and after 
each break in the 3.6 and 4.5~$\mu$m data for downlink.  This ramp in the flux values has been previously noted for 
IRAC 8.0~$\mu$m observations (e.g. \citet{knu07, knu09a}; \citet{agol10}) as well as 3.6 and 4.5~$\mu$m observations (e.g. \citet{bee11}; \citet{tod12}).
While the precise nature of this ramp is not well-understood, it can be at least partially attributed to thermal settling of the 
telescope at a new pointing, which contributes an additional drift in position.  However, we see that the ramp often persists beyond this initial drift in 
position, and we therefore speculate that there is an additional component analogous to the effect observed at long wavelengths due to 
charge-trapping in the array.
We tested several functional forms to describe this ramp, 
including quadratic, logarithmic, and linear functions of time, but find that functional form that gives us the best correction is given 
by the formulation presented in \citet{agol10}:
\begin{equation}\label{ramp_eq}
F'/F=1\pm a_1 e^{-t/a_2} \pm a_3 e^{-t/a_4}
\end{equation}
where $F'$ is the measured flux, $F$ is the stellar flux incident on the array, $t$ is the time from the start of the observation in days, 
and $a_1-a_4$ are free parameters in the fit.

We find that in the 3.6 and 4.5~$\mu$m data the asymptotic shape of the ramp converges on timescales less than an hour for 
most of the data segments.   Instead of including parameters in our fits for this ramp behavior, we instead elect to simply 
trim the first hour at the start of each observation and subsequent downlinks.  This reduces the complexity of our fits while 
avoiding possible correlations between the shape of the underlying phase curve of HAT-P-2b and the ramp function.  We find that 
the ramp persists beyond the one hour mark at the start of the 3.6~$\mu$m observations, which affects the shape of the first observed 
secondary eclipse and therefore include a ramp correction for that data segment.  For the 5.8~$\mu$m data, we find that the flux asymptotically 
decays from the start of the observation (Figure \ref{ch3_raw}).  This decrease in 
the 5.8~$\mu$m flux is also apparent in the derived background flux values and has been previously noted 
by studies such as \citet{ste11}.  As seen in Figure~\ref{ch4_raw}, the 8.0~$\mu$m 
data exhibit the well know asymptotic increase in flux \citep[e.g.][]{knu07,knu09a,agol10}.

In order to select the best functional form for the ramp correction in each data set we use the Bayesian Information Criterion (BIC), defined as 
\begin{equation}\label{bic}
BIC=\chi^2+k\ln(n)
\end{equation}
where $k$ is the degrees of freedom in the fit and $n$ is the total number of points in the fit \citep{lid07}.  This allows us to determine if we
are `over-fitting' the data by including additional parameters to describe the ramp correction in our fits.  We find that for the 3.6~$\mu$m 
data set using only a single exponential gives us the lowest BIC value.  
For the 8.0~$\mu$m data, using 
all the terms ($a_1-a_4$) in Equation~\ref{ramp_eq} with no trimming of the data near the start of the observation gives us the lowest BIC value.
For the 5.8~$\mu$m data, we find using a single decaying exponential gives the lowest BIC if we trim data within 30 minutes of the start of the observation.  
Trimming significantly more or less than than this amount ether gives a higher BIC or reduces the amount of out of eclipse data to less than the 
eclipse duration.

\subsection{Transit and Eclipse Fits}\label{trans_sec_fit}

We model our transit and eclipse events using the equations of \citet{man02} modified to account for 
the orbital eccentricity, $e$, and argument of periapse, $\omega$, of the HAT-P-2 system. For an eccentric 
system the normalized separation of the planet and star centers, $z$, is given by 
\begin{equation}\label{z0}
z=\frac{r(t)}{R_{\star}}\sqrt{1-(\sin{i}*\sin{(\omega+f(t))})^2}
\end{equation}
where $r(t)$ is the radial planet-star distance as a function of time, $R_*$ is the stellar radius, $i$ is the 
orbital inclination of the planet, and $f(t)$ is the true anomaly as a function of time.  The $r(t)/R_{\star}$ term in 
Equation \ref{z0} is calculated as 
\begin{equation}\label{rt}
\frac{r(t)}{R_{\star}}=\frac{a}{R_{\star}}\frac{1-e^2}{1+e\cos{(f(t))}}.
\end{equation}
The true anomaly angle ($f$) that appears in both Equations \ref{z0} and \ref{rt} is determined using 
Kepler's equation \citep{mur99} and is a function of $e$, $\omega$, and the orbital period of the planet, $P$.
Because of the degeneracies that exist between $e$, $\omega$ and $P$ in determining $f(t)$ we elect to not 
use $P$ as a free parameter in our fits.  Instead we fix $P$ to the value reported in \citet{pal10}, 5.6334729 d.
We further minimize correlations in our solutions for  
$e$ and $\omega$ by solving for the Lagrangian orbital elements $k\equiv e\cos\omega$ and $h\equiv e\sin\omega$.

In addition to the parameters that define the orbit of HAT-P-2b, we solve for the fractional planetary 
radius, $R_p/R_{\star}$.  Because $R_p/R_{\star}<0.1$ for HAT-P-2b \citep{pal10, bak07a}, there exists 
a strong correlation between $i$ and $a/R_{\star}$ in the transit solution \citep{win07a,pal08}.  We 
instead solve for the parameters $b^2$ and $\zeta/R_{\star}$ as suggested by \citet{bak07b} and \citet{pal10} 
where
\begin{equation}
b=\frac{a}{R_{\star}}\frac{1-e^2}{1+h}\cos i
\end{equation}  
and
\begin{equation}
\frac{\zeta}{R_{\star}}=\frac{a}{R_{\star}}\frac{2\pi}{P}\frac{1}{\sqrt{1-b^2}}\frac{1+h}{\sqrt{1-e^2}}.
\end{equation}
For the transit portion of the light curves we use four parameter nonlinear limb-darkening 
coefficients for each bandpass calculated by \citet{sin10}, where we assume a stellar atmosphere with $T_{eff}=6290$ K, $\log(g)=4.138$, 
and [Fe/H]=+0.14 \citep{pal10}.  For the secondary eclipse portion of the light curves we treat the planet as a 
uniform disk and scale the ingress and egress to match the amplitude of the phase curve (\S\ref{phase_fit}), which varies 
significantly over the duration of the eclipse and is therefore poorly approximated by a constant value.  
The secondary eclipse depth is defined by the average of the ingress and egress amplitudes.   

For the 3.6 and 4.5~$\mu$m datasets, we use nine free parameters to constrain the properties of the 
planetary orbit, transit, and secondary eclipse.  The 8.0~$\mu$m observations only include 
one secondary eclipse and therefore only require eight free parameters to constrain the same
system properties.  
The 5.6~$\mu$m data set includes only the secondary eclipse portion of the light curve.  We therefore elect to only allow the 
secondary eclipse depth and timing to vary for the 5.6~$\mu$m data and fix $a/R_{\star}$, $i$, $e$, $\omega$, 
and $R_p/R_{\star}$ to the average values from the 3.6, 4.5, and 8.0~$\mu$m data sets.  

\subsection{Phase Curve Fits}\label{phase_fit}

The functional form of the phase curve for a planet on an eccentric orbit is not well defined.  Unlike close-in, 
tidally locked planets on circular orbits, eccentric planets experience time-variable heating and 
non-synchronous rotation rates.  Previous studies by \citet{lan08} and \citet{cow11} have investigated 
theoretical light curves for planets on eccentric orbits using two-dimensional hydrodynamic simulations 
and semi-analytic model atmospheres respectively.  We also developed a three-dimensional 
atmospheric model for HAT-P-2b that couples radiative and dynamical processes to further investigate possible 
phase curve functional forms that will be presented in a future paper.  The functional forms for the phase curves described here 
all provide a reasonable fit to the theoretical light curves presented in \citet{lan08}, \citet{cow11}, and from our three-dimensional
atmospheric model.

To first order, the flux from the planet is proportional to the inverse square of the distance between the planet
and host star, $r(t)$.  This assumes that the planet has a constant albedo and responds instantaneously and uniformly to changes in the incident stellar flux.  
We know that there must exist a lag in the peak of the incident stellar flux and the peak of the planet's temperature 
since atmospheric radiative timescales are finite \citep{lan08, iro10, lew10, cow11}.  Our first functional form for the phase 
variation of HAT-P-2b is a simple $1/r(t)^2$ with a phase lag:
\begin{equation}\label{dist_sq}
F(f)=F_0+c_1(1+\cos(f-c_2))^2
\end{equation}
where $f$ is the true anomaly and $c_1-c_2$ are free parameters in the fit.  The $1+\cos(f-c_2)$ is a simplified 
form of the denominator of Equation \ref{rt} for $r(t)$.  We also test a simpler form of Equation \ref{dist_sq} given by
\begin{equation}\label{cosf}
F(f)=F_0+c_1\cos(f-c_2)
\end{equation}
where $c_1-c_2$ are free parameters in the fit.
This functional form of the phase curve is similar to a simple sine or cosine of the orbital phase angle, $\lambda$ or $\xi$, used 
in the case of a circular orbit.  

We also find that a Lorentzian function of time provides a reasonable representation of the expected shape 
of the orbital phase curve for HAT-P-2b.  This is not surprising given that we expect the flux from the planet to 
vary as $1/r(t)^2$ with a time lag between the minimum of the planetary distance and the peak of the planetary flux.  
We test both symmetric and asymmetric Lorentzian functions of the time from periapse passage, $t$, given by
\begin{equation}\label{lor}
F(t)=F_0+\frac{c_1}{u(t)^2+1}
\end{equation}
where 
\begin{equation}\label{lor_sym}
u(t)=(t-c_2)/c_3
\end{equation}
in the symmetric case and 
\begin{equation}\label{lor_asym}
u(t)=
\begin{cases} 
(t-c_2)/c_3 & \text{if $t<c_2$;}  \\ 
(t-c_2)/c_4 & \text{if $t>c_2$} 
\end{cases}
\end{equation}
in the asymmetric case.  In Equations \ref{lor}, \ref{lor_sym}, and \ref{lor_asym} $c_1-c_4$ are free parameters in 
the fit.  The Lorentzian functional form for the phase curve is especially useful since the $c_2$ parameter gives the offset 
between the time of periapse passage and the peak of the planet's flux and the $c_3-c_4$ parameters gives an estimate 
of relevant atmospheric timescales. 

We also find that the preferred functional form for the phase curve of planets on circular orbits 
described in \citet{cow08} provides a reasonable fit to our theoretical light curves if the orbital phase angle, $\xi$, is replaced 
by the true anomaly, $f$.  In this case 
\begin{eqnarray}\label{cowan_eq}
F(\theta)=F_0+c_1\cos(\theta)+c_2\sin(\theta)\\ \nonumber
+c_3\cos(2\theta)+c_4\sin(2\theta)
\end{eqnarray}
where $c_1-c_4$ are free parameters in the fit and $\theta=f+\omega+\pi$ such that transit occurs at $\theta=-\pi/2$ and secondary eclipse
occurs at $\theta=\pi/2$.  

In addition to the functional forms for the phase curve presented above, we also 
test a flat phase curve, $F(t)=F_0$, to make sure that we are not over-fitting the data with the phase curve parameters.
In all cases we tie the $F_0$ parameter to the secondary eclipse depth to give the appropriate 
average value of the phase curve during secondary eclipse.  We also set any portion of the phase curve that falls below the 
secondary eclipse depth to zero.  During secondary eclipse we no longer see flux from the planet, only the star, therefore 
the combined star and planetary flux should always be greater than or equal to the flux at the bottom of the secondary eclipse.  

We choose the optimal phase curve solution to be the one that 
gives the lowest BIC value (Equation~\ref{bic}) using a Levenberg-Marguardt 
non-linear least squares fit to the data \citep{mar09}.  For the 3.6 and 4.5~$\mu$m data sets, we also compare the BIC values from each of the three data segments
separated by the downlink periods to make sure that the solution is robust at all orbital phases.  For the 3.6, 4.5, and 8.0~$\mu$m data sets, 
we find that all of the functional forms presented Section \ref{phase_fit} provide a significant improvement in the BIC values 
over the `flat' phase curve BIC values (BIC$_{3.6~\mu\mathrm{m}}$=1,282,575; BIC$_{4.5~\mu\mathrm{m}}$=1,459,141; BIC$_{8.0~\mu\mathrm{m}}$=11,183). 
For the 3.6~$\mu$m 
data we obtain the lowest BIC value with a functional form for the phase curve defined by either Equation~\ref{cosf} 
or Equation~\ref{cowan_eq} with $c_3-c_4$ fixed at zero (BIC=1,278,521).  This is not surprising since basic 
trigonometric identities make Equation~\ref{cowan_eq} equivalent to Equation~\ref{cosf} if $c_3-c_4$ can be 
assumed to be zero.  For the 4.5 $\mu$m data we obtain the lowest BIC value with a functional form for the 
phase curve given by Equation~\ref{cowan_eq} with the $c_1$ and $c_3$ terms fixed at zero (BIC=1,458,842).

The phase curve model given by Equation~\ref{cowan_eq} also gives us a reasonable improvement in the BIC 
for our 8.0~$\mu$m data set.  However, we find that the second harmonic in Equation~\ref{cowan_eq} is highly degenerate with 
our ramp correction at 8.0~$\mu$m such that we cannot reliably constrain the significance of this harmonic in the fit.  We instead use the 
phase curve functional form given by Equation~\ref{dist_sq}, 
which gives us the lowest BIC for the 8.0~$\mu$m data set (BIC=10,945).  For the 5.8~$\mu$m data, which only span ten hours, we find no 
statistical difference between solutions with and without parameterizations for planetary phase variations. 

\subsection{Stellar Variability}
HAT-P-2 is a rapidly rotating F star ($v\sin i$=20.7~km~s$^{-1}$).   Measurements of 
the Ca II H \& K lines of HAT-P-2 by \citet{knu10} do not detect any significant emission in 
the line cores, which would suggest a chromospherically quiet stellar host for HAT-P-2b.  However, 
this indicator provides relatively weak constraints on chromospheric activity for F stars, which 
have strong continuum emission in the wavelengths of the Ca II H \& K lines.  
Ground-based monitoring of HAT-P-2 in the Str\"{o}mgren $b$ and $y$ bands over a period of more 
than a year indicates that it varies by less than 0.13\% at visible wavelengths.  We would expect the 
star to vary by substantially less than this amount in the infrared, where the flux contrast of spots and 
other effects is correspondingly reduced.  
These observations rule out both periodic variability that 
could be associated with the rotation rate of HAT-P-2 or longer-term trends (G. Henry 2010, private communication).  
Previous observations find the rotation rate of the star to be on the order of 3.8~d \citep{win07} based on 
the line-of-sight stellar rotation velocity ($v\sin i_{\star}$) and assuming $\sin i_{\star}=1$, which is roughly 0.6 times the orbital 
period of HAT-P-2b.  We do not expect stellar variability to be significant on the timescales of our observations, 
but for the sake of completeness we also check this assumption directly using our {\it Spitzer} data.

We employ two models for stellar variability.   The first model is a simple linear function of time given by 
\begin{equation}
F_0(t) = d_1 t
\end{equation}
where $t$ is measured from the predicted center of transit in each observation and $d_1$ is a free parameter in the fit.  The second model 
we test has the form
\begin{equation}\label{sine_star}
F_0(t)=d_1\sin((2\pi/d_2)t-d_3)
\end{equation}
where $t$ is measured from the predicted center of transit in each observation and $d_1-d_3$ are free parameters in the fit.  Equation \ref{sine_star} 
attempts to capture stellar variability that is associated with star spots that rotate in and out of view with the $d_2$ parameter representing the rotation 
rate of the star.  

We find that the linear model for stellar variability does not improve the $\chi^2$ of the fits significantly and in fact increases the 
BIC (Equation~\ref{bic}).  Inclusion of the stellar model given by Equation \ref{sine_star} 
does improve both the $\chi^2$ and BIC in our fits.  
However, we find that the solutions for the `sine-curve' model for stellar variability are often degenerate with our models 
for the phase curve and the residual ramp at the start of the 3.6 $\mu$m observations.  We also find that the stellar rotation 
rate predicted by the $d_2$ term in Equation \ref{sine_star} differs significantly between the 3.6 and 4.5~$\mu$m observations
with a rotation rate of 4.3~d preferred for the 3.6~$\mu$m data and 3.9~d preferred for the 4.5~$\mu$m data.  Although these 
predicted rotation periods are near to the expected 3.8~d rotation period of HAT-P-2, the amplitude of the 
predicted stellar variations in the mid-infrared seem spurious.  The amplitude of the predicted stellar flux variations at 3.6 and 4.5~$\mu$m 
are on the order of $\sim0.1\%$, which is comparable to our upper limit on stellar variability at visible wavelengths ($0.13\%$).
We would expect star spots to display much larger variations in the visible than at mid-infrared wavelengths.  The amplitude of these 
stellar variations are also similar to the amplitude of our predicted phase variations.  We therefore conclude that our best-fit 
solutions for the stellar variability are physically implausible, and likely the result from a degeneracy with the other terms 
in our fits.  As we have no convincing evidence for stellar variability, we assume that the star's flux is constant in our final fits.

\subsection{Radial Velocity Measurements}

Since the discovery of HAT-P-2b by \citet{bak07a}, 
the California Planet Search (CPS) team has continued to obtain regular radial velocity (RV) measurements of this system
using the HIRES instrument \citep{vogt94} on Keck.  
We used the CPS pipeline \citep[see, e.g.][]{how09} to measure precise RVs from the high-resolution 
spectra of HAT-P-2 using a superposed molecular iodine spectrum as a Doppler reference and point spread function calibrant.
Previous studies of this system \citep{bak07a, win07, pal10} have 
included HIRES radial velocity measurements through May 2008 (BJD = 2454603.932112).
Here we present 16 additional RV measurements from the HIRES 
instrument on Keck for a total of 71 data points spanning 6 years (Table~\ref{rv_data}).  We exclude measurements obtained during transit 
\citep{win07}  from our fits in order to avoid measurements affected by the Rossiter-McLaughlin effect.  

\begin{deluxetable}{lcc}
\tabletypesize{\scriptsize}
\tablecaption{Radial Velocities for HAT-P-2b from Keck\label{rv_data}}
\tablewidth{0.5\textwidth}
\tablehead{
\colhead{BJD - 2400000} & \colhead{RV} & \colhead{Uncertainty}\\
\colhead{(days)} & \colhead{(m s$^{-1}$)} & \colhead{(m s$^{-1}$)}}
\startdata
        53981.777500  &      -19.17  &        7.46  \\
        53982.871700  &     -310.87  &        7.64  \\
        53983.814865  &      538.73  &        7.81  \\
        53984.894980  &      855.62  &        7.94  \\
        54023.691519  &      698.64  &        7.94  \\
        54186.998252  &      696.78  &        7.72  \\
        54187.104158  &      684.36  &        6.84  \\
        54187.159878  &      717.17  &        6.59  \\
        54188.016885  &      757.70  &        7.00  \\
        54188.159622  &      774.61  &        6.80  \\
        54189.010378  &      651.86  &        6.59  \\
        54189.088911  &      635.16  &        6.48  \\
        54189.157721  &      619.70  &        7.22  \\
        54216.959395  &      722.59  &        8.04  \\
        54257.756431  &       27.91  &        5.79  \\
        54257.758677  &       35.49  &        6.17  \\
        54257.760702  &       24.33  &        6.18  \\
        54257.794116  &      -11.70  &        5.48  \\
        54257.796779  &      -19.01  &        5.02  \\
        54257.799452  &      -20.94  &        5.10  \\
        54257.802148  &      -23.48  &        4.74  \\
        54257.804926  &      -21.89  &        5.15  \\
        54257.807634  &      -35.65  &        4.91  \\
        54257.810342  &      -24.40  &        5.08  \\
        54257.813143  &      -40.82  &        5.22  \\
        54257.815817  &      -37.08  &        5.45  \\
        54257.818490  &      -23.98  &        5.21  \\
        54258.024146  &     -313.00  &        4.87  \\
        54258.027167  &     -310.40  &        4.89  \\
        54258.030292  &     -311.56  &        4.37  \\
        54258.033278  &     -326.04  &        4.58  \\
        54258.042630  &     -314.19  &        5.10  \\
        54258.045488  &     -336.27  &        4.92  \\
        54258.048393  &     -342.10  &        5.07  \\
        54258.051483  &     -351.92  &        4.88  \\
        54258.054828  &     -356.80  &        4.94  \\
        54258.058161  &     -356.72  &        4.65  \\
        54258.061472  &     -360.49  &        4.85  \\
        54258.064666  &     -374.47  &        4.73  \\
        54258.099110  &     -432.29  &        5.27  \\
        54258.102836  &     -433.53  &        4.97  \\
        54258.106679  &     -446.91  &        4.74  \\
        54279.876893  &      371.51  &        8.26  \\
        54285.823854  &      137.15  &        5.91  \\
        54294.878702  &      728.72  &        6.61  \\
        54304.864982  &      588.92  &        6.07  \\
        54305.870122  &      737.55  &        6.23  \\
        54306.865216  &      731.39  &        7.95  \\
        54307.912379  &      456.62  &        6.39  \\
        54335.812619  &      549.60  &        6.64  \\
        54546.098175  &     -684.07  &        7.79  \\
        54547.115700  &      536.15  &        7.19  \\
        54549.050468  &      754.51  &        7.26  \\
        54602.916550  &      271.30  &        6.47  \\
        54603.932112  &      662.48  &        5.74  \\
        54839.166895  &     -369.59  &        8.09  \\
        55015.871081  &      679.68  &        8.89  \\
        55350.945695  &     -513.11  &        8.45  \\
        55465.742817  &      606.08  &        6.99  \\
        55703.872995  &      623.50  &        7.70  \\
        55704.836273  &      377.02  &        8.02  \\
        55705.853346  &     -535.77  &        8.12  \\
        55706.831645  &     -257.19  &        7.57  \\
        55707.844353  &      504.50  &        7.68  \\
        55808.759715  &      323.97  &        8.49  \\
        55850.695496  &      557.65  &        7.37  \\
        55932.161110  &     -241.35  &        9.48  \\
        55945.128011  &      595.28  &        8.57  \\
        55992.018498  &      401.30  &       10.08  \\
        56147.753065  &      553.56  &        7.95  \\
        56149.738596  &      381.71  &        8.37  \\
\enddata
\end{deluxetable}

We simultaneously fit for the RV semi-amplitude ($K$), zero-point ($\gamma$), and long term velocity drift ($\dot{\gamma}$) along with the orbital, transit, eclipse, 
and phase parameters in the 3.6, 4.5, and 8.0~$\mu$m data sets separately.  We also test for possible curvature in the RV signal ($\ddot{\gamma}$), but find that our derived 
values for $\ddot{\gamma}$ were consistent with zero.
As discussed in \citet{win10},  the transit to secondary eclipse timing strongly constrains the 
$e\cos\omega$ term in our fits, but the $e\sin\omega$ term is better constrained by the inclusion of these RV data.  
Rapidly rotating stars are known to have an increased scatter in their RV velocity distribution beyond the reported 
internal errors \citep[as discussed in][]{bak07a, win07}.
We therefore estimate a stellar jitter term ($\sigma_{\rm jitter}$) as described in Section~\ref{hat2_results}.

\section{Results}\label{hat2_results}

\begin{deluxetable*}{lcccc}
\tabletypesize{\scriptsize}
\tablecaption{Global Fit Parameters\label{hat2_fits}}
\tablewidth{0pt}
\tablehead{
\colhead{Parameter} & \colhead{3.6 $\mu$m} & \colhead{4.5 $\mu$m} &  \colhead{5.8 $\mu$m} & \colhead{8.0 $\mu$m}}
\startdata
\emph{Transit Parameters} & & & \\
$R_p/R_{\star}$ & 0.06821$\pm$0.00075  &  0.07041$\pm$0.00060 &  0.06933\tablenotemark{d} & 0.0668$\pm$0.0016 \\
$b^2$ & 0.122$^{+0.066}_{-0.078}$ &  0.345$\pm$0.042   &  .... &  0.238$^{+0.096}_{-0.131}$ \\
$i$ ($^{\circ}$) &  87.37$^{+1.34}_{-0.81}$ &  84.91$\pm$0.47 & 85.97\tablenotemark{d} &  86.0$^{+1.5}_{-1.0}$ \\
$a/R_{\star}$  & 9.53$\pm$0.35 &  8.28$\pm$0.24  & 8.70\tablenotemark{d} & 8.83$^{+0.67}_{-0.53}$  \\
$T_c$ (BJD-2400000)\tablenotemark{a} & 55288.84988$\pm$0.00060 &   55756.42520$\pm$0.00067 & ... &  54353.6911$\pm$0.0012\\
$\zeta/R_{\star}$ (d$^{-1}$) & 12.221$\pm$0.058 &  12.286$\pm$0.057   & ... &  12.21$\pm$0.11  \\
$T_{14}$ (d)\tablenotemark{b} & 0.1770$\pm$0.0011 &  0.1813$\pm$0.0013 & ...  & 0.1789$\pm$0.0023 \\   
$T_{12}$ (d)\tablenotemark{b} & 0.0128$\pm$0.0010 &  0.0177$\pm$0.0012 & ...  & 0.0144$\pm$0.0021\\  
&&&&\\
\emph{Secondary Eclipse Parameters} & & & \\
$\phi_{sec}$  & 0.19177$\pm$0.00029 & 0.19310$\pm$0.00025  & ...  &  0.19253$\pm$0.00048 \\
$T_{14}$ (d)\tablenotemark{b} & 0.1550$\pm$0.0027 & 0.1651$\pm$0.0023 &  & 0.1610$\pm$0.0043 \\
$T_{12}$ (d)\tablenotemark{b} & 0.01090$\pm$0.00075 &  0.01444$\pm$0.00083 &  &  0.0121$\pm$0.0016 \\
1$^{st}$ Eclipse Depth & 0.080\%$\pm$0.012\% &  0.1009\%$\pm$0.0084\%  &  ... & ... \\
$T_c$ (BJD-2400000)\tablenotemark{a} & 55284.2967$\pm$0.0014 & 55751.8795$\pm$0.0011 & ... & ... \\
2$^{nd}$ Eclipse Depth & 0.0993\%$\pm$0.0090\% & 0.1057\%$\pm$0.0090\%  & 0.071\%$^{+0.029\%}_{-0.013\%}$  & 0.1392\%$\pm$0.0095\% \\
$T_c$ (BJD-2400000)\tablenotemark{a} & 55289.9302$\pm$0.0014 & 55757.5130$\pm$0.0011  &   54906.8561$^{+0.0076}_{-0.0062}$  &  54354.7757$\pm$0.0022 \\
&&&&\\
\emph{Orbital \& RV Parameters} & & & \\ 
$e\cos\omega$ & -0.50539$\pm$0.00057 & -0.50301$\pm$0.00051  & ...  & -0.50419$\pm$0.00088 \\
$e\sin\omega$ & -0.0741$\pm$0.0072 & -0.0729$\pm$0.0054 & ... &  -0.0685$\pm$ 0.0059 \\
$e$ & 0.51081$\pm$0.00092 &  0.50829$\pm$0.00068 & 0.50910\tablenotemark{d}  &  0.50885$\pm$0.00097 \\
$\omega$ ($^{\circ}$) & 188.34$\pm$0.80 & 188.25$\pm$0.62  & 188.09\tablenotemark{d}  &  187.74$\pm$0.66 \\
$T_p$ (BJD-2400000) & 55289.4734$\pm$0.0079 &  55757.05194$\pm$ 0.0061 & ...  &  54354.3109$\pm$0.0065 \\
$K$ (m s$^{-1}$) & 927.0$\pm$5.9 &  923.0$\pm$5.8  & ...  &  923.1$\pm$6.0 \\
$\gamma$ (m s$^{-1}$) & 247.4$\pm$3.7 &  248.0$\pm$3.6  &  ...  &  248.0$\pm$3.6  \\
$\dot{\gamma}$ (m s$^{-1}$ d$^{-1}$) & -0.0890$\pm$0.0050  & -0.0881$\pm$0.0049  & ... &  -0.0886$\pm$0.0058\\
&&&&\\
\emph{Phase Curve Parameters} & & &\\
Functional Form &  Eq. (\ref{cowan_eq})  & Eq. (\ref{cowan_eq})   &  Flat   & Eq. (\ref{dist_sq}) \\
$c_1$ & 0.0379\%$\pm$0.0015\% & 0 (fixed)                   &    ... &  0.0555\%$\pm$0.0032 \\
$c_2$ & 0.0422\%$\pm$0.0025\% &  0.0293\%$\pm$0.0042\%    &  ...   & 41.8$^{\circ}\pm$2.7$^{\circ}$  \\ 
$c_3$ & 0 (fixed)                                &  0 (fixed)                   &     &    ... \\
$c_4$ & 0 (fixed)                                &  0.0163\%$\pm$ 0.0035\%    &  ...   &    ...  \\
Amplitude & 0.114\%$\pm$0.010\% & 0.079\%$\pm$0.013\% & ...  & ... \\
Minimum Flux & 0.00014\%$^{+0.00927\%}_{-0.00014\%}$ & 0.0372\%$^{+0.0086\%}_{-0.0096\%}$  &    ...   &   ...\\
Minimum Flux Offset (h)\tablenotemark{c} & -18.36$^{+0.18}_{-1.00}$ & 6.71$\pm$0.43 & ...  &  ...  \\
Maximum Flux & 0.1139\%$\pm$0.0089\% &  0.1162\%$^{+0.0089\%}_{-0.0071\%}$  & ... &  0.1888\%$\pm$0.0072\%\\
Maximum Flux Offset (h)\tablenotemark{c} & 4.39$\pm$0.28 &  5.84$\pm$0.39  & ...  & 4.68$\pm$0.37 \\
&&&&\\
\emph{Ramp Parameters} & & &\\
Functional Form  & Eq. (\ref{ramp_eq}) & None & Eq. (\ref{ramp_eq}) & Eq. (\ref{ramp_eq})\\
$a_1$ & -0.00134$^{+0.00105}_{-0.00054}$ & ... &  +0.00683$^{+0.00086}_{-0.00054}$  &  -0.00537$^{+0.00088}_{-0.00069}$ \\  
$a_2$ & 0.078$^{+0.050}_{-0.029}$ & ... &  0.095$\pm$0.015   &     0.0195$^{+0.0067}_{-0.0051}$  \\  
$a_3$ &  0 (fixed) & ... &  ...   &  -0.01765$^{+0.00053}_{-0.00075}$ \\
$a_4$ &  0 (fixed) & ... &  ...   &  0.2812$^{+0.0155}_{-0.0091}$\\
&&&&\\
\emph{Noise Parameters} & & & & \\
$\sigma_{\rm phot}$ & 0.0042209$\pm$0.0000024  &  0.0057064$\pm$ 0.0000032 & 0.015380$\pm$0.000034 & 0.002164$\pm$0.000015   \\
$\sigma_{\rm jitter}$ (m s$^{-1}$)  & 26.0$\pm$2.1 & 25.7$\pm$2.2 &  ...    &    25.5$\pm$2.1 
\enddata
\tablenotetext{a}{We list all time in BJD$\_$UTC for consistency with other studies; to convert to BJD$\_$TT add 66.184 s.}
\tablenotetext{b}{$T_{14}$ is the total transit or eclipse duration. $T_{12}$ is the ingress duration, which equivalent to 
the egress duration ($T_{34}$) to within error.}
\tablenotetext{c}{Minimum flux offset is measured relative to the center of transit time ($T_c$). Maximum flux 
offset is measured relative to the time of periapse passage ($T_p$).} 
\tablenotetext{d}{Represents average value from 3.6~$\mu$m, 4.5~$\mu$m and 8~$\mu$m analyses.  Value held fixed in 5.8~$\mu$m analysis.}
\end{deluxetable*}

We perform a simultaneous fit and calculate uncertainties 
for the relevant transit, secondary eclipse, phase curve, radial velocity, flux ramp, and intrapixel 
sensitivity correction parameters in our data sets using a Markov Chain Monte Carlo (MCMC) 
method \citep{for05}.  For the 3.6, 4.5, and 8.0~$\mu$m data sets fit parameters include $b^2$, $\zeta/R_{\star}$, $e\cos\omega$, 
$e\sin\omega$, $R_p/R_{\star}$, transit time, the secondary eclipse depth(s), the phase function coefficients $c_1-c_4$, 
a photometric noise term ($\sigma_{phot}$), 
and RV parameters $K$, $\gamma$, $\dot{\gamma}$, $\sigma_{jitter}$.  For the 3.6 and 8.0~$\mu$m data sets we additionally fit 
for the ramp correction coefficients $a_1-a_4$.   The free parameters in the fit to the 5.8~$\mu$m data set 
are the eclipse time, eclipse depth,  ramp correction coefficients $a_1-a_4$, and $\sigma_{phot}$.
The value of the stellar 
flux, $F_0$, is inherently accounted for with our intrapixel sensitivity correction method described in the Appendix. 
Because we do not apply intrapixel sensitivity corrections to our 5.8 and 8.0~$\mu$m data sets we additionally 
fit for $F_0$ in those cases.  

The only parameter held fixed in our analysis is the orbital period ($P$), the value for 
which we take from \citet{pal10}.  All other orbital, planetary, phase, and ramp parameters are allowed to vary freely.  
Initial attempts at fitting just the 3.6 and 4.5~$\mu$m transits simultaneously produce a 
value for $P$ within 1$\sigma$ of the \citet{pal10} value, and we are therefore confident that adopting the 
\citet{pal10} value for $P$ has not introduced any significant errors into our analysis.

We initially attempted to fit for the wavelength independent parameters $b^2$, $\zeta/R_{\star}$, $e\cos\omega$, 
$e\sin\omega$, and $T_c$ simultaneously for the RV, 3.6, 4.5, and 8.0~$\mu$m data sets.   However, given 
the size of our data sets (over 2.5 million data points) and the time required to create our `pixel-maps' for the 
3.6 and 4.5~$\mu$m data (see Appendix),
we found it computationally infeasible.  It is possible that further improvements to our analysis code, including 
parallelization, could make the problem 
more computationally tractable.  Such improvements are left for future iterations of our analysis methods.  

\begin{figure}
\centering
  \includegraphics[width=0.5\textwidth]{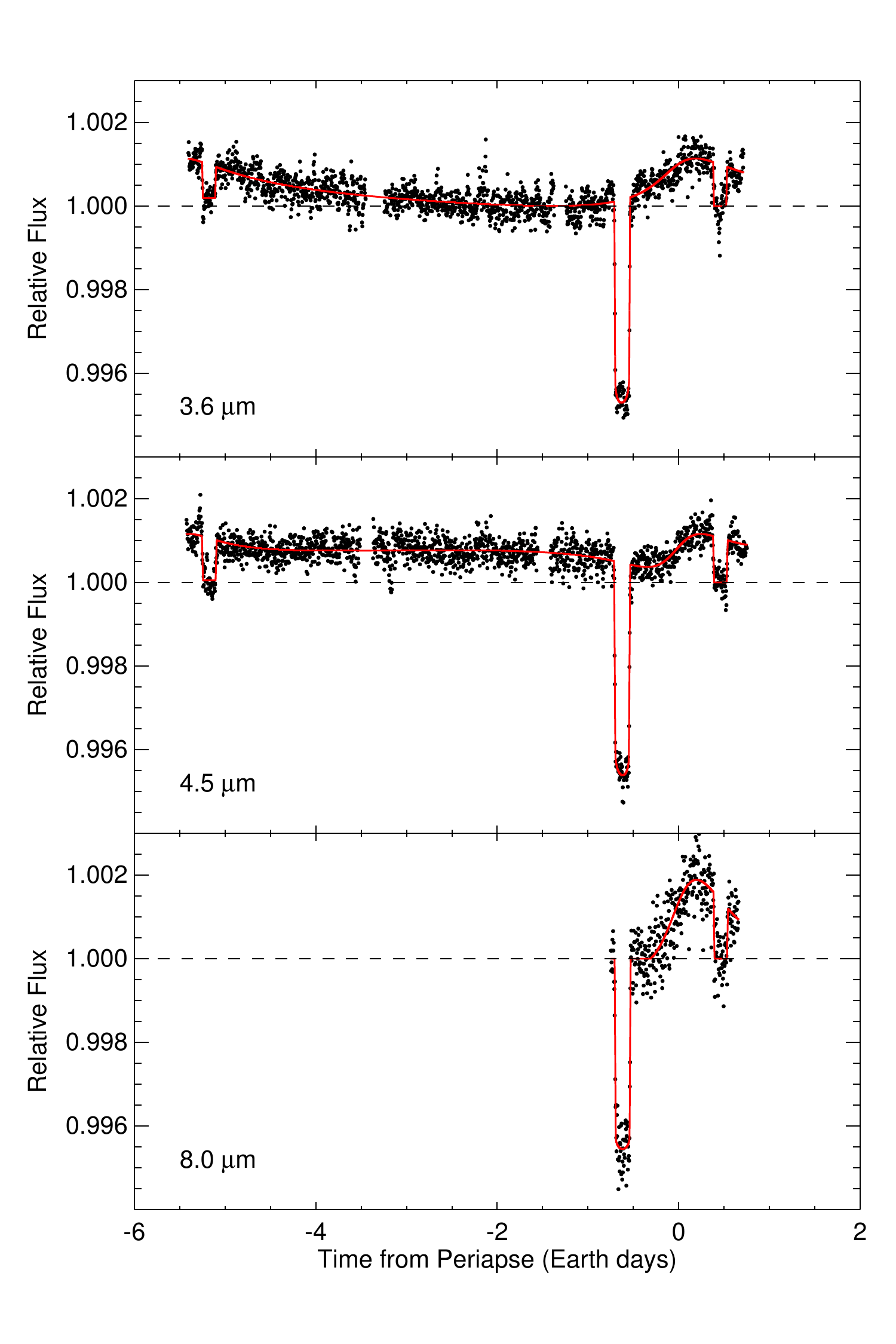}
  \caption{Final 3.6 (top), 4.5~$\mu$m (middle), and 8.0~$\mu$m photometry (filled circles) after correcting for intrapixel sensitivity variations (3.6 and 4.5~$\mu$m) 
  and the ramp-like behavior of the flux with time (3.6 and 8.0~$\mu$m), binned 
  into five minute intervals.  The best-fit phase, transit and secondary eclipse curves are overplotted as a red line. The dashed line represents the stellar flux level.}\label{hat2_phase_curves}
\end{figure}

\begin{figure}
\centering
  \includegraphics[width=0.5\textwidth]{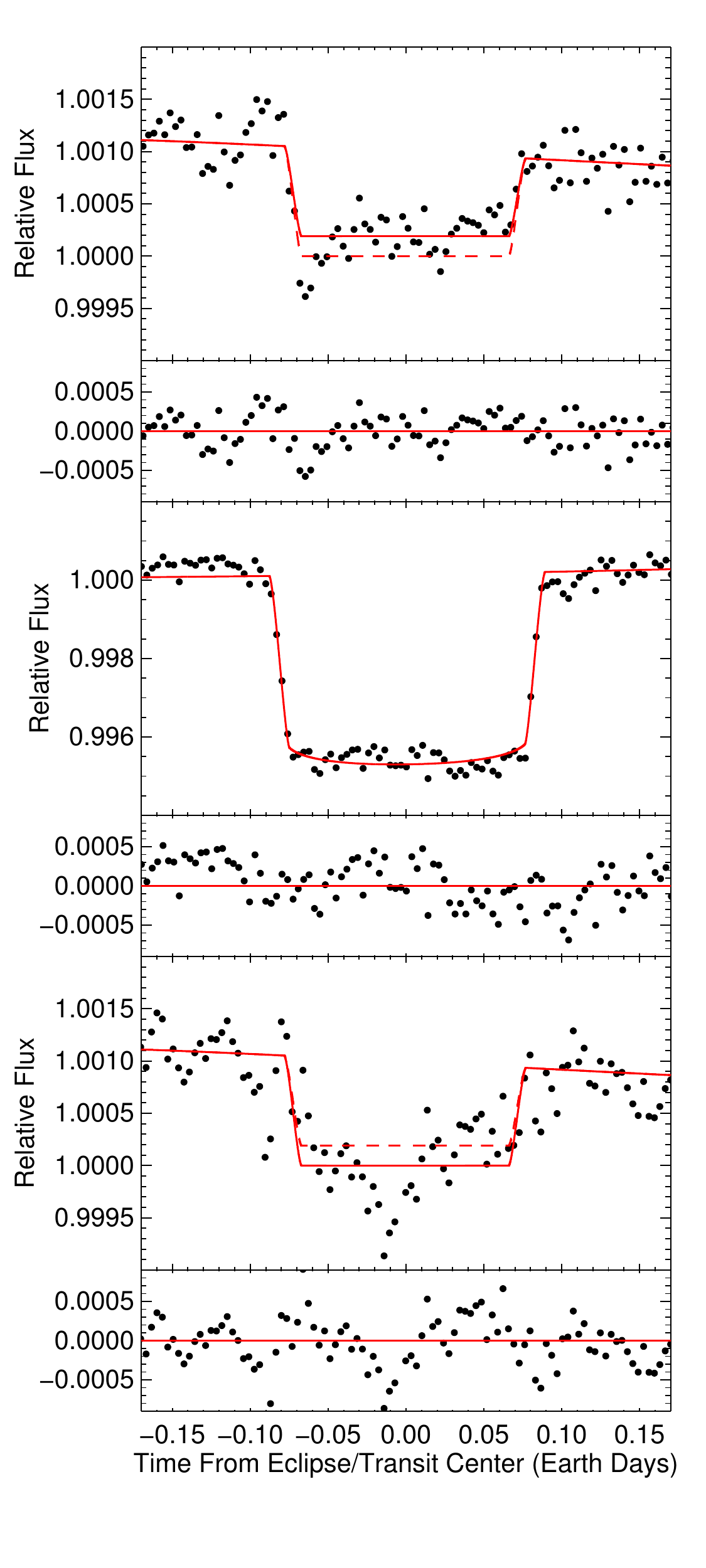}
  \caption{Best-fit transit (middle panel) and secondary eclipse (top and bottom panels) light curves (red lines) for the 
  	3.6~$\mu$m observations (black filled circles).  The data have been binned by five minute intervals.  Residuals for each of the fits are
	presented just below each transit or secondary eclipse event.  The dashed red lines in the secondary eclipse panels 
	show the best fit light curve corresponding to the other secondary eclipse. }\label{ch1_sec_trans}
\end{figure}

\begin{figure}
\centering
  \includegraphics[width=0.5\textwidth]{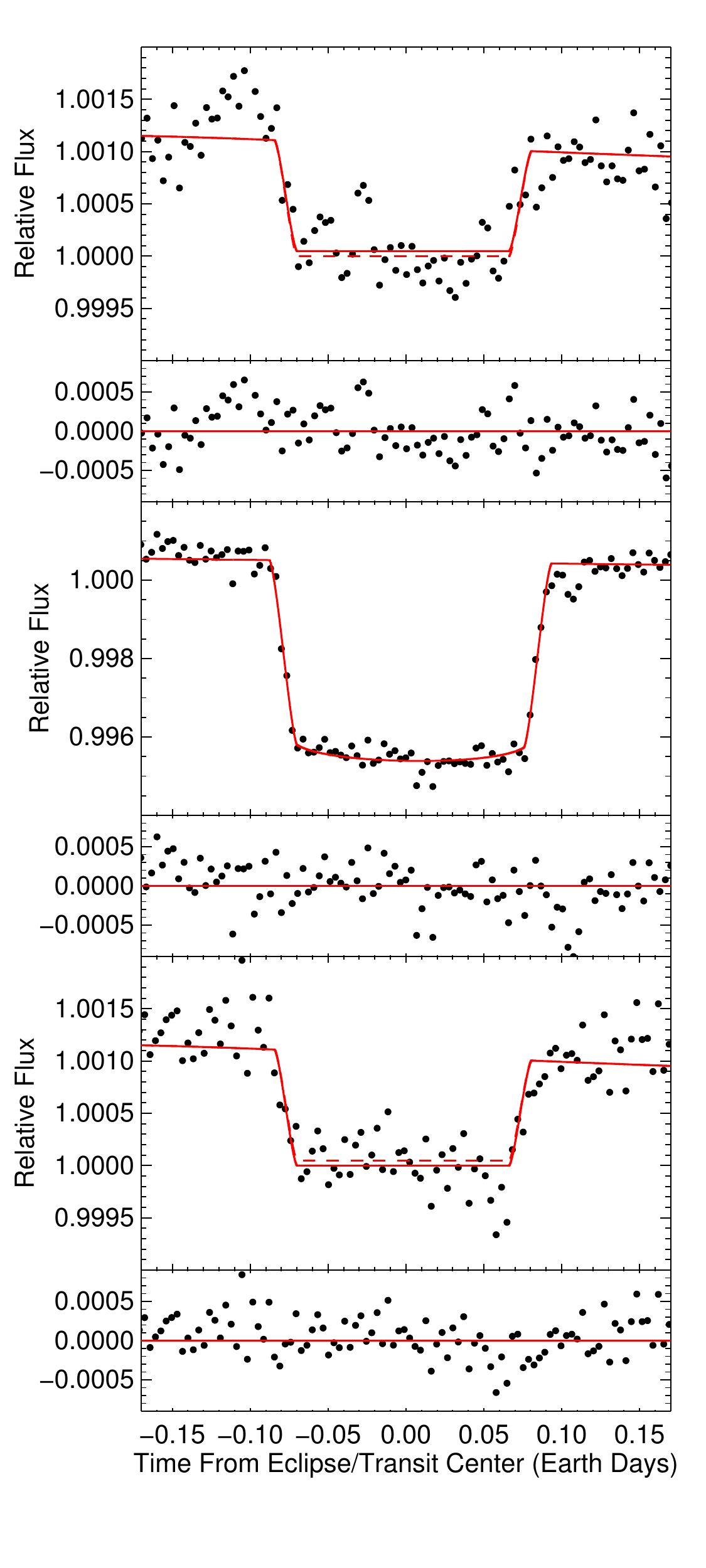}
  \caption{Best-fit transit (middle panel) and secondary eclipse (top and bottom panels) light curves (red lines) for the 
  	4.5~$\mu$m observations (black filled circles).  The data have been binned by five minute intervals.  Residuals for each of the fits are
	presented just below each transit or secondary eclipse event.  The dashed red lines in the secondary eclipse panels 
	show the best fit light curve corresponding to the other secondary eclipse.}\label{ch2_sec_trans}
\end{figure}

\begin{figure}
\centering
  \includegraphics[width=0.5\textwidth]{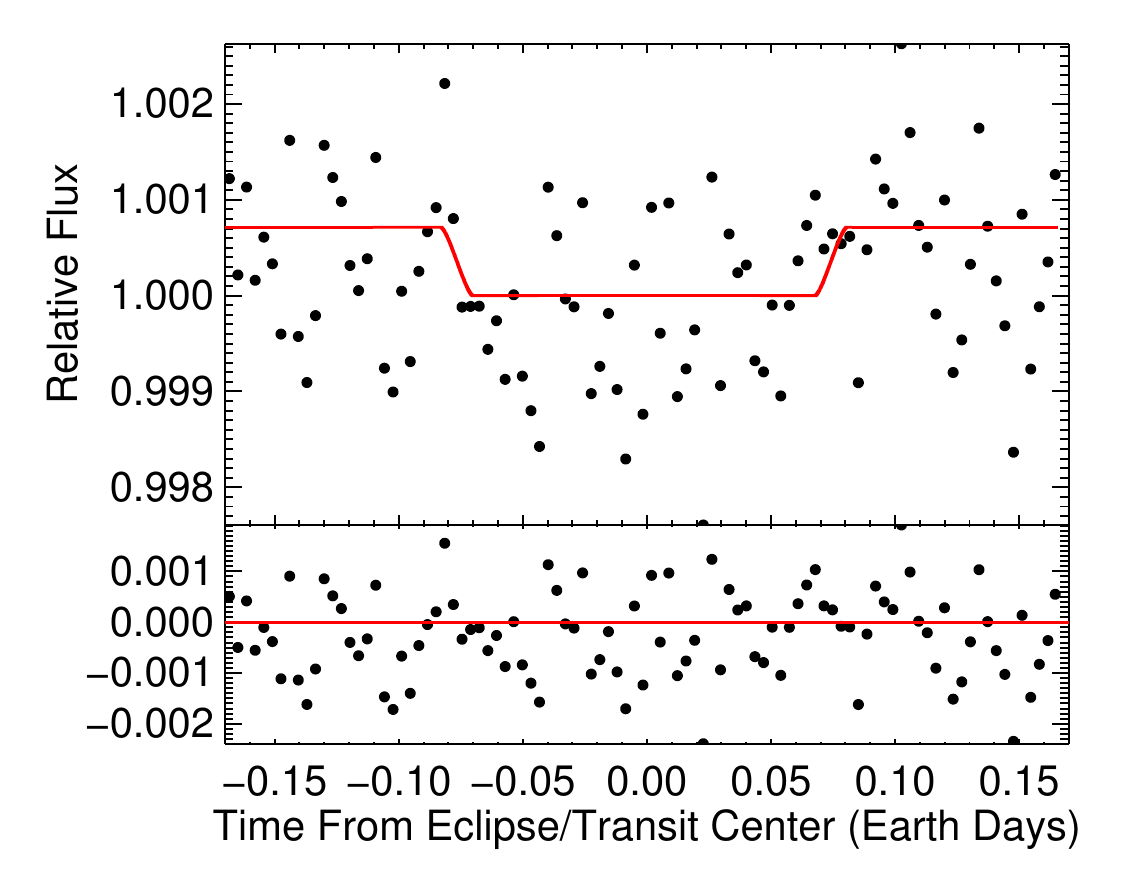}
  \caption{Best-fit secondary eclipse light curve (red line) for the 
  	5.8~$\mu$m observations (black filled circles).  The data have been binned by five minute intervals.  Residuals to the fit are
	presented just below the secondary eclipse event.}\label{ch3_sec}
\end{figure}

\begin{figure}
\centering
  \includegraphics[width=0.5\textwidth]{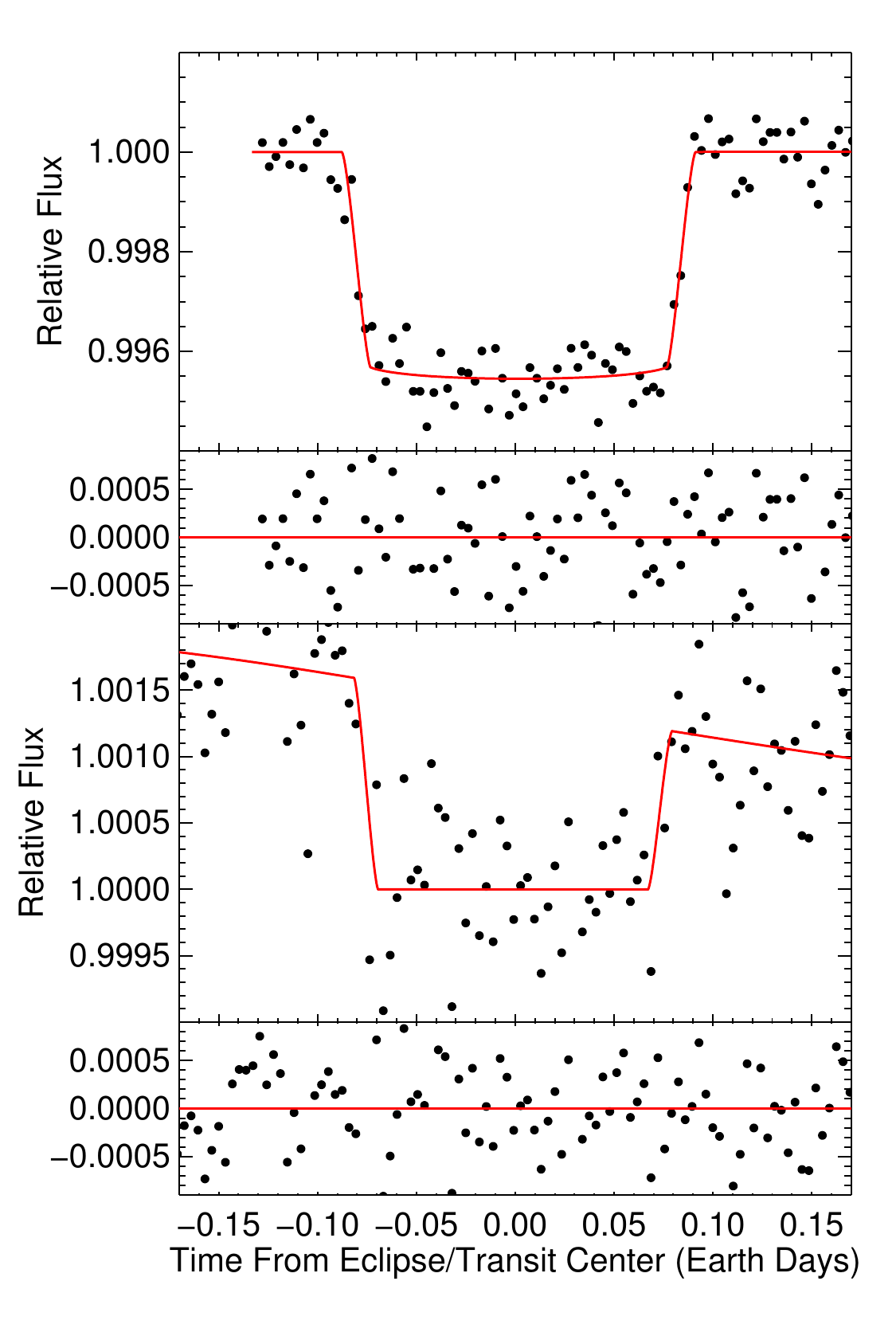}
  \caption{Best-fit transit (top panel) and secondary eclipse (bottom panel) light curves (red lines) for the 
  	8.0~$\mu$m observations (black filled circles).  The data have been binned by five minute intervals.  Residuals for each of the fits are
	presented just below each transit or secondary eclipse event.}\label{ch4_sec_trans}
\end{figure}

\begin{figure}
\centering
  \includegraphics[width=0.5\textwidth]{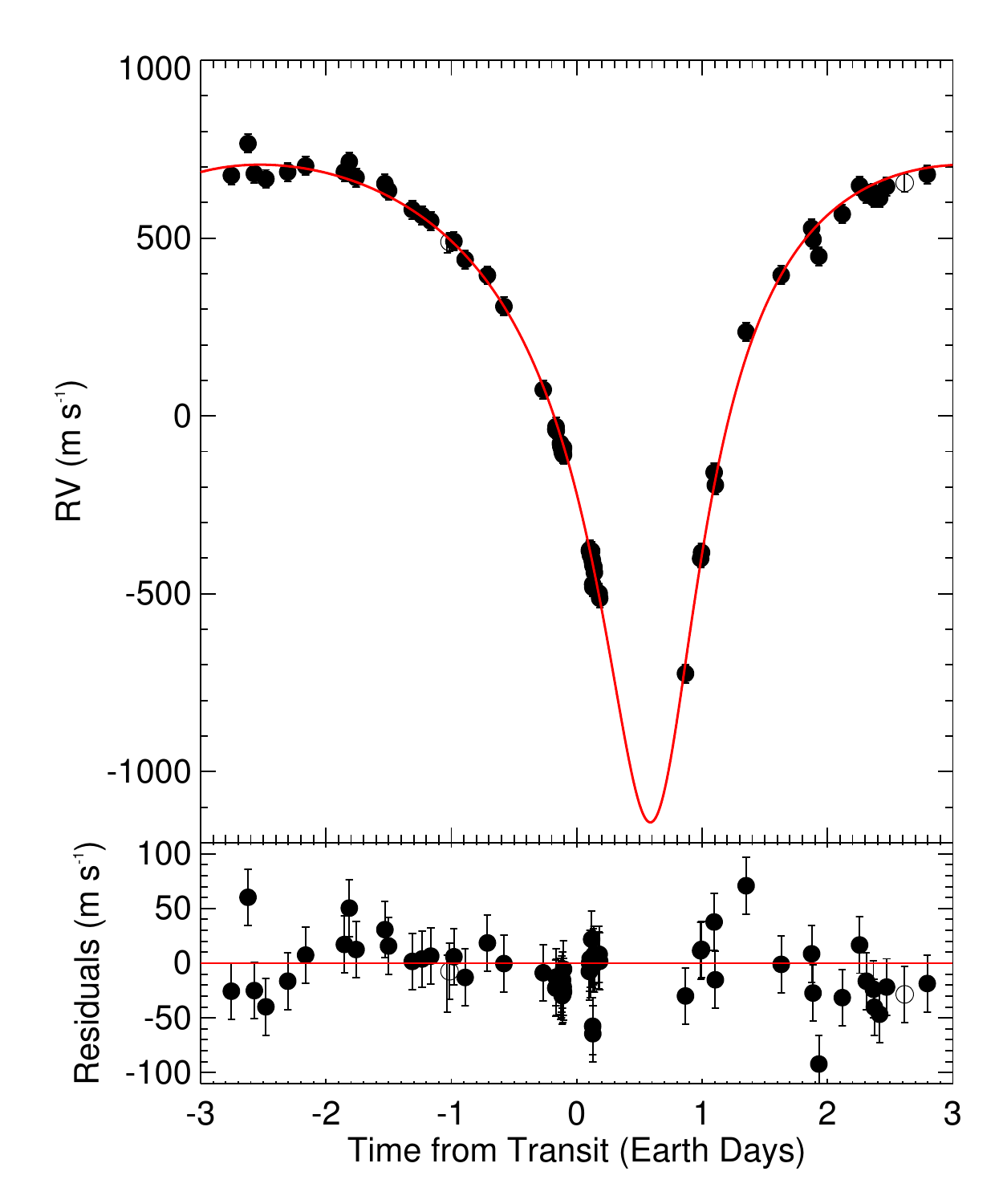}
  \caption{Top Panel:  RV measurements for HAT-P-2 presented in Table~\ref{rv_data} (black circles) folded with 
  an orbital period equal to 5.6334729 d along with the best fit RV solution (red line) with long term variation in the RV signal due 
  to substellar object c removed.  Open circles represent RV measurements taken after the completion of the analysis.  Bottom Panel:  Residuals between the 
  RV measurements and best fit solution.  Error bars on the data points include the best-fit stellar jitter term ($\sigma_{\rm jitter}$).}\label{RV_fit}
\end{figure}

We plot the normalized 
time series for the 3.6, 4.5, and 8.0~$\mu$m data sets after the best-fit intrapixel sensitivity variations and ramp corrections 
have been removed in Figure~\ref{hat2_phase_curves}.  The regions near secondary eclipse and transit for each best fit 
solution are presented in Figures~\ref{ch1_sec_trans}, \ref{ch2_sec_trans}, \ref{ch3_sec}, and \ref{ch4_sec_trans} for the 3.6, 4.5, 5.8, and 
8.0~$\mu$m data sets respectively.  We also present our best fit solution to the RV data in Figure~\ref{RV_fit}.

We use a total of five independent chains with $10^5$ steps per chain in our MCMC analysis.  Each chain is initialized 
at a position in parameter space determined by randomly perturbing the best-fit parameters from a Levenberg-Marquardt 
non-linear least squares fit to the data \citep{mar09}.
Instead of using a standard $\chi^2$ minimization scheme, we instead opt to maximize 
the $\log$ of the likelihood ($L$) given by
\begin{equation}
\log(L)=\sum\left(-\log(2\pi\sigma^2)-(\textrm{data}-\textrm{model})^2/(2\sigma^2)\right), 
\end{equation}
where $\sigma$ is the relevant error term.  This allows us to simultaneously solve for 
the noise terms $\sigma_{\rm phot}$ and $\sigma_{\rm jitter}$ with our other fit parameters \citep[see][for further discussion of 
the maximum likelihood method as applied to exoplanet transit observations]{car09}.  
After each chain has reached $10^5$ steps, we find the the point where $\log(L)$ first surpasses the median $\log(L)$ 
value and discard all steps up to that point.  We then combine the results from our independent 
chains and find the range about a median value that contains 68\% of the values for a given 
parameter.  We set our best-fit parameters equal to this median value and use this distribution to initially determine the
 1$\sigma$ uncertainties in each of our parameters.  For most parameters our error distribution was close to being 
 symmetric about the median value.  In the cases
where the distribution was significantly asymmetric, we have noted both 
the positive and negative uncertainties in the parameter value.

We find that there is (`red') correlated noise in our data even after the best fit intrapixel sensitivity variations have been removed.
To account for correlated noise in our data we first employed the wavelet-based MCMC method described in \citet{car09}.  
However, we found that our correlated noise could not be treated as a stationary noise parameter.  Often we found that 
fits to the transit and secondary eclipse portions of our phase curves were degraded to introduce `red' noise consistent 
with other portions of the light curve.  As a result, we instead opt to use the `residual-permutation' or `prayer-bead' 
method to estimate the errors in our parameters in the presence of correlated noise \citep[see, for example][]{jen02, sou08, bea08, win08}. 
The `prayer-bead' errors are typically 1.5-3$\times$ larger than the errors determined from our MCMC analysis.  The largest 
increases in the uncertainty using the `prayer-bead' method were for the planet-star ratio and secondary eclipse depths.  In 
some cases the `prayer-bead' errors are slightly smaller ($\sim0.9\times$) than the errors from the MCMC analysis.  In those 
cases we report the larger errors from the MCMC analysis.  Table~\ref{hat2_fits} presents the best-fit parameters for our data sets 
and their 1$\sigma$ error bars.  We find that our photometric errors, $\sigma_{\rm phot}$, are 
1.05, 1.11, 1.11, and 1.15 times higher than the predicted photon noise limit at 3.6, 4.5, 5.8, and 8.0~$\mu$m respectively.

\section{Discussion}\label{hat2_discussion}

In the following sections we discuss the implications of these results for our 
understanding of the HAT-P-2 system and the atmospheric properties of HAT-P-2b.  
We compare our results with the previous studies of the HAT-P-2 system from \citet{bak07a}, \citet{win07}, \citet{loe08},  
and \citet{pal10}, which were limited to ground based transit and radial velocity data.  We also 
compare our results to predictions from one-dimensional radiative transfer and 
semi-analytic models of HAT-P-2b's atmosphere.

\subsection{Orbital and RV Parameters}

\begin{deluxetable*}{lcccc}
\tabletypesize{\scriptsize}
\tablecaption{Results from Previous HAT-P-2 Studies\label{pre_data}}
\tablewidth{0pt}
\tablehead{
\colhead{Parameter} & \colhead{\citet{bak07a}} & \colhead{\citet{win07}} &  \colhead{\citet{loe08}} & \colhead{\citet{pal10}} }
\startdata
\emph{Transit Parameters} & & & & \\
$R_p/R_{\star}$ & 0.0684$\pm$0.0009 & 0.0681$\pm$0.0036\tablenotemark{a} &  0.06891$^{+0.00090}_{-0.00086}$ & 0.07227$\pm$0.00061\\
$i$ ($^{\circ}$) &  $>$84.6 &  $>$86.8 & 90.0$^{+0.85}_{-0.93}$ & 86.72$^{+1.12}_{-0.87}$ \\
$a/R_{\star}$  &  9.77$^{+1.10}_{-0.02}$ & 9.90$\pm$0.39\tablenotemark{a} & 10.28$^{+0.12}_{-0.19}$ & 8.99$^{+0.39}_{-0.41}$ \\
$T_c$ (BJD-2400000)  &  54212.8563$\pm$0.0007 & 54212.8565$\pm$0.0006 & ... & 54387.49375$\pm$0.00074 \\
$T_{14}$ (d)   &  0.177$\pm$0.002 & ... & ... & 0.1787$\pm$0.0013 \\
$T_{12}$ (d)   &  0.012$\pm$0.002 & ... & ... & 0.0141$^{+0.0015}_{-0.0012}$ \\
&&&&\\
\emph{Secondary Eclipse Parameters} & & & &\\
$\phi_{sec}$  &  0.1847$\pm$0.0055\tablenotemark{a} & 0.1969$\pm$0.0040\tablenotemark{a} & 0.1896$\pm$0.0016\tablenotemark{a} & 0.1868$\pm$0.0019 \\
$T_{14}$ (d) & ... & ... &  ... & 0.1650$\pm$0.0034 \\
$T_c$ (BJD-2400000) & ... & ... & ... & 54388.546$\pm$0.011\\
&&&&\\
\emph{Orbital \& RV Parameters} & & & &\\ 
$P$ (d) & 5.63341$\pm$0.00013 & 5.63341 (fixed) & 5.63341 (fixed) & 5.6334729$\pm$0.0000061 \\
$e\cos\omega$ & ... &  ... & ... &  -0.5152$\pm$0.0036 \\
$e\sin\omega$ &  ... & ...  & ...  & -0.0441$\pm$0.0084 \\
$e$ &  0.520$\pm$0.010 & 0.501$\pm$0.007 & 0.5163$^{+0.0025}_{-0.0023}$ & 0.5171$\pm$0.0033 \\
$\omega$ ($^{\circ}$) &  179.3$\pm$3.6 & 187.4$\pm$1.6 &  189.92$^{+1.06}_{-1.20}$ & 185.22$\pm$0.95 \\
$T_p$ (BJD-2400000) &  54213.369$\pm$0.041 & ... & 54213.4798$^{+0.0053}_{-0.0030}$ & ... \\
$K$ (m s$^{-1}$) &  1011$\pm$38 & 883$\pm$57\tablenotemark{a} & 966.9$\pm$8.3 & 983.9$\pm$17.2 \\
&&&&\\
\emph{Planetary Parameters} & & & &\\
$M_p$ ($M_J$) & 9.04$\pm$0.50 & 8.04$\pm$0.40 & 8.62$^{+0.39}_{-0.55}$ &   9.09$\pm$0.24 \\
$R_p$ ($R_J$) & 0.982$^{+0.038}_{-0.105}$ &  0.98$\pm$0.04 & 0.951$^{+0.039}_{-0.053}$ & 1.157$^{+0.073}_{-0.062}$ \\
$\rho_p$ (g cm$^{-3}$) & 11.9$^{+4.8}_{-1.6}$ & 10.60$\pm$0.55\tablenotemark{a} & 12.5$^{+2.6}_{-3.6}$ & 7.29$\pm$1.12 \\
$g_p$ (m s$^{-2}$) &  227$^{+44}_{-16}$ &  207$\pm$20\tablenotemark{a} & 237$^{+30}_{-41}$ & 168$\pm$17 \\
a (AU) & 0.0677$\pm$0.0014 &  0.0681$\pm$0.0014\tablenotemark{a} & 0.0677$^{+0.0011}_{-0.0017}$ &  0.06878$\pm$0.00068 \\
&&&&\\
\emph{Noise Parameters} & & & & \\
$\sigma_{jitter}$ (m s$^{-1}$) & 60 & 31 & 17 & ... \\
\enddata
\tablenotetext{a}{Parameter value not directly quoted in reference, but calculated from quoted parameter values and errors}
\end{deluxetable*}

We find that our estimates for the orbital and RV parameters listed in Table \ref{hat2_fits} fall within the 
range of the values from previous studies presented in Table~\ref{pre_data}.  We note that there is often 
a more than 3$\sigma$ discrepancy between orbital parameters for the HAT-P-2 system presented in previous 
studies (Table~\ref{pre_data}).  We conclude that either previous studies underestimate their error bars 
for the discrepant parameters, or the planet's orbital properties are varying in time.  If we compare the orbital parameters we derived from the 3.6, 
4.5, and 8.0~$\mu$m data sets, we find that our estimates are within 3$\sigma$ of each other.  The previous 
studies presented in Table~\ref{pre_data} included only ground-based transit and radial velocity data.  By the 
inclusion of secondary eclipse in our data we were able to improve the estimate of the orbital eccentricity 
of HAT-P-2b by an order of magnitude over the value presented in \citet{pal10}.

From our orbital and RV parameters we can estimate the mass of HAT-P-2b using
\begin{equation}
M_p=\frac{2\pi}{P}\frac{K\sqrt{1-e^2}}{G\sin i}\left(\frac{a}{R_{\star}}\right)^2 R_{\star}^2, 
\end{equation}
where the orbital period ($P$) and stellar radius ($R_{\star}$) are assumed to be the values 
presented in \citet{pal10}, $5.6334729\pm0.0000061$~d and $1.64^{+0.09}_{-0.08}R_{\odot}$ respectively.  Values 
for the RV semi-amplitude ($K$), eccentricity ($e$), inclination ($i$), and normalized semi-major axis ($a/R_{\star}$) 
are taken as the error-weighted average of the values from the 3.6, 4.5, and 8.0~$\mu$m observations, which are presented in 
Table~\ref{ave_val}.   We estimate the mass of HAT-P-2b to be $8.00\pm0.97$~$M_J$, which is within 1$\sigma$ of 
the previous estimates of $M_p$ for HAT-P-2b (Table~\ref{pre_data}).
 
\begin{deluxetable}{lc}
\tabletypesize{\scriptsize}
\tablecaption{HAT-P-2b parameters from a weighted average of the values from the 3.6, 4.5, and 8.0~$\mu$m fits \label{ave_val}}
\tablewidth{0pt}
\tablehead{
\colhead{Parameter} & \colhead{Value}}
\startdata
$R_p/R_{\star}$  &    0.06933$\pm$0.00045 \\
$E_{\rm transit}$ (BJD) &    2455288.84923$\pm$0.00037\\
$a/R_{\star}$  &   8.70$\pm$0.19 \\
$i$ ($^{\circ}$) &    85.97$^{+0.28}_{-0.25}$\\
$e$  &   0.50910$\pm$0.00048 \\
$\omega$  &  188.09$\pm$0.39\\
$\phi_{sec}$  &   0.19253$\pm$0.00018\\
$E_{\rm sec}$  (BJD) &    2455289.93211$\pm$0.00066\\
$E_{\rm periapse}$ (BJD) &       2455289.4721$\pm$0.0038\\
$M_{p}$ ($M_J$) & 8.00$\pm$0.97\\
$R_{p}$ ($R_J$) &  1.106$\pm$0.061\\
$\rho_p$ (g cm$^{-3}$) & 7.3$\pm$1.6 \\
$g_p$ (m s$^{-2}$) &  162$\pm$27 \\
$a$ (AU) & 0.0663$\pm$0.0039
\enddata
\end{deluxetable}

\subsection{Linear RV Trend}

\begin{figure}
  \centering
  \includegraphics[width=0.45\textwidth]{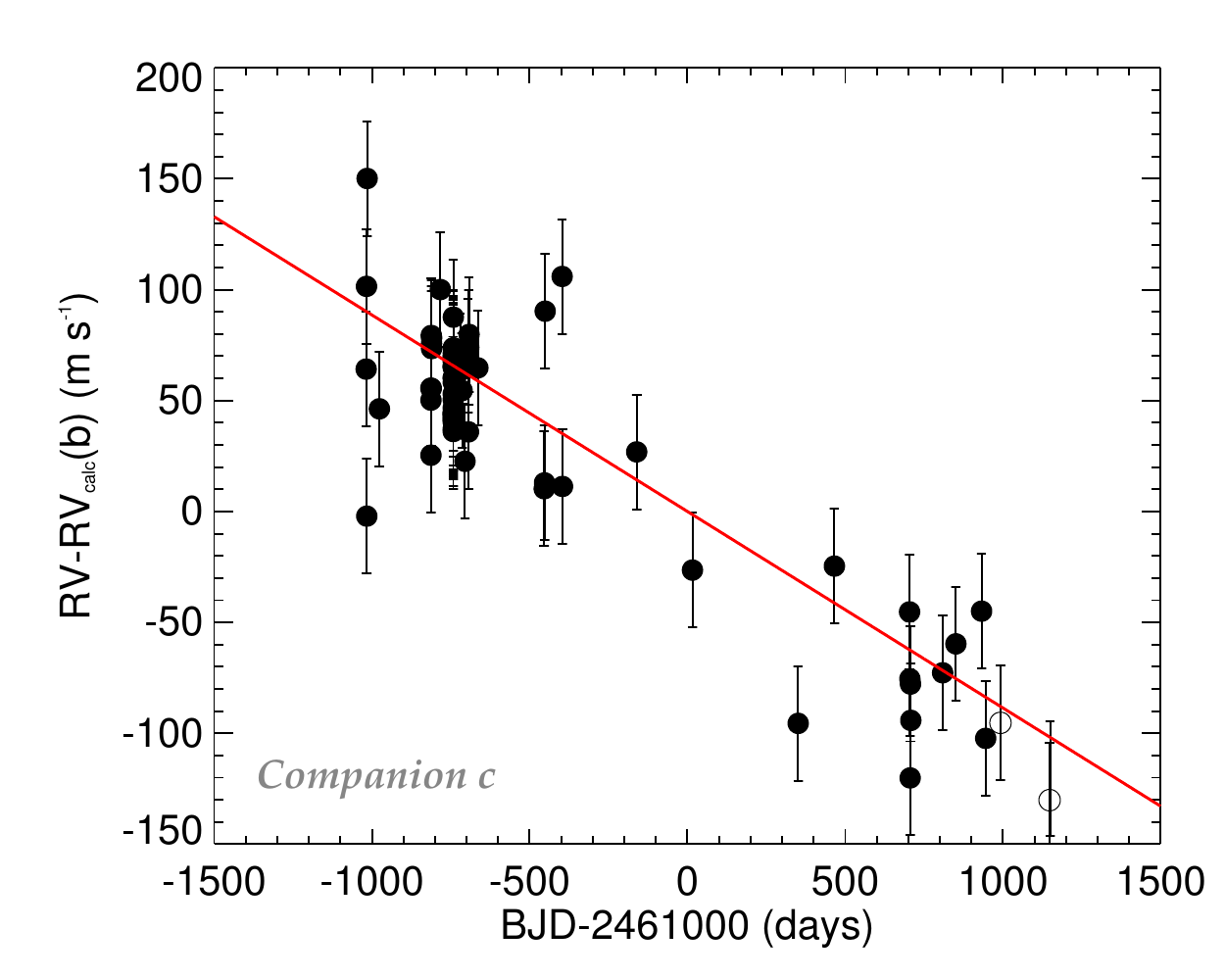}
  \includegraphics[width=0.45\textwidth]{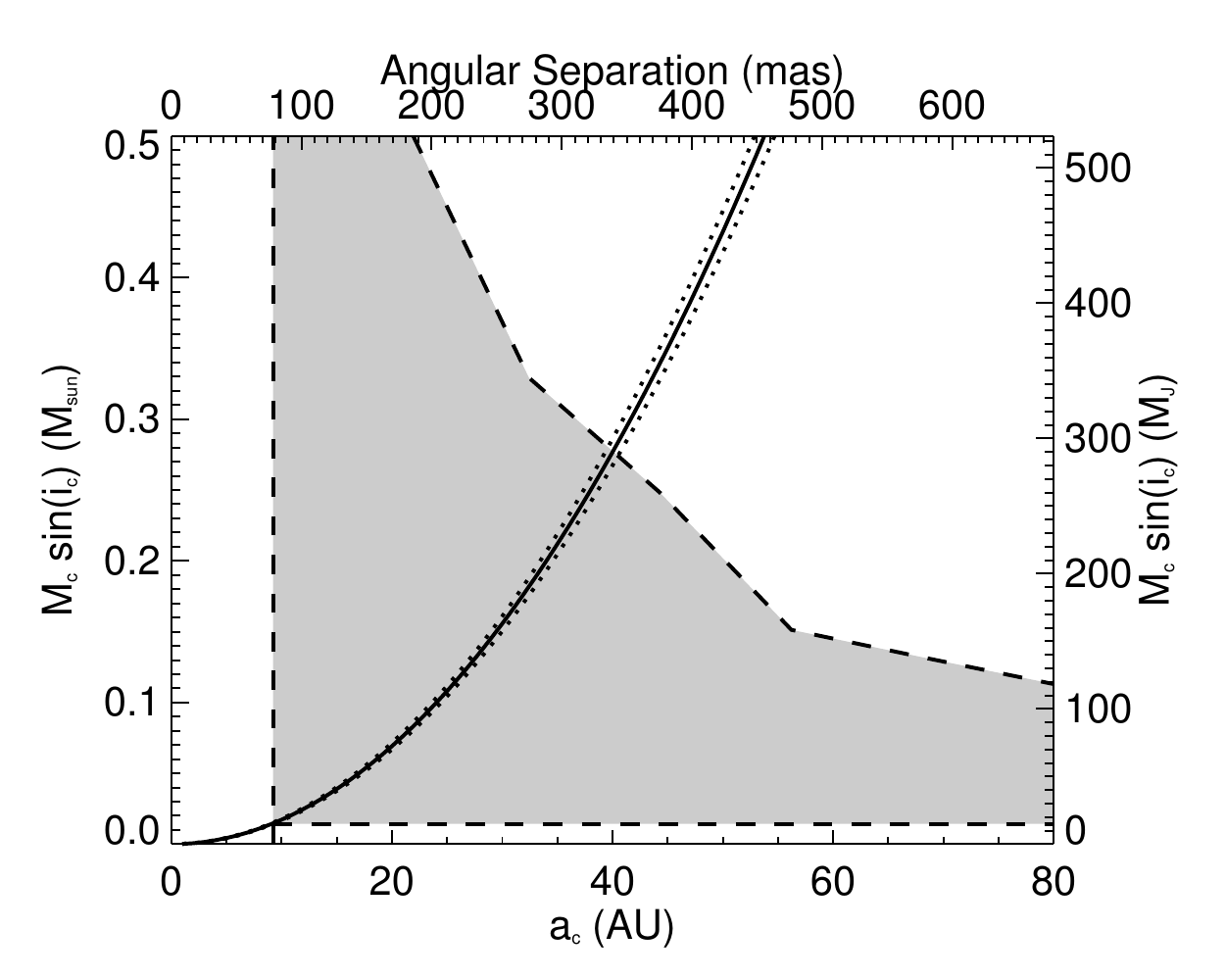}
  \caption{Top:  RV variation as a function of BJD after subtracting the calculated variations due to HAT-P-2b.  Red line shows our best-fit 
                  solution for the linear trend ($\dot{\gamma}$) in the data that result from companion `c'.  Open circles represent RV measurements 
                  taken after the completion of the analysis, which conform with our measured linear trend.
                  Bottom:  Range of $M\sin i$ and semi-major axis ($a$) 
                  for companion `c' (solid line) as estimated from the long term drift in the RV data ($\dot{\gamma}$). 
                  Dotted lines estimate the range in $M\sin i$ and $a$ given the error in $\dot{\gamma}$.  The shaded region gives the acceptable range 
                  of parameter space for companion `c' given our upper and lower bounds on $M_c\sin i$ vs. $a_c$ (dashed lines).}\label{gamma_dot}
\end{figure}

Our RV data spans a period of nearly six years, which allows us to test for long term trends in the RV signal.  
We find a non-zero value for the linear term in our RV fit ($\dot{\gamma}$), which indicates 
the presence of a second body orbiting in the system.  If we assume that 
this second companion to HAT-P-2 (`c') is on a circular orbit and is much less massive than its host star,  
we can relate $\dot{\gamma}$ to the mass and orbital semi-major axis of `c' by the expression presented in \citet{win09}
\begin{equation}
\dot{\gamma}\approx\frac{G M_c \sin i_c}{a_c^2}.
\end{equation}
Figure~\ref{gamma_dot} shows the range of values for $M_c\sin i_c$ and $a_c$ that are allowed for the HAT-P-2 system 
given that $\dot{\gamma}=-0.0886\pm0.0030$.  We know the orbital period of `c' must be significantly longer than six years 
since we do not detect any significant curvature in our long term RV trend, so we set a lower limit of 
$M_c\sin i_c\sim15~M_J$ and $a_c\sim10$~AU based on an orbital period for `c' of 24~years.  We further employ 
adaptive optics imaging to search for `c' and set an upper limit on the values of $M_c\sin i_c$ and $a_c$.  

\subsubsection{Adaptive Optics Imaging}

In an attempt to directly image the body responsible for causing the
linear RV trend, we observed HAT-P-2 on May 29, 2012 using NIRC2 (PI: Keith
Matthews) and the Keck II adaptive optics (AO) system at Mauna Kea \citep{wiz00}. 
Our observations consist of dithered images
taken with the K' ($\lambda_c=2.12 \mu$m) filter. Using the narrow
camera setting, which provides fine spatial sampling of the NIRC2
point-spread-function (10~mas~$\rm pix^{-1}$), we acquired 9 frames
each having 13.2~seconds of on-source integration time. The seeing was
estimated to be 0.4$\arcsec$ at visible wavelengths at the time of
observations using AO wavefront sensor telemetry. We note that
significant wind-shake degraded the quality of correction for some
images but by only a marginal amount.

\begin{figure}
\centering
  \includegraphics[trim = 0.5in 0.5in 0.5in 0.5in, clip, width=0.45\textwidth]{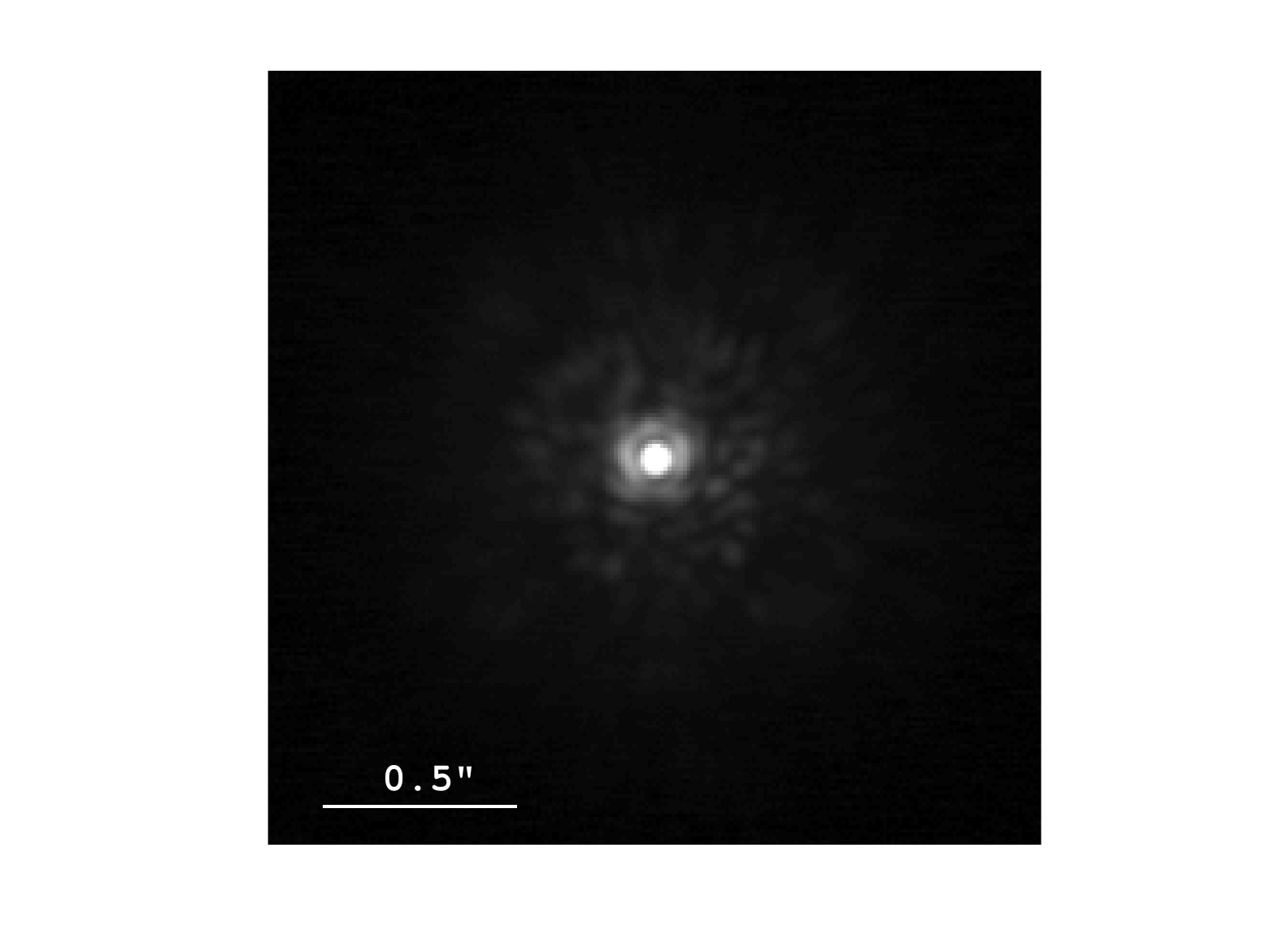}
  \includegraphics[width=0.45\textwidth]{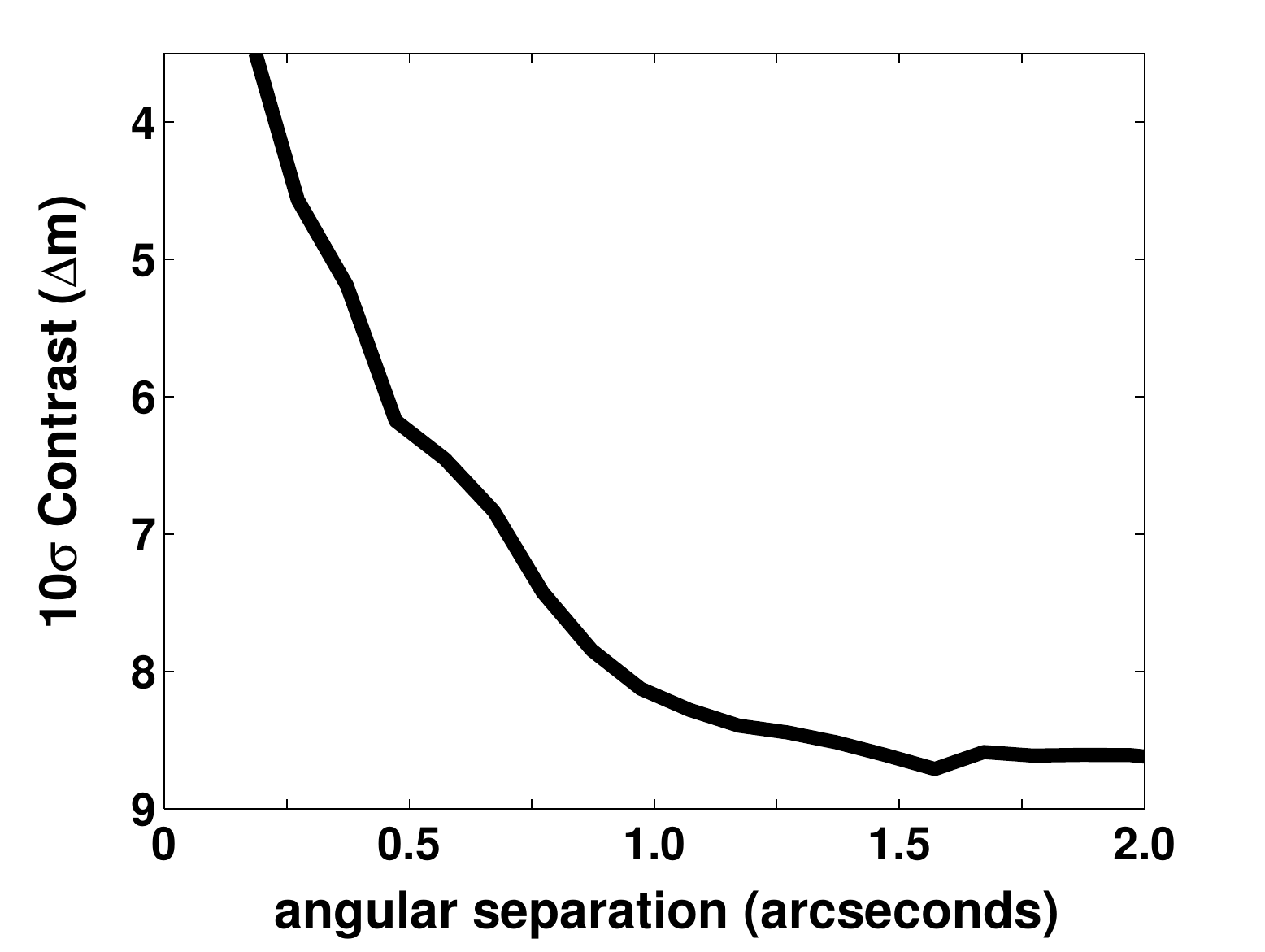}
  \caption{Left:  Keck AO/NIRC2 K band image of the HAT-P-2 system.  Image displayed using a square root scaling.
                  Right:  10$\sigma$ contrast limit for companion detection as a function of angular separation from which define the upper 
                  bounds on $M_c\sin i$ vs $a_c$.}\label{ao_image}
\end{figure}

The data were processed using standard techniques to correct for hot
pixels, remove background radiation from the sky and instrument
optics, flat-field the array, and align and co-add individual frames.
Figure~\ref{ao_image} shows the final reduced AO image along with the corresponding
($10\sigma$) contrast levels achieved as a function of angular
separation. No companions were detected in raw or processed frames.

We can use the limits from a non-detection to rule out the presence of
companions as a function of $M_c \sin (i_c)$ and $a_c$.  Using HAT-P-2's 
parallax distance of $119\pm8 pc$ and estimated age of $2.6\pm0.5$ Gyr
as determined by \citet{pal10} through a combined isochrone, light curve, 
and spectroscopic analysis, we find that our
diffraction-limited observations are sensitive to companions on the
hydrogen-fusing boundary for separations beyond $\approx 1\arcsec$.
Interior to this region, the combination of imaging and RV data
eliminates most low-mass stars, though late-type M-dwarf tertiaries
located at $\sim40$ AU could cause the long-term Doppler drift yet
simultaneously evade direct detection (Figure~\ref{gamma_dot})

\subsubsection{Orbital Evolution}

The likely presence of an M/L/T/Y dwarf at an orbital distance, $a_c$, of 10 to 40 AU from HAT-P-2b lends credence 
to the possibility that HAT-P-2b owes is current orbit to a history of Kozai cycling \citep{koz62} combined with long-running 
orbital decay generated by tidal friction \citep{egg98, wu03, fab07}.
At first glance, this combination of Kozai Cycling and Tidal Friction (KCTF) seems more plausible than disk migration for producing the 
current orbital configuration of HAT-P-2b and indeed, long-distance disk migration for HAT-P-2b seems quite problematic. The protostellar disk itself 
would have needed to be exceptionally massive and long lived, and an ad-hoc mechanism must be invoked to explain the large observed 
orbital eccentricity of HAT-P-2b.  Furthermore, HAT-P-2b has planetary and orbital parameters that place it significantly outside the well-delineated 
population of `conventional' hot Jupiters with $P\sim3$ days, $M \sim M_{\rm Jup}$, and $e\sim0$.

In the context of the KCTF process, HAT-P-2b is envisioned to have formed well outside its current orbit, perhaps via gravitational 
instability in the original protostellar disk. KCTF can operate if the mutual orbital inclination between companion `c' and planet b 
was larger (or better, substantially larger) than the Kozai critical angle, $i_c = \arccos[(3/5)^{1/2}]\sim 40^{\circ}$ 
(assuming $M_{c} \gg M_{b}$, and an initially circular orbit for $b$). In the event that KCTF did operate, planet b experienced 
periodic cycling between successive states of high orbital eccentricity and high mutual inclination. The timescale for these cycles was
\begin{equation}
\tau_{\rm Kozai}=\frac{2 P_{c}^{2}}{3\pi P_{\rm b}}\frac{M_{\star}+M_{\rm b}+M_{\rm c}}{M_{\rm c}}(1-e_{\rm c}^2)^{3/2} \, .
\end{equation}
With plausible values of $P_c=25$ years, $e_c=0.5$, and $M_c=20M_{\rm Jup}$, $\tau_{Kozai}=300$Kyr. By contrast, the general 
relativistic precession rate for HAT-P-2b is currently
\begin{equation}
\dot{\omega_{\rm GR}}=\frac{3 G^{3/2}(M_{\star}+M_{b})^{3/2}}{a_{\rm b}^{5/2}c^2(1-e_{b}^2)}\, ,
\end{equation}
which generates a full $2\pi$ circulation of the apsidal line in 20 Kyr ($\tau_{\rm GR}$). Because $\tau_{\rm Kozai} \gg \tau_{\rm GR}$, the magnitude 
of the Kozai cycling is strongly suppressed at present. 

To good approximation, tidal friction is currently the dominant contributor to the orbital evolution of HAT-P-2b, and it is plausible that the 
current state of the system has undergone dissipative evolution from an earlier epoch where $\tau_{\rm Kozai} = \tau_{\rm GR}$. If we assume 
that the tidal evolution has roughly conserved HAT-P-2b's periastron distance, $d_{\rm peri}=a_{\rm b}(1-e_{\rm b})=0.033$AU, then in order
for $\tau_{\rm Kozai} = \tau_{\rm GR}$ implies that HAT-P-2b has evolved to its current orbit from an orbit with an initial period, $P_{o_{min}}$, given by 
\begin{equation}
P_{o_{min}}=(1-e_{\rm c}^{2})^{3/4}\frac{P_{\rm c}M_{\star}}{c} (\frac{G}{M_{\rm c}d_{\rm peri}})^{1/2}, 
\end{equation}
or
\begin{equation}
P_{o_{min}}\sim 33\,{\rm d}\,\,(\frac{P_{\rm c}}{25\, {\rm yr}})(\frac{20\,M_{\rm Jup}}{M_{\rm c}})^{1/2}\, .
\end{equation}
Additionally, the radial velocity-derived constraint on the unseen companion `c' indicates that $M_{\rm c}\sim P_{\rm c}^{4/3}$, which allows us to simply write
\begin{equation}
P_{o_{min}} \sim 33\,{\rm d}\,(\frac{P_{\rm c}}{25 \,{\rm yr}})^{1/3} \, .
\end{equation}
Given that direct imaging constrains the maximum period for companion `c'  to be of order 250 years, the KCTF scenario gives 
a lower limit on the possible initial periods for HAT-P-2b to be $30\,{\rm days}<P_{o_{min}}<60\,{\rm days}$.

The KCTF process delivers planets into orbits in which the planetary orbital angular momentum and the stellar spin angular momentum are 
initially largely uncorrelated. As mentioned earlier in the text, initial measurements of the projected spin-orbit misalignment angle, $\lambda$, 
suggested that HAT-P-2b's orbit lies in the stellar equatorial plane. Recently, however, an reassessment by \citet{alb12} reports 
$\lambda=9^{\circ} \pm 5^{\circ}$, indicating a modest projected misalignment. In addition, HAT-P-2b's parent star is almost precisely on the 
$T_{\rm eff}=6250$K boundary at which stars empirically appear to be able to maintain misalignment over multi-Gyr time scales \citep{alb12}. 

\subsection{Transit Timing Variations}

\begin{figure}
\centering
  \includegraphics[width=0.5\textwidth]{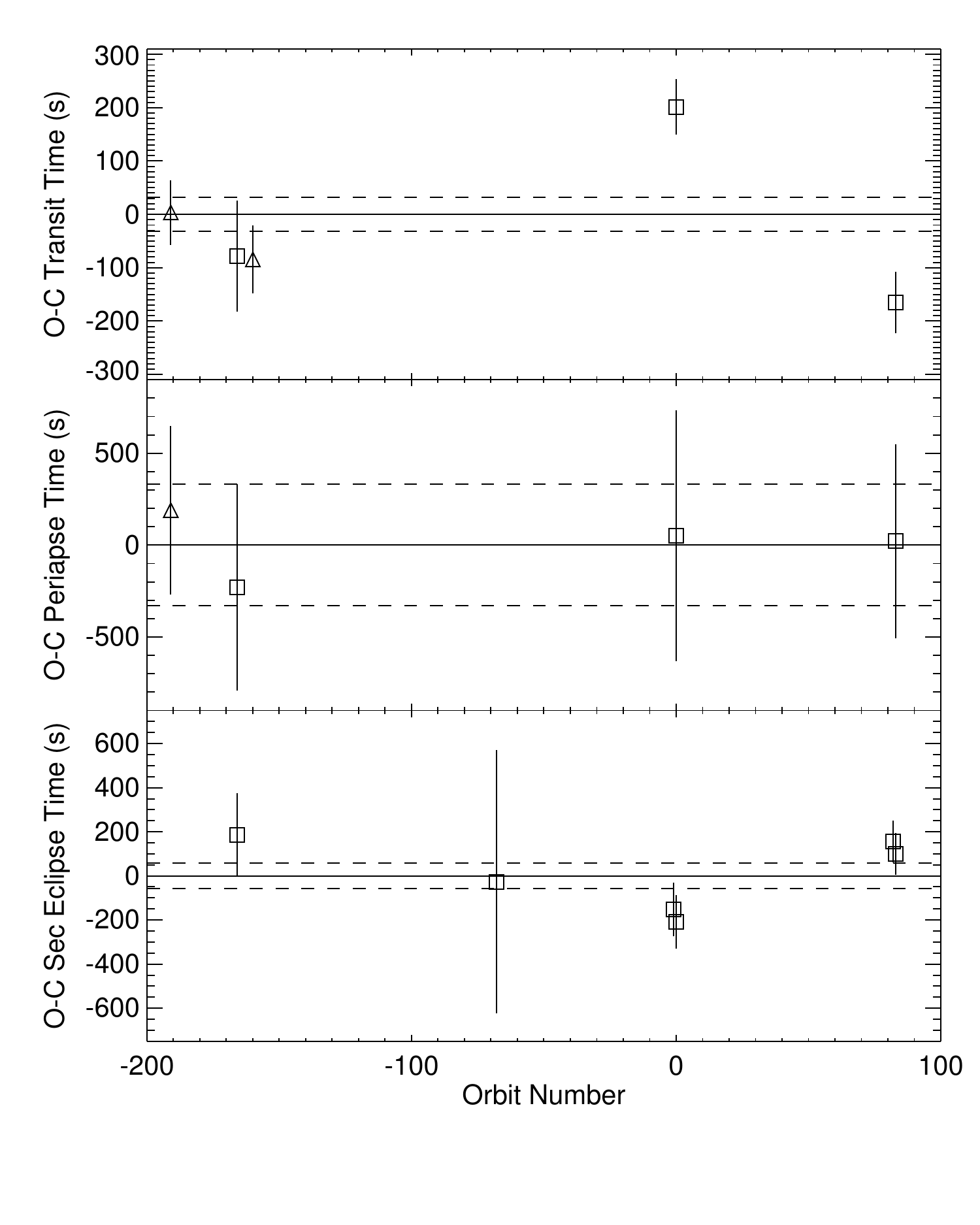}
  \caption{Observed minus calculated transit (top panel), periapse passage (middle), and secondary eclipse (bottom panel) times from data presented in 
  Tables~\ref{hat2_fits} (squares) and \ref{pre_data} (triangles).  Dashed lines indicate the 1$\sigma$ uncertainties in the predicted transit, periapse,  
  and eclipse times.  The variation that we see in the transit times}\label{oc_plot}
\end{figure}

We calculate ephemeris for the HAT-P-2 system given the center of transit and eclipse times presented in Table~\ref{hat2_fits} 
as:
\begin{equation}
T_c(n)=T_c(0)+n\times P
\end{equation}
where $T_c$ is the predicted transit, eclipse, or periapse time and $n$ is the number of orbits that have elapses since $T_c(0)$.  From our transit data 
we calculate $T_c(0)=2455288.84923\pm0.00037$~BJD and $P=5.6334754\pm0.0000026$~d, which is consistent with the orbital 
period for HAT-P-2b presented in \citet{pal10}.  We calculate $T_c(0)=2455289.93211\pm0.00066$~BJD and 
$P=5.6334830\pm0.0000086$~d from our secondary eclipse data and $T_c(0)=2455289.4721\pm0.0038$~BJD and 
$P=5.633479\pm0.000026$~d from our estimated times of periapse passage.   The orbital periods estimated from the secondary 
eclipse and periapse passage timings are within 1$\sigma$ of the value derived from our transit timings.
Figure~\ref{oc_plot} presents our transit, secondary eclipse, and periapse times compared to our derived constant ephemeris.
We define the center of transit/eclipse to occur when the projected planet-star distance given by Equation~\ref{z0} is minimized.  
The definition of transit/eclipse center is not always clearly stated in studies for eccentric systems.  \citet{pal10} estimate a difference 
between RV and photometric transit centers for the HAT-P-2 system of nearly two minutes.  Some of the spread in the measured vs. predicted 
transit times could be accounted for by inconsistent definitions of transit center.  

The separate visits to HAT-P-2b for the 3.6, 4.5, 5.8, and 8.0 $\mu$m observations allow constraints to be placed on possible transit 
timing variations (TTV) for HAT-P-2b.  
Previous transit timing measurements for this system \citep{bak07a, win07,pal10} had relatively low timing 
resolution ($\Delta t\sim 50 \rightarrow 500\,$s) and appeared to be consistent with a constant ephemeris \citep{pal10}.
Somewhat surprisingly, the {\it Spitzer} 
data from orbits 0 and 83 appear to be inconsistent with a constant ephemeris at the $\sim$3.5$\sigma$ level, with a typical deviation of 150~s. 
The deviations are anti-correlated 
between primary transit and secondary eclipse, and they switch sign on the two successive visits. 
This could be due to either an astrophysical cause, or an as-yet unmodeled systematic error. 
In the sections below we consider two potential astrophysical 
explanations for the TTVs:  a planetary satellite or an external perturber.

\subsubsection{TTVs generated by a planetary satellite}

HAT-P-2b is expected to be in pseudo-synchronous rotation, in which its spin period is close to the orbital frequency in 
the vicinity of periapse. Given $P_{\rm orbit}=5.63347 \, {\rm d}$, \citet{hut81}'s treatment gives
\begin{equation}
P_{\rm spin}=\big[\frac{\frac{5 \text{e}^6}{16}+\frac{45 \text{e}^4}{8}+\frac{15
   \text{e}^2}{2}+1}{\left(1-\text{e}^2\right)^{3/2} \left(\frac{3 \text{e}^4}{8}+3
   \text{e}^2+1\right)}\big]^{-1}P_{\rm orbit} = 1.89\,{\rm d}\, .
\end{equation}

In order to maintain orbital dynamical stability, a prospective moon for HAT-P-2b must have have an orbital semi-major 
axis, $a_{\rm sat}$, such that $a_{\rm sat} \lesssim F_{crit} R_{\rm Hill}$, where $R_{\rm Hill}$ is the Hill Sphere radius at periapse,
\begin{equation}
R_{\rm Hill}\sim a_{\rm pl}(1-e)(\frac{M_{\rm pl}}{3M_{\star}})^{1/3} = 0.0042\, {\rm AU}\, ,
\end{equation}
and $F_{crit}\lesssim0.5$ (see, e.g. \citet{bob02} for a discussion of satellite stability for planets with low orbital eccentricity). At 
distance $a_{\rm sat} < F_{crit} R_{\rm Hill}$ from HAT-P-2b, the orbital period of the satellite is $P_{\rm sat}=0.385\,$d, which is substantially 
shorter than the $P_{\rm spin}=1.89\,$d spin period of the planet. Tidal evolution will therefore cause the satellite's orbit to gradually spiral 
in to meet the planet \citep{sbo12}.

Based on the fairly uniform satellites-to-planet mass ratios exhibited by the regular satellites of the Jovian planets in the solar system, and 
in the absence of any concrete information regarding exomoons, it is not unreasonable to expect $M_{\rm sat}=M_{\rm pl}/10^{4} \sim 0.28 M_{\oplus}$. 
The characteristic time for a satellite's orbital decay, assuming an equilibrium theory of tides with a frequency-independent 
tidal quality factor, $Q_{\rm p}$, is \citep{mur99}
\begin{equation}
\tau = \frac{2}{13}(F_{crit} R_{\rm Hill})^{13/2}\frac{Q_{\rm pl}}{3k_{2p}M_{\rm s}R_{\rm pl}^5}(\frac{M_{\rm pl}}{G})^{1/2}\, .
\end{equation}
Adopting $Q_{\rm pl}=10^{6}$, and $k_{2p}=0.1$ we find $\tau\sim2\,$Gyr, which is comparable to the estimated 
stellar age, $\tau_{\star}=2.7\,$Gyr \citep{pal10}. The small value for the apsidal Love number, $k_{2p}$, stemming from the 
fact that HAT-P-2b at 8$M_{\rm Jup}$ is quite centrally condensed in comparison to Jupiter 
\citet[see][for a discussion of the relation between $k_{2p}$ and interior models]{blm01, bbl09}. 
Given the substantial range of possible values 
for $Q_{\rm pl}$, it would therefore not be unreasonable to find that HAT-P-2b harbors a fairly massive moon.

A moon with the above qualities would produce transit 
timing variations of magnitude \citep{kip09}
\begin{equation}
\delta_{TTV_{\rm sat}}=\frac{a_{\rm pl}^{1/2} F_{crit} R_{\rm Hill} M_{\rm sat}}{M_{\rm sat}+M_{\rm pl}{(2GM_{\star})}^{1/2}\Lambda}\, ,
\end{equation}
where the geometric factor $\Lambda$ appropriately amplifies or damps the transit timing variations in accordance with the eccentricity 
and orientation of the  elliptical orbit of the planet about the parent star
\begin{equation}
\Lambda=\cos\left[ \tan^{-1}(\frac{-e \cos(\omega)}{1+e\sin\omega})\right]\left(\frac{2(1+e\sin\omega)}{1-e^2}-1\right)^{1/2} \, .
\end{equation}
Substituting the various values discussed above, we find 
\begin{equation}
\delta_{TTV_{\rm sat}}=0.15\,{\rm s}\,.
\end{equation}
 which is several orders of magnitude too small to be of current interest. So while (somewhat surprisingly) it is distinctly 
 possible that HAT-P-2b could currently harbor a massive satellite, any such satellite cannot be responsible for transit timing 
 variations of the size that appear to be present.
 
 \subsubsection{TTVs generated by an external perturber with a long period}
 
 Perturbations from an as-yet undetected perturbing body present another potential explanation for the apparent transit timing 
 variations. After the signature of HAT-P-2b's orbit has been removed from the existing RV observations, a secular 
 acceleration that can be attributed to a distant companion remains. The constancy of the acceleration implies an orbital period for the 
 companion of at least several years; for example, a body with mass, $M_{c}\sim18 M_{\rm Jup}$, and semi-major axis, 
 $a_{c}=10\,{\rm AU}$, would suffice. For a configuration in which $a_c \gg a_{pl}$ \citep{hm05}, 
\begin{equation}
\delta_{TTV_{\rm perturber}} \sim \frac{45\pi}{16}\frac{M_c}{M_{\star}}P_{\rm pl} \alpha_e^3(1-\sqrt{2}\alpha_{e}^{3/2})^{-2}\, .
\end{equation}
where $\alpha_{e}=a_{\rm pl}/(a_c(1-e_c))$. For a perturber with $M_{c}\sim18 M_{\rm Jup}$, $e_c=0.3$ and $a_{c}=10\,{\rm AU}$, 
we find $\delta_{TTV_{\rm perturber}}=0.1\,{\rm s}$. In addition, direct numerical integrations indicate that for external perturbing planets 
that are consistent with the observed secular acceleration, the expected TTVs are invariably very small, and furthermore, do not exhibit the 
rapid variation shown by the timing reversal observed between orbit 0 and orbit 83.

\subsubsection{A low-mass resonant perturber?}

There are an effectively infinite number of stable orbits for perturbing bodies in mean-motion resonance with HAT-P-2b, and 
the diversity of such orbits is extended by HAT-P-2b's substantial eccentricity. A body in low-order resonance with HAT-P-2b can 
readily induce transit timing variations of the magnitude that are apparently observed, and {\it may} be able to produce the curious 
structure exhibited by the timing variations of the secondary eclipses and the primary transits. Exploratory calculations are currently 
underway to evaluate this possibility. If we assume that the observed TTV are generated by a gravitational perturber, this appears to 
be the most promising approach.  However, we note that 
the significance of the reported transit timing variations is less than 4$\sigma$.
Further high-precision transit and/or secondary eclipse observations along with additional RV measurements 
would be needed to confirm the presence of these TTVs and constrain the properties of the perturbing body.

\subsection{Transit and Secondary Eclipse Depths}\label{trans_sec_mod}

From our three transit observations we obtain planet-star radius ratios of 0.06821$\pm$0.00075, 0.07041$\pm$0.00060, 
and 0.0668$\pm$0.0016 at 3.6, 4.5, and 8.0~$\mu$m respectively.  
Our estimates of the planet/star radius ratio, $R_p/R_{\star}$, are significantly smaller than the value presented in \citet{pal10}, 
but are fairly well aligned with the values presented in \citet{bak07a}, \citet{win07}, and \citet{loe08}.  Although \citet{pal10} incorporate 
the $I$ band observations from \citet{bak07a} they also include follow up observations that 
utilize the $z$ and Str\"{o}mgren $b+y$ bandpasses.  While the $I$ and $z$ bands probe a similar 
wavelength range ($\sim0.8-0.9\mu$m), the Str\"{o}mgren $b$ and $y$ bands probe a slightly 
shorter wavelength range ($\sim0.5\mu$m) where atmospheric scattering may become important.  

The difference between our values of $R_p/R_{\star}$ at 3.6, 4.5, and 8.0~$\mu$m could point to enhanced atmospheric opacity in the 
4.5~$\mu$m bandpass due to CO, but our values of $R_p/R_{\star}$ differ by less than 2$\sigma$.  Given the large value 
of the average gravitational acceleration of HAT-P-2b (162$\pm$27~m~s$^{-2}$, Table~\ref{ave_val}), we would 
expect atmospheric scale heights to be small, which supports our wavelength independent values for 
the planetary radius.
If we assume a value of $R_{\star}=1.64^{+0.09}_{-0.08} R_{\odot}$ \citep{pal10}, then we find an average radius 
value for HAT-P-2b of $R_p=1.106\pm0.061 R_J$.  The radius of planet, given its mass, is in line with models of the 
thermal evolution of massive strongly irradiated planets.  Planetary radii of 1.13 to 1.22~$R_J$ for 8~$M_J$ planets 
are expected for 1-10 Gyr ages \citep{for07}.

\begin{figure}
\centering
  \includegraphics[width=0.5\textwidth]{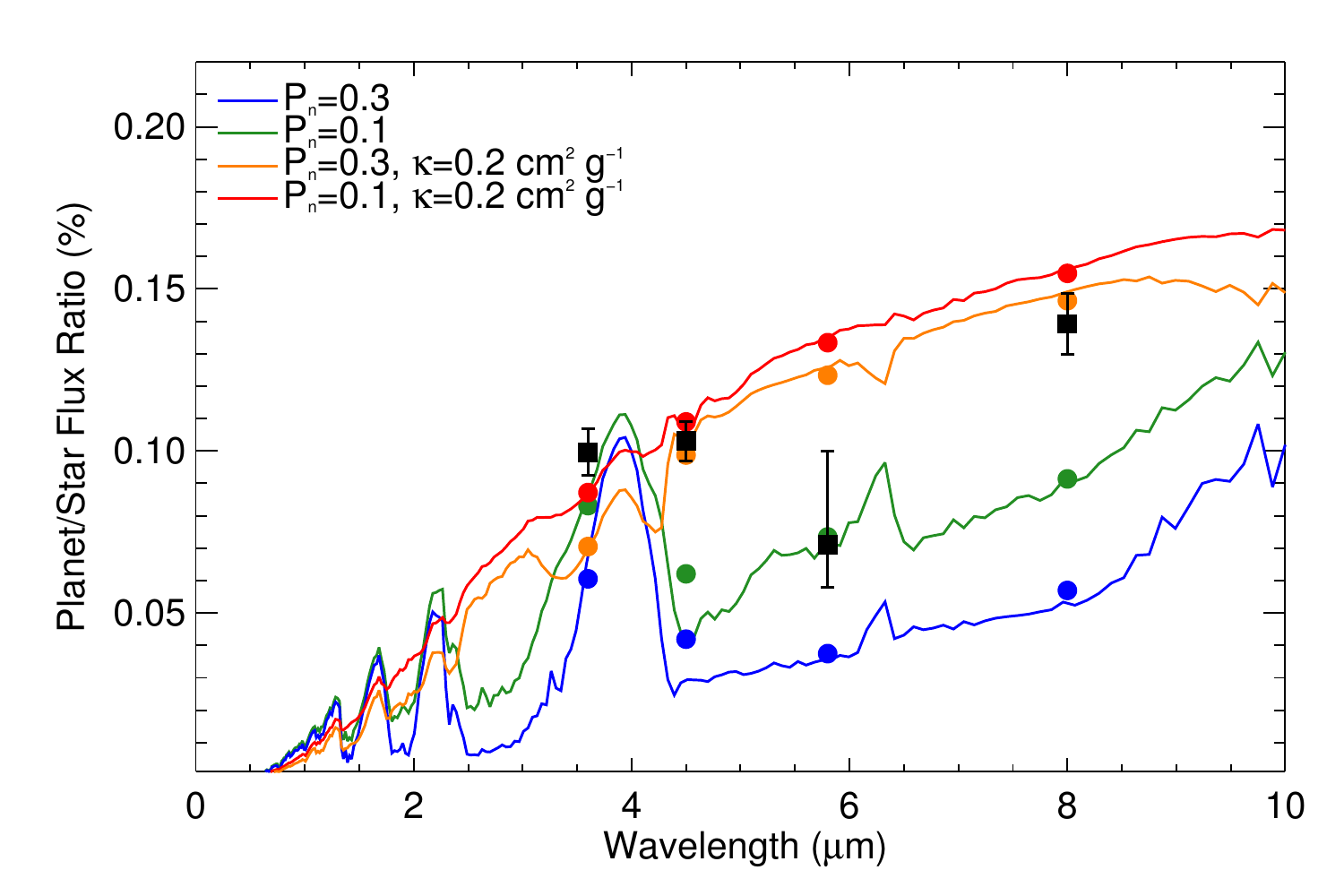}
  \caption{Secondary eclipse depths (squares) at 3.6, 4.5, 5.8, and 8.0~$\mu$m 
  compared to one-dimensional atmosphere models for HAT-P-2b's day side following \citet{bur08}
  with the planet at a distance of 0.0478 A.U.  The models presented here incorporated values of the parameterized 
  recirculation parameter ($P_n$) of both 0.1 and 0.3 and allowed for the presence of a modest high-altitude 
  absorber ($\kappa= 0.2$~cm$^{2}$~g$^{-1}$).   Filled circles represent the band integrated planet/star flux ratio for each of the IRAC bands.  
  Models with an inversion best match data at 3.6, 4.5, and 8.0~$\mu$m.  
  }\label{bur_mod}
\end{figure}

The estimates for the secondary eclipse depths at 3.6, 4.5, 5.8, and 8.0~$\mu$m presented here are the 
first secondary eclipse measurements for HAT-P-2b.  
We compare these results to the predictions of atmospheric models for this planet to better understand 
the atmospheric properties of HAT-P-2b.  Figure~\ref{bur_mod} presents predicted day side emission as 
a function of wavelength from a one-dimensional radiative transfer model similar to those described in \citet{bur08}. 
These models assume a solar-metallicity atmosphere in local thermochemical equilibrium and include
a parameterization for the redistribution of heat from the day side to the night side of the planet ($P_n$) 
and the possible presence of a high-altitude optical absorber ($\kappa$) 
It is important to note that these one-dimensional models assume instantaneous radiative equilibrium and therefore 
do not account for the expected phase lag between the incident flux and planetary temperatures for 
eccentric planets. 

Models that also include an additional high-altitude optical absorber to produce a dayside inversion 
best match our 3.6, 4.5, and 8.0~$\mu$m data points.  We find some difficulty in matching the 5.8~$\mu$m 
data point.  We note that even with our best attempts to remove the ramp from this data that the secondary eclipse depth 
appears to be slightly underestimated (Figure~\ref{ch3_sec}).  Given the large uncertainty in the 5.8~$\mu$m secondary eclipse 
depth it is within 2.5$\sigma$ of the flux ratio predicted by our models that include an inversion.  Both the 4.5 and 8.0~$\mu$m 
secondary eclipse depths are more than 14$\sigma$ larger than those predicted by atmospheric models without an inversion.  
Even if the planet were assumed to be $\sim$100~K warmer to account for the phase lag in planet-wide temperatures, models 
without an inversion would still underestimate the planetary flux in the 4.5 and 8.0~$\mu$m bands.  

We calculate brightness temperatures in each band using a PHOENIX stellar atmosphere model for the star \citep{hau99}.  The 3.6~$\mu$m 
eclipse depths have an average value of $0.0996\%\pm0.0072\%$, which corresponds to a brightness temperature of $2392\pm84$~K
We find an average 4.5~$\mu$m eclipse depth of $0.1031\%\pm0.0061\%$ and a corresponding brightness temperature of $2117\pm65$~K.
Our eclipse depth at 5.8~$\mu$m corresponds to a brightness temperature 
of $1613^{+335}_{-150}$~K , while our eclipse depth at 8.0~$\mu$m corresponds to a brightness temperature of $2258\pm106$~K.
Our secondary eclipse measurements in the 3.6 and 4.5~$\mu$m bandpasses agree at the 1$\sigma$ level, and therefore 
do not provide any convincing evidence for variability in the planet's flux.  Further 
observations of HAT-P-2b at 3.6 and 4.5~$\mu$m could place better constraints on possible orbit-to-orbit variability 
in the planet's thermal structure.  

\subsection{Phase Curve Fits}

The overall shape and timing of the maximum and minimum of the planetary flux from our phase curves 
reveals a great deal about the atmospheric properties of HAT-P-2b.  For planets on circular orbits, 
phase curve observations are generally related to day/night brightness contrasts on the planet.  In the 
case of HAT-P-2b, the phase variations in the planetary flux are more indicative of thermal variations 
that result from the time-variable heating of the planet.  In both the circular and eccentric cases the 
thermal phase amplitude and phase lag are determined by the radiative and advective timescales 
of the planet.  By comparing the 3.6, 4.5, and 8.0~$\mu$m phase variations we can gain further 
insights into the thermal, wind, and possible chemical structure of HAT-P-2b's atmosphere.  

We find that the peak of the observable planetary flux at 3.6~$\mu$m occurs $4.39\pm0.28$ hours after 
periapse passage with a peak value of $0.1139\%\pm0.0089\%$, which corresponds to a brightness 
temperature of  $2556\pm100$~K.  The exact timing and level of the minimum planetary flux at 3.6~$\mu$m 
is much more uncertain.  We find the the planetary flux at 3.6~$\mu$m drops below observable levels  
for a period of $\sim$1~day 18.36$^{+0.18}_{-1.0}$~hours before transit.  As such we can only put an upper limit of 1040~K on the 
minimum brightness temperature of HAT-P-2b at 3.6~$\mu$m.  

The observed 4.5~$\mu$m HAT-P-2b phase curve exhibits a peak in the observable planetary 
flux $5.84\pm0.39$ hours after periapse passage with a value of $0.1162\%^{+0.0089\%}_{-0.0071\%}$ 
and corresponding brightness temperature of $2255^{+92}_{-74}$~K.  We detect emission from HAT-P-2b 
over the entirety of its orbit at 4.5~$\mu$m with a minimum in the planet/star flux ratio of  $0.0372\%^{+0.0086\%}_{-0.0096\%}$, 
which corresponds to a brightness temperature of $1345^{+119}_{-133}$~K.  This minimum in the planetary 
flux occurs $6.71\pm0.43$ hours after transit.  Roughly 
speaking, the shift of the minimum of the observed phase curve at 4.5~$\mu$m away from region between 
apoapse and transit points to a minimum in the planetary 
temperature that is shifted west from the antistellar longitude and/or that the nightside of the planet is still cooling even after the 
transit event.  We would expect the planet to have a temperature minimum east of the anti-stellar point near the sunrise 
terminator assuming a super-rotating flow as shown for HD~189733b and HD~209458b in \citet{sho09}.  This 
dip in the planetary flux after transit is indeed puzzling and will be investigated further in a future study.

Our 8.0~$\mu$m observations cover only the portion of HAT-P-2b's orbit between transit and secondary eclipse, 
so we can only constrain the behavior of the 8.0~$\mu$m phase curve near the peak of the planetary flux.  We 
find that the peak in the planetary flux at 8.0~$\mu$m occurs 4.64$\pm$0.33 hours after the predicted 
time for periapse passage.  The maximum of the planetary flux at 8.0~$\mu$m is $0.1888\pm0.0072\%$, 
which corresponds to a brightness temperature of 2806$\pm$79~K.  

It is significant that we obtain a good fit to the data using Equation~\ref{cowan_eq}, which is similar to the functional form used to fit the 
light curves of HD~189733b and WASP-12b in \citet{knu12} and \citet{cow12a} respectively and produce a longitudinal map of the planet's thermal variations. 
The ``longitudinal" direction of thermal phase maps, $\phi$,  must be understood to mean ``local stellar zenith angle", $\Phi$. 
Thermal phase variations probe the diurnal cycle, $T(\Phi)$. A tidally-locked planet has 1-to-1 correspondence between local stellar 
zenith angle and longitude (e.g.: $\phi = \Phi$), but phase maps can be made regardless of rotation rate 
\citep[e.g., for Earth][]{cow12b}.

Nonetheless, there are two reasons that phase mapping should not work for eccentric planets: 1) the time-variable incident flux makes
the brightness map time-variable ($T(\Phi, t)$), and 2) the time-variable orbital angular velocity of the planet dictates that longitudinally sinusoidal 
variations in brightness on the planet would not correspond to sinusoidal variations in the light curve.  Equation~\ref{cowan_eq} is sinusoidal in the 
true anomaly ($f$) rather than in time, which implicitly accounts for 2).

The fact that Equation~\ref{cowan_eq} fits the phase variations of HAT-P-2b implies that the diurnal brightness profile of the planet is constant 
throughout the orbit: $dT(\Phi)/dt = 0$.  This is entirely different from claiming that the longitudinal brightness profile of the planet is 
constant: since the planet is on an eccentric orbit, there is no fixed correspondence between longitude and stellar zenith angle.  Rather, 
our data seem to indicate that the planet maintains a constant brightness as a function of stellar zenith angle, in marked disagreement with 
expectations for such an eccentric planet.  This is almost certainly due to a geometric coincidence, however: HAT-P-2b shows us its day-side 
shortly after periapse. The day-side brightness of the planet likely changes throughout its orbit, but we are not privy to it.

\begin{figure*}
\centering
 \includegraphics[width=0.45\textwidth]{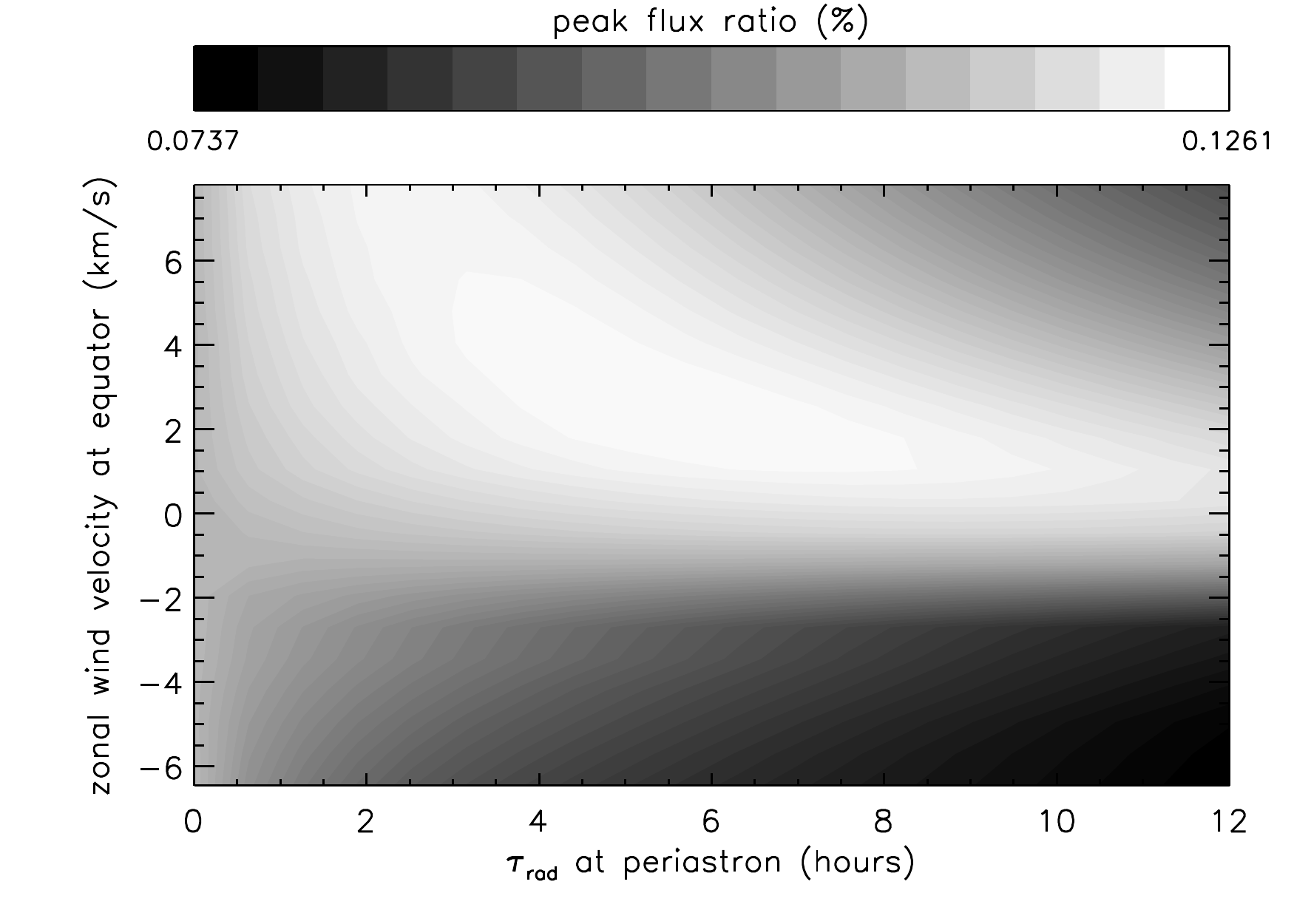}
 \includegraphics[width=0.45\textwidth]{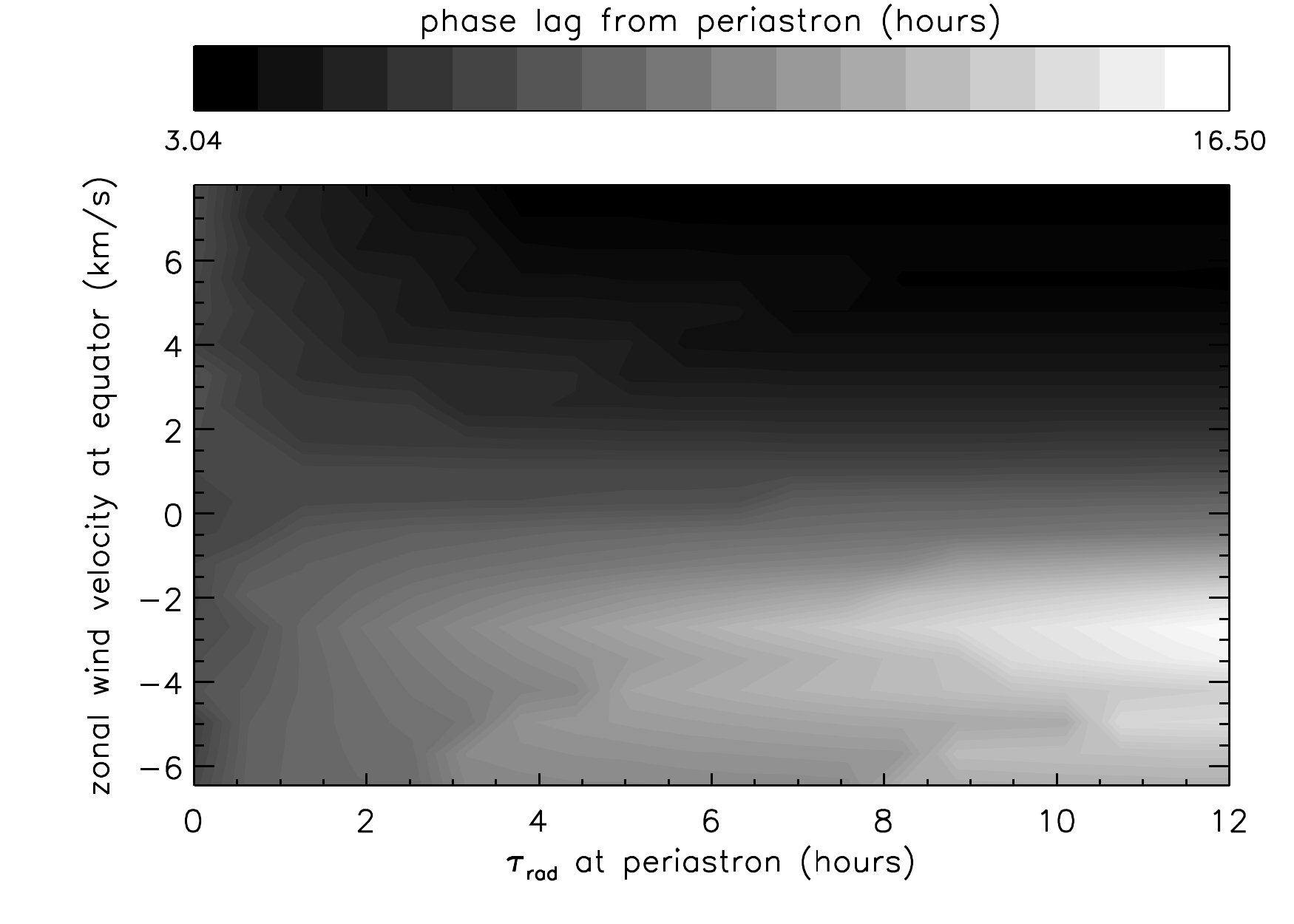}\\
 \includegraphics[width=0.45\textwidth]{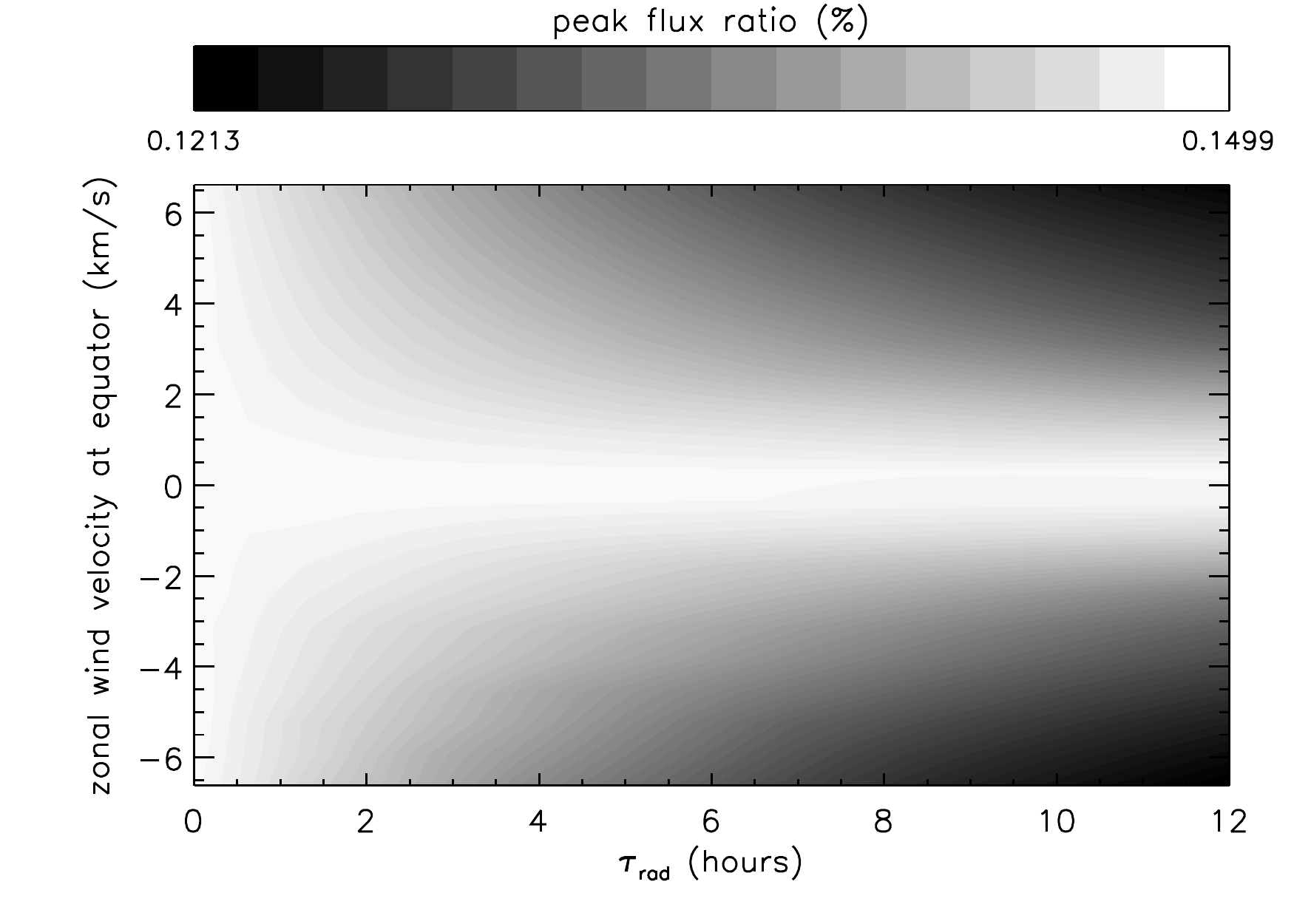}
 \includegraphics[width=0.45\textwidth]{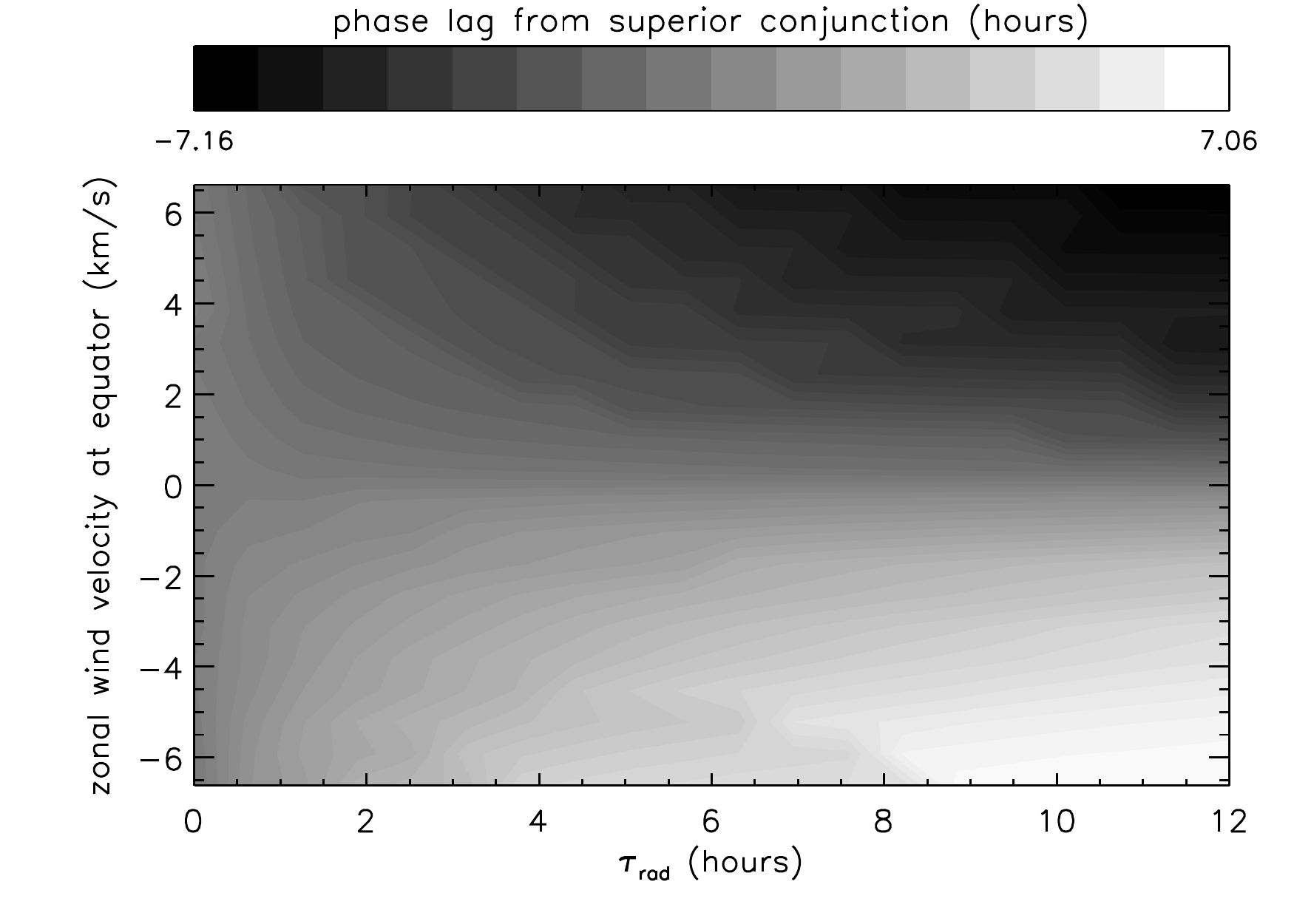}
 \caption{The dependence of peak flux ratio and phase lag at 4.5~$\mu$m on radiative timescale and zonal wind speed. The top panels 
                 are for HAT-P-2b; the bottom panels are for a hypothetical twin on a circular orbit with semi-major axis fixed at the 
                 periapse separation of HAT-P-2b. The plots were generated by computing 4.5~$\mu$m lightcurves on a $20\times20$ grid in 
                 $\tau_{\rm rad}$ and $v_{\rm wind}$ using the energy balance model of \citet{cow11}.}\label{cowan_ch2}
\end{figure*}

\subsubsection{Interpreting the Flux Maximum}

We use the semi-analytic model developed in \citet{cow11} to interpret the peak amplitudes and phase lags of the 
planet's thermal brightness variations.  This is essentially a two-dimensional, one-layer energy balance model where the 
user specifies Bond albedo, radiative timescale at the sub-stellar point at periapse, and the characteristic 
zonal\footnote[1]{zonal refers to the east/west direction} wind velocity.

The Bond albedo is assumed to be 0.1 for all of the simulations shown here. Measured albedos of hot Jupiters range from 0.025 for TrES-2b
\citep{kip11} to 0.32 for Kepler-7b \citep{dem11}.  Since HAT-P-2b is viewed equator-on, its albedo is degenerate with poleward heat 
transport, which we do not explicitly model. Increasing either albedo or meridional\footnote[2]{meridional refers to the north/south direction}  
energy transport has the effect of uniformly decreasing the planet's flux throughout its orbit.

The sub-stellar radiative timescale is specified at periapse, and is assumed to 
scale as $\tau_{\rm rad} \propto T^{-3}$ \citep{sho02, sho10}, 
as one would expect for blackbody parcels of gas. The radiative relaxation time therefore varies throughout the orbit and is also a function 
of the location of a parcel of gas on the planet.


Observationally, the zonal wind speeds on a gas giant like HAT-P-2b are degenerate with its rotation rate. In our model we 
therefore specify the angular velocity of gas parcels in an inertial frame, rather than in the usual rotating planet frame. By adopting 
the \citet{hut81} prescription for pseudo-synchronous rotation of binary stars, we convert the inertial angular frequency into an equatorial 
zonal wind speed in the rotating planet's frame.  If another prescription is more appropriate 
for the planet's rotation rate \citep[e.g.][]{iva07}, then the equatorial wind velocities 
presented here are off by a uniform offset.

Both zonal wind speeds and albedo are assumed to be constant throughout the planet's orbit. The constant zonal wind velocity is likely the 
most problematic assumption 
given that three-dimensional simulations of \citet{kat12} predict that zonal wind speeds at the mid-IR photosphere (pressures of 0.1-1~bar) 
change by tens of percent throughout the orbit of a hot Jupiter with an eccentricity of 0.5.
It is also likely that the amount of equator-to-pole heat transport, which is degenerate with albedo in our model, will vary throughout HAT-P-2b's 
orbit.   Our assumption of a constant albedo for the planet could also be limiting given that clouds could form near the apoapse of HAT-P-2b's orbit, 
then dissipate near periapse \citep{kan10}.  By focusing on the region near periapse we limit the possible influence of temporal changes in 
albedo and wind speeds on our results.

\subsubsection{Model of HAT-P-2b and Circular Analog}
The top panels of Figure~\ref{cowan_ch2} shows how $\tau_{\rm rad}$ and $v_{\rm wind}$ affect the 4.5~$\mu$m peak flux ratio 
and phase lag for HAT-P-2b. The dependencies are qualitatively similar for the other two wavebands. 
For comparison, we also show in bottom panels of Figure~\ref{cowan_ch2} the same dependencies for a hypothetical 
circular twin to HAT-P-2b with semi-major axis fixed at the actual planetÕs periapse separation.
There are a number of conclusions we can draw from these models:
\begin{enumerate}

\item For circular planets, the radiative time scale and wind velocity are entirely degenerate: both the peak flux ratio and the phase 
lag depend solely on the product $\tau_{\rm rad}v_{\rm wind}\propto \epsilon$, the `advective efficiency' of \citet{cow11}. Furthermore, 
the peak flux does not depend on the direction of zonal winds. The phase lag, on the other hand, would have the same amplitude but 
opposite sign for eastward vs.\ westward zonal winds.

\item The peak flux ratio for the eccentric planet HAT-P-2b depends approximately on the advective efficiency, $\epsilon$, as for 
circular planets, but the dependence is no longer monotonic. The peak flux ratio exhibits a maximum for radiative times of $\sim6$~hours 
and wind velocities of 2~km~s$^{-1}$. 
This is because eastward advection of heat brings the hot spot into view shortly after periapse, as discussed in \citet{cow11}.
There is a local minimum in peak flux for $\tau_{\rm rad} = 0$, the radiative equilibrium case. The global 
minimum, however, still occurs in the limit of long radiative times and rapid winds, which results in a planet with zonally uniform, time-constant 
temperature, as in the circular case.

\item Comparing the top-right and bottom-right panels of Figure~\ref{cowan_ch2} shows that zonal winds have qualitatively the same effect, regardless of 
orbital eccentricity: eastward winds make the peak flux occur early, while westward winds cause a delay in the peak flux.  The major difference between 
the two geometries is the phase-lag expected in the absence of winds, the null hypothesis.  For a circular, tidally-locked planet there is no phase lag in 
this limit, whereas the eccentric planet exhibits a large positive phase lag in the absence of winds. As a result of this different zero-point, the amplitude 
of phase lag \emph{from periapse} decreases nearly monotonically for increasing $\epsilon$ in the eccentric model, exactly the opposite behavior from a 
circular planet. Such counter-intuitive behavior is precisely why it is more useful to compare phase lags to the windless scenario rather than quote 
absolute phase lags from periapse \citep{cow11}.
\end{enumerate}

\begin{figure}
\centering
    \includegraphics[width=0.5\textwidth]{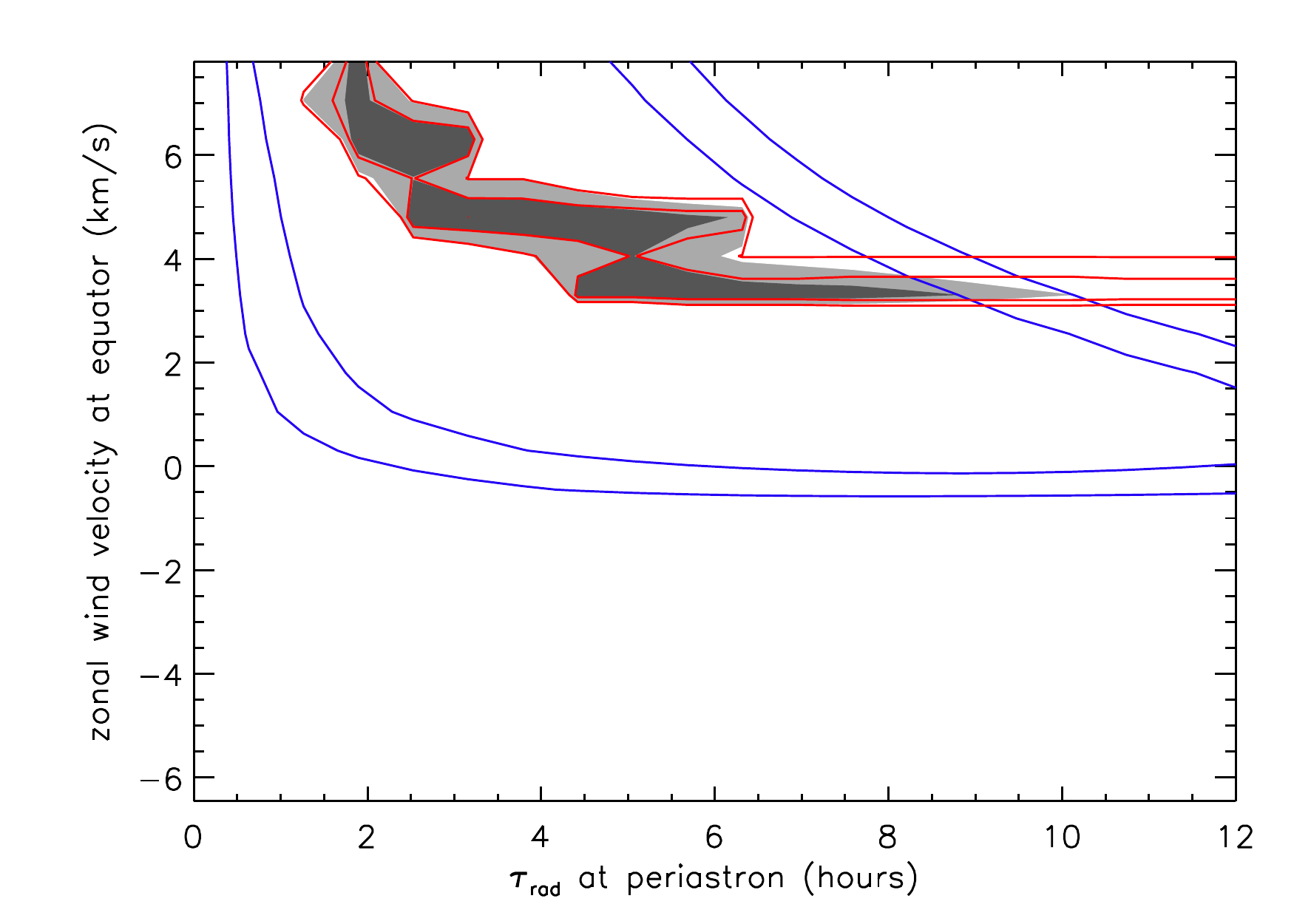}\\
    \includegraphics[width=0.5\textwidth]{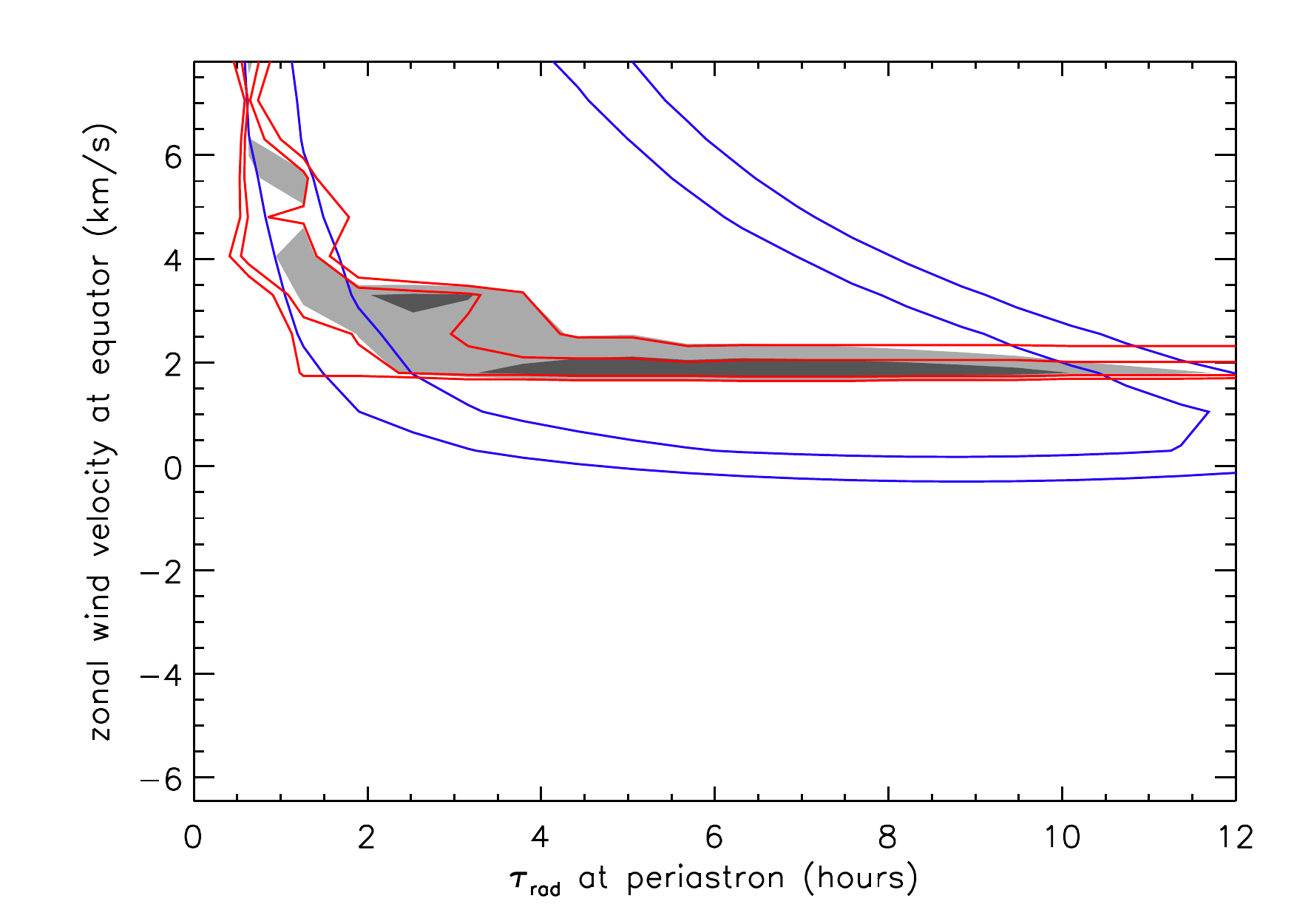}\\
    \includegraphics[width=0.5\textwidth]{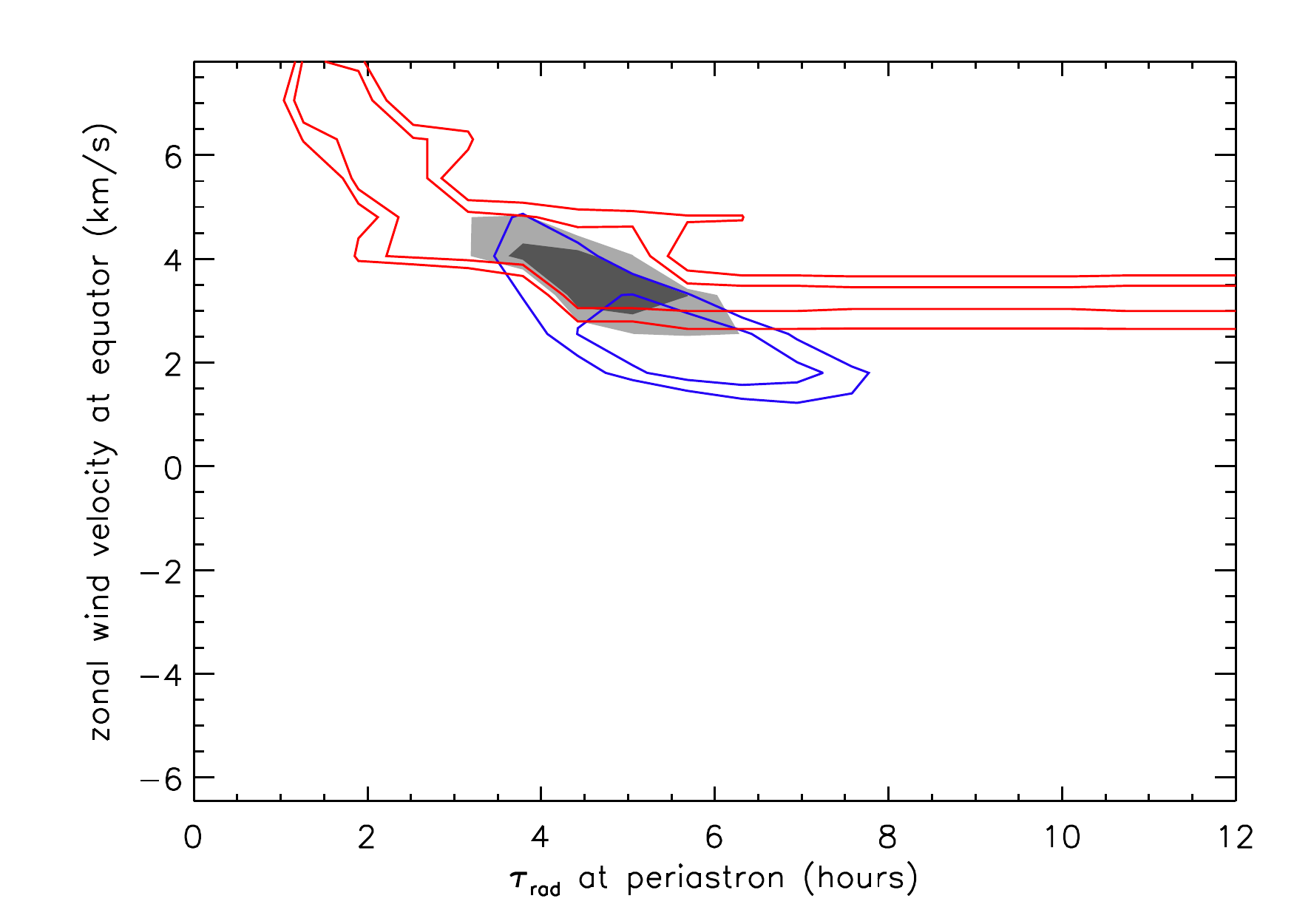}
    \caption{Exclusion diagrams for 3.6, 4.5, and 8.0~$\mu$m phase peaks.  The blue lines show the one and two 
      sigma confidence intervals based on the observed peak flux.  Red lines show the same for the observed phase lag.  Grayscale 
      shows the combined constraint.  The plots were generated by computing lightcurves on a $20\times20$ grid in 
      $\tau_{\rm rad}$ and $v_{\rm wind}$ using the energy balance model of \citet{cow11}.}\label{cowan_all}
\end{figure}

\subsubsection{Constraints on Model Parameters}
In Figure~\ref{cowan_all} we show exclusion diagrams in the 
$v_{\rm wind}$ vs.\ $\tau_{\rm rad}$ plane for each waveband.  Since we use a one-layer model, this can roughly 
be thought of as constraining the properties of the photospheres at each waveband, with the understanding that the 
mid-IR vertical contribution functions span approximately a factor of ten in pressure \citep[e.g.][]{knu09b, sho09}.  
Blue lines in Figure~\ref{cowan_all} show the 1 and 2$\sigma$ 
confidence intervals from the peak flux value.  Red lines show the same for the phase offset. Grayscale shows the 
combined confidence intervals. 

At 3.6 and 4.5~$\mu$m, the peak fluxes are consistent with a broad range of eastward zonal wind scenarios. Only very high 
values of $\epsilon$ (top-right of plot) and most westward winds (bottom of plot) are excluded.
At 8~$\mu$m, only the highest peak flux values are favored, but even these are a very poor fit. If we adopt a Bond albedo of zero, 
the 8~$\mu$m peak flux still disagrees with any models by $>5\sigma$. We therefore conclude that the high flux at 8~$\mu$m 
is not due solely to atmospheric dynamics, but also to chemistry and the planet's vertical temperature profile. This waveband falls 
within a water vapor absorption feature and is therefore expected to originate higher up in the atmosphere than either the 3.6 and 4.5~$\mu$m flux.  
The high flux in the 8.0~$\mu$m channel therefore suggests a temperature inversion, which is also supported by our measured 
secondary eclipse depths in Section~\ref{trans_sec_mod}.

The constraints from the phase lag of peak flux (red lines in Figure~\ref{cowan_all}) are stronger, 
albeit still degenerate.  The data favor a narrow range of advective efficiencies ($\tau_{\rm rad}v_{\rm wind}$).  The bottom panel of Figure~\ref{cowan_all} 
gives the impression that the peak flux and phase lag at 8~$\mu$m combine to give a nice constraint on the model parameters, but the peak flux 
constraints should be taken with a grain of salt, as described above.  It is important to 
note the decaying exponential-like trend of the preferred values of the zonal wind speeds with $\tau_{\rm rad}$ in Figure~\ref{cowan_all} 
that pairs short values for $\tau_{\rm rad}$ with higher values for $v_{\rm wind}$ and longer values for $\tau_{\rm rad}$ with lower values for $v_{\rm wind}$. 
Three-dimensional models that consistently couple radiative and advective processes will help to 
further constrain the relevant timescales in HAT-P-2b's atmosphere over the full range of its orbit.


\section{Conclusions}\label{hat2_conclusions}

In this paper we present the first secondary eclipse and phase curve observations for the eccentric hot Jupiter HAT-P-2b 
in the {\it Spitzer} 3.6, 4.5, 5.8, and 8.0~$\mu$m bandpasses.  These data include two full-orbit 
phase curves at 3.6 and 4.5~$\mu$m, a partial-orbit phase curve at 8.0~$\mu$m, three transit events, and 
six secondary eclipse events.  The timing between 
transit and secondary eclipse during a single planetary orbit combined with radial velocity measurements allows us to better constrain the 
eccentricity ($e=0.50910\pm0.00048$) and argument of periapse ($\omega=188.09^{\circ}\pm0.39^{\circ}$) of HAT-P-2b's orbit.   
Long-term linear trends in the RV data indicate the presence of a third body in the system.  

A comparison of our secondary eclipse depths with a one-dimensional model for the day-side 
emission of the planet suggests the presence of a day-side inversion in HAT-P-2b's atmosphere.  
The timing and magnitude of the peak in the planetary flux at 3.6, 4.5, and 8.0~$\mu$m 
are explained by a range of advective and radiative parameters in our two-dimensional energy balance model, 
but suggest the presence of strong eastward equatorial winds ($\sim2-6$~km~s$^{-1}$) and short radiative 
timescales ($\sim2-8$~hours) at mid-IR photospheric levels near periapse.

Further work is needed to fully explain our observations of HAT-P-2b.  Three-dimensional atmospheric models that 
couple radiative and advective processes and include a range of compositions would help to 
further explain the phase variations we observe for HAT-P-2b, especially outside of the orbital 
region near periapse.  Additionally, low-resolution spectroscopy taken during secondary eclipse 
would help to better constrain the atmospheric chemistry of HAT-P-2b.
Exoplanets on highly eccentric orbits like HAT-P-2b present observers 
and modelers with a unique opportunity to disentangle the contributions of radiative, advective, 
and chemical processes at work in hot Jupiter atmospheres.  By refining our understanding of exoplanets like 
HAT-P-2b we will be able to use that knowledge to better understand the atmospheric processes 
at work in other exoplanet atmospheres.

\acknowledgments

This work is based on observations made with the {\it Spitzer Space Telescope}, which 
is operated by the Jet Propulsion Laboratory, California Institute of Technology, under 
a contract with NASA.  Support for this work was provided by JPL/Caltech.
N.K.L was further supported by NASA Headquarters under the NASA Earth and Space Science 
Fellowship Program (NNX08AX02H), Origins Program (NNX08AF27G), and in part under contract with the California Institute of 
Technology (Caltech) funded by NASA through the Sagan Fellowship Program 
executed by the NASA Exoplanet Science Institute.
Some of the data presented herein were obtained at the W.M. Keck
Observatory, which is operated as a scientific partnership among the
California Institute of Technology, the University of California and
the National Aeronautics and Space Administration. The Observatory was
made possible by the generous financial support of the W.M. Keck Foundation.
N.K.L wishes to thank B. K. Jackson and J. A.  Carter for many useful discussions 
during the preparation of this manuscript and the anonymous referee for their helpful 
suggestions.

\appendix

\section{Noise Pixel Parameter}\label{noise_pix}

\begin{figure}
\centering
 \includegraphics[width=0.24\textwidth]{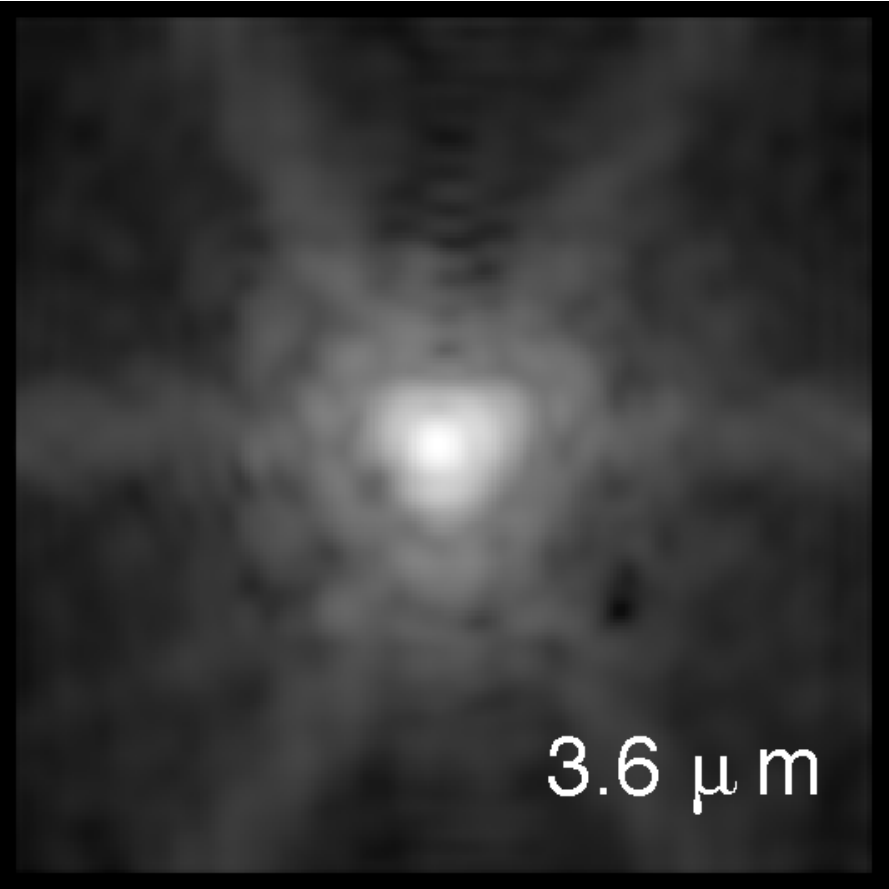}
 \includegraphics[width=0.24\textwidth]{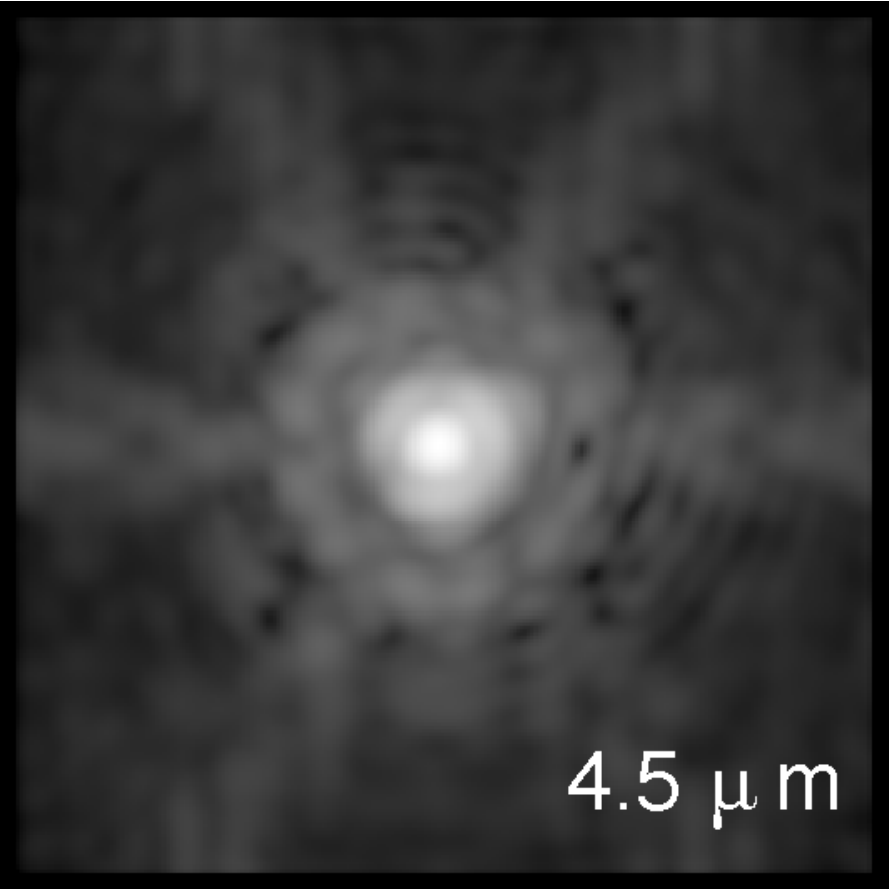}
 \includegraphics[width=0.24\textwidth]{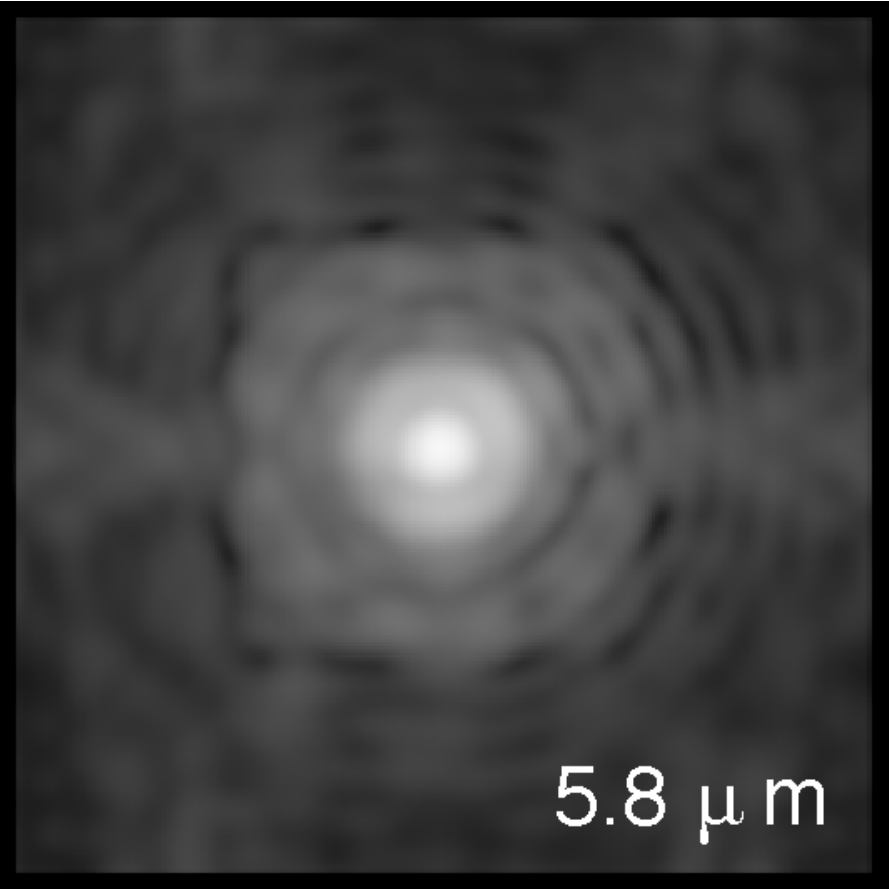}
 \includegraphics[width=0.24\textwidth]{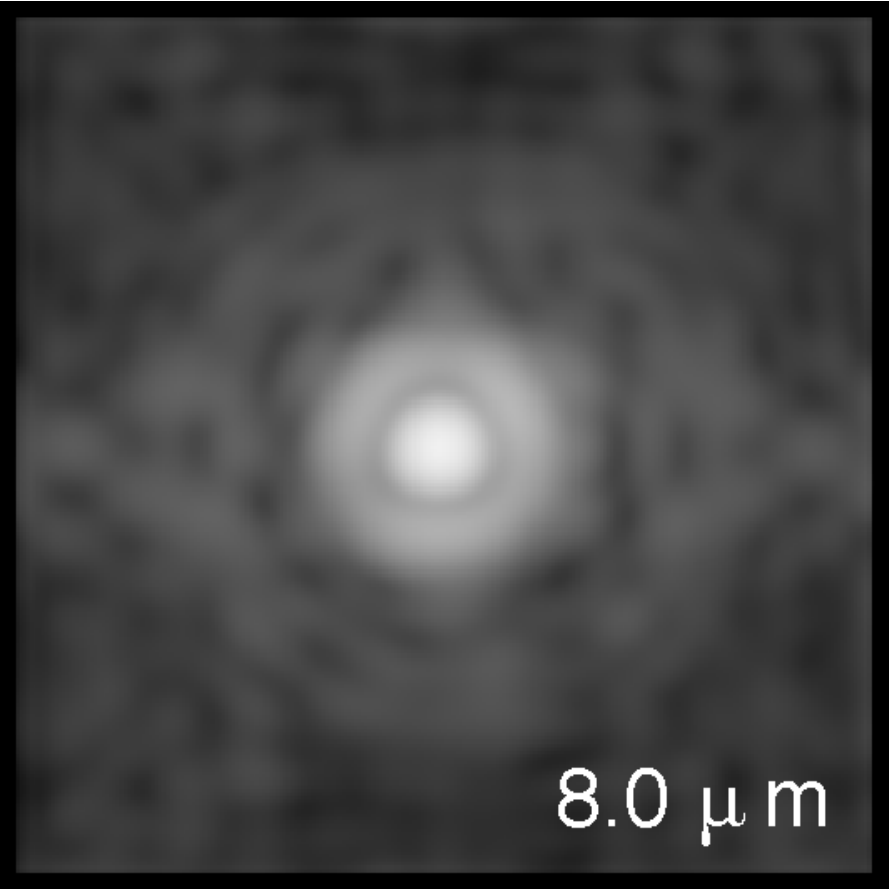}
\caption{{\it Spitzer} IRAC point response functions (PRFs) at 3.6, 4.5, 5.8, and 8.0~$\mu$m.
                 The PRFs were generated by the IRAC team from bright calibrations stars observed at several epochs.  The PRFs are displayed using 
                 a logarithmic scaling to highlight the differences in the PRFs in each bandpass.}\label{prf_fig}
\end{figure}

The IRAC Point Response Function (PRF) results from the convolution of the stellar Point Spread Function (PSF) 
with the Detector Response Function (DRF).   The shape of the PRF is not constant 
and varies with the DRF at each stellar centroid position and with changes in the stellar PSF determined by the optics. 
The shape of the PRFs differs substantially in each channel of the IRAC instrument as shown in Figure~\ref{prf_fig}.  
Given this change in the shape of the PRF with wavelength, we expect that different methods to determine the stellar centroid position 
and correct for systematic effects may be needed in each of the four IRAC bands.    
Both the 3.6 and 4.5~$\mu$m channels of the IRAC instrument are shortward of the  
{\it Spitzer} 5.5~$\mu$m diffraction limit \citep{geh07} and exhibit undersampled PRFs whose shape changes 
as the stellar centroid moves from the center to the edge of a pixel causing intrapixel sensitivity variations.  
The PRFs in the 5.8 and 8.0~$\mu$m channels of the IRAC instrument exhibit a more Gaussian-like shape that is better 
sampled by the detector resulting in only small intrapixel sensitivity variations.

Ideally, we would like to account for these changes in 
the PRF in our photometric measurements.  We cannot make a direct determination of 
the exact shape of the PRF as a function of pixel position, but we can measure the normalized 
effective-background area of the PRF ($\tilde{\beta}$), which is also called noise pixels by the IRAC Instrument Team.  
If we assume the measured flux in a given pixel ($F_i$) is given by 
\begin{equation}
F_i=F_0P_i
\end{equation}
where $F_0$ is the point source flux and $P_i$ is PRF in the $i^{th}$ pixel.  The noise pixel parameter, $\tilde{\beta}$, is given by
\begin{equation}\label{prf_eq}
\tilde{\beta}=\frac{(\sum{F_i})^2}{\sum(F_i^2)}=\frac{(\sum{F_0P_i})^2}{\sum(F_0P_i)^2}=\frac{(\sum{P_i})^2}{\sum(P_i^2)}
\end{equation}
since $F_0$ can be assumed to be constant.   The numerator in Equation \ref{prf_eq} is simply the square of 
the PRF volume ($V$) and the denominator is the effective background area ($\beta$).  These quantities are
related to the sharpness parameter, $S_1$, first introduced by \citet{mul74} for constraining 
adaptive optics corrections by
\begin{equation}\label{prf_beta}
\tilde{\beta}=\beta V^2=\frac{1}{S_1}.
\end{equation}
The $S_1$ parameter describes the `sharpness' of the PRF and can range from zero (flat stellar image) 
to one (all the stellar flux in the central pixel).  

As discussed in \citet{mig05}, the $S_1$ parameter
is related to the standard deviation ($\sigma$) of the PRF by
\begin{equation}\label{prf_sig}
S_1=\frac{1}{C_1\sigma^2}
\end{equation}
where $C_1$ is a constant that depends on the sampling of the PRF on the detector.  From Equations
\ref{prf_sig} and \ref{prf_beta} it can be shown that
\begin{equation}
\sigma \propto \sqrt{\tilde{\beta}}.
\end{equation}
For both the 3.6 and 4.5~$\mu$m data we measure $\tilde{\beta}$ with the same circular aperture sizes 
used to determine the stellar centroid position.  

We find that $\tilde{\beta}$ can serve two purposes in improving the signal-to-noise
of the 3.6 $\mu$m observations.  First, using a circular aperture with a radius proportional to $\sqrt{\tilde{\beta}}$ reduces 
the overall variations in raw unbinned flux from $\sim$5\% to $\sim$2\%.  
\citet{har90}  and \citet{pri81} note that the optimal
aperture radius for a circularly-symmetric Gaussian with a standard deviation of $\sigma$
is $r_0 \approx 1.6\,\sigma$, which is similar to our optimal aperture radius $r_0\approx\sigma\approx\sqrt{\tilde{\beta}}$.  
Second, we find that using $\sqrt{\tilde{\beta}}$ as an additional 
spatial constraint in the intrapixel sensitivity correction at 3.6 $\mu$m can improve the accuracy, as defined by the 
standard deviation of the residuals, in our final solution by $\sim$1-2\%.  We find that using $\tilde{\beta}$ 
as an additional constraint for the 4.5 $\mu$m observations does not significantly improve the accuracy 
of our results.  This is not surprising since the IRAC 4.5 $\mu$m channel is closer to the 
{\it Spitzer} diffraction limit of 5.5 $\mu$m \citep{geh07}.  We also find that $\tilde{\beta}$ does not vary 
with stellar centroid position in the 5.8 and 8.0~$\mu$m observations, which are longward of the {\it Spitzer} 
of the diffraction limit.  

\section{Intrapixel Sensitivity Correction}\label{intra_pix}

The 3.6 and 4.5 $\mu$m channels of the {\it Spitzer} IRAC instrument both exhibit 
variations in the measured flux that are strongly correlated with the position of the 
star on the detector (Figures \ref{ch1_raw} and \ref{ch2_raw}).  These flux variations are the 
result of well documented intrapixel sensitivity variations that are exacerbated by an 
undersampled PRF (e.g., \citet{rea05}, \citet{cha05,cha08}, \citet{mor06}, \citet{knu08}).  
The most common method used to correct intrapixel sensitivity induced flux variations is to fit a 
polynomial function of the stellar centroid position \citep{rea05, cha05, cha08, mor06, knu08}.  This method works reasonably 
well on short timescales ($<$ 10 hours) where the variations in the stellar centroid position are small ($<$ 0.2 pixels).  Recently, 
studies by \citet{bal10} and \citet{ste11} have implemented non-parametric corrections for intrapixel sensitivity variations 
by creating pixel `maps', which give a smaller scatter in the residuals compared with parametric models in most cases.  

The pixel mapping method of \citet{bal10} uses a Gaussian low-pass spatial filter to estimate 
the weighted sensitivity function given by  
\begin{equation}\label{bal10_eq}
W(x_i,y_i)=\frac{ \sum_{j\neq i}^n K_i(j)\times F_{0,j}}{\sum_{j\neq i}^n K_i(j)}
\end{equation}
where
\begin{equation}\label{wfunc_bal10}
K_i(j)=\exp\left(- \frac{(x_j-x_i)^2}{2\sigma_x^2} - \frac{(y_j-y_i)^2}{2\sigma_y^2} \right)
\end{equation}
$F_{0,j}$ is the stellar flux measured in the $j$th frame, $x_j$/$y_j$ and $x_i$/$y_i$ are the stellar centroid 
position in the $j$th and $i$th image respectively,  and $\sigma_x$ and $\sigma_y$ are the widths of the 
Gaussian filter in the $x$ and $y$ directions.  In the \citet{bal10} study, they assumed that all observations 
obtained outside of planetary transit ($F_{0,j}$) represented the intrinsic stellar flux ($F_{0}$) 
convolved with the intrapixel sensitivity function.  In our study, we must account for possible variations in 
the flux from the HAT-P-2 system due to phase variations in the flux from HAT-P-2b.  This requires us to 
iteratively solve Equation~\ref{bal10_eq}, where $F_{0,j}$ is determined at each step by 
removing a model for the planetary transit, eclipse, and phase variations from the observed flux ($F_j$).   

The main challenge in applying the \citet{bal10} method to the HAT-P-2 data set
is that as the number of data points, $n$, becomes large the 
time required to compute Equation \ref{bal10_eq} over the full data set becomes prohibitively 
long.  We must iteratively solve Equation~\ref{bal10_eq} to constrain the phase variations of the planet, 
which requires the summation over $n$ data points $n$ times for each iteration.  
One solution is to bin the data into a manageable number of points as was done in \citet{bal10}.  
We find that binning the data degrades the precision of our final solution.  Binning the data results in 
average values for the measured flux and stellar centroid position that are not necessarily representative 
of the true variations in the stellar flux due to intrapixel sensitivity effects.  We also find that the 
optimal $\sigma_x$ and $\sigma_y$ used in Equation \ref{wfunc_bal10} varies with the stellar centroid 
position.  \citet{bal10} used fixed values for $\sigma_x$ and $\sigma_y$ that empirically produces the 
lowest scatter.  Using values for $\sigma_x$ and $\sigma_y$ that vary with centroid position gives us 
a lower scatter in the residuals compared to using fixed values for $\sigma_x$ and $\sigma_y$.

Here we present an enhanced version of the \citet{bal10} pixel mapping method that allows for 
a large number of data points and optimized values of $\sigma_x$ and $\sigma_y$ without 
being computationally prohibitive.  In Equation \ref{wfunc_bal10}, points that are outside $\sim 6\sigma_{x/y}$ 
of the position of the $i$th data point will contribute negligibly to the weighted sensitivity function $W(x_i,y_i)$.  Given 
this fact, we reduce $n$ in Equation \ref{bal10_eq} by only summing over a fixed number of nearest 
spatial neighbors.  We determine the nearest neighbors to the $i$th flux measurement by the distance, $r_i$, 
given by
\begin{equation}\label{dist}
r_i=\sqrt{a\left(x_j-x_i\right)^2+b\left(y_j-y_i\right)^2+
c\left(\tilde{\beta_j}^{1/2}-\tilde{\beta_i}^{1/2}\right)^2}
\end{equation}
where $x_j$, $y_j$, and $\tilde{\beta_j}$ are the position and noise pixel estimates 
for the $j$th image.  

By including $\tilde{\beta}$ in Equation~\ref{dist} we ensure that the nearest 
neighbors to the $i$th flux measurement share the same systematic effects.  
Although $\tilde{\beta}$ is not strictly a spatial parameter, as discussed in the Appendix 
variations in $\tilde{\beta}$ incorporate systematics that affect the shape of the PRF including and beyond 
intrapixel sensitivity variations.  Our use of $\tilde{\beta}$ is similar to the incorporation of parameters for the PSF width and 
elongation in the correction of systematic variations as discussed in \citet{bak10} for HATnet data.  
The factors $a$, $b$, and $c$ in Equation \ref{dist} can be adjusted 
to give more weight to flux variations in a given spatial direction. For our 3.6~$\mu$m observations 
we find that the optimal value of $\sqrt{1/b}$ is 0.75 with $a=c=1$.  This difference in the $a$ and $b$ parameters 
accounts for the asymmetric shape of the IRAC PSF in the 3.6~$\mu$m bandpass \citep{geh07}.  
For our 4.5~$\mu$m observations we find that $a=b=c=1$ is optimal although the results are similar if we assume $c=0$.  

\begin{figure}
\centering
  \includegraphics[width=0.45\textwidth]{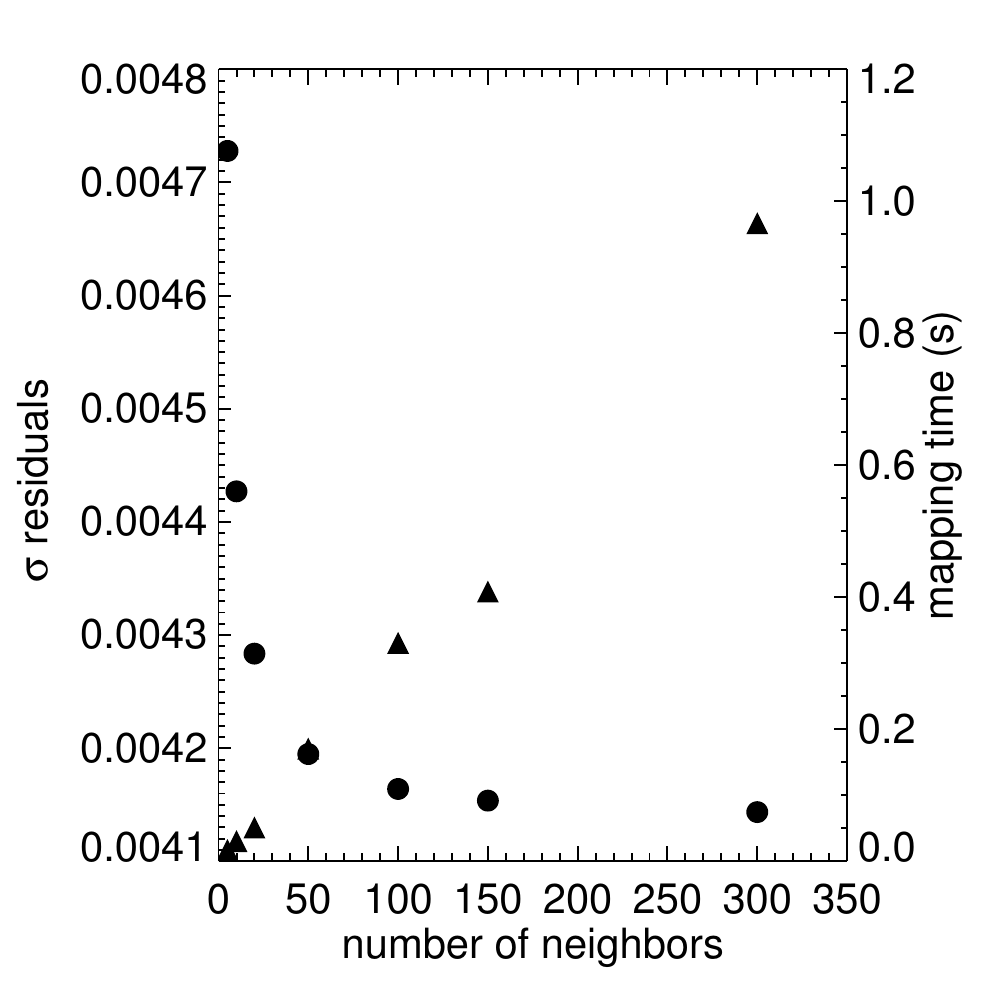}
  \includegraphics[width=0.45\textwidth]{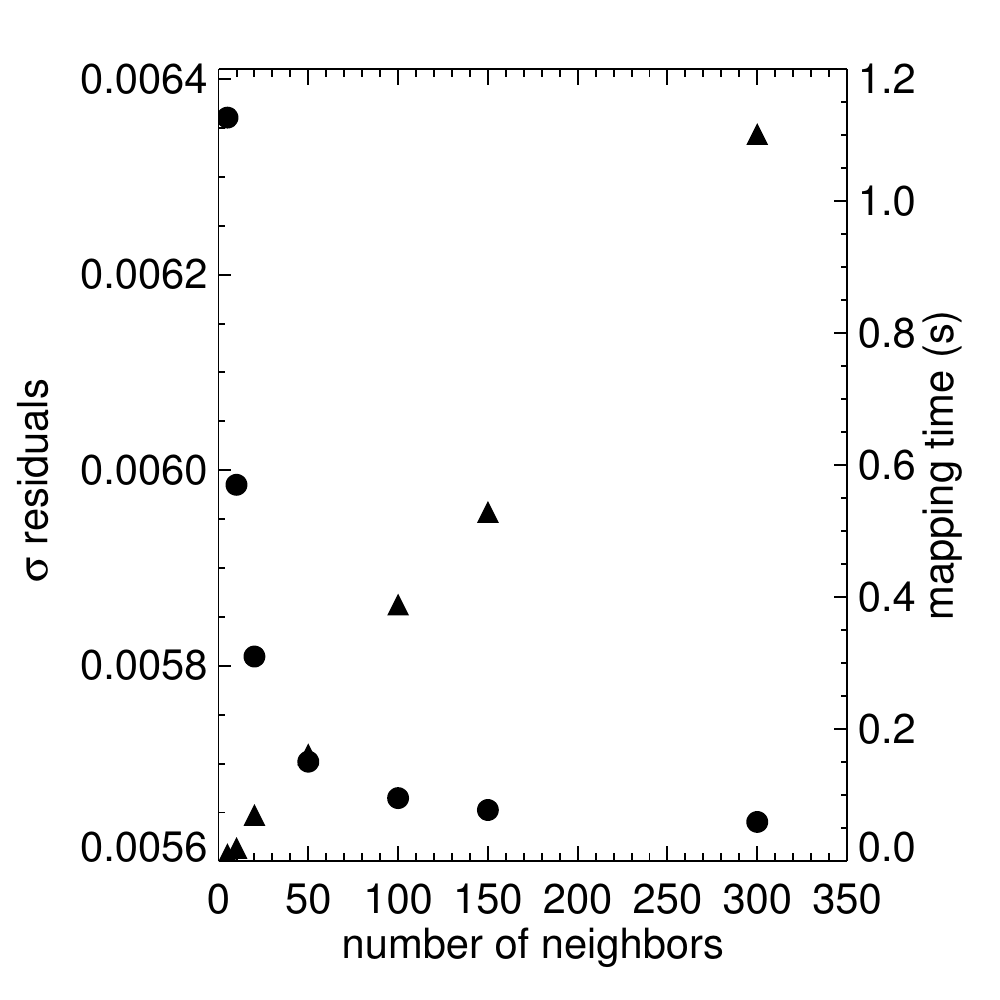}
  \caption{Standard deviation of the residuals (circles) and intrapixel sensitivity mapping time (triangles) as a function 
  of the number of nearest neighbors included in the Gaussian weighting function (Equation \ref{wfunc}) for 
  the 3.6 $\mu$m (top) and 4.5 $\mu$m (bottom) observations.   The standard deviation of the residuals from our fits decreases 
  rapidly as the number of nearest neighbors considered is increased from 5 to 50 after which the gains in the precision of the fit are 
  negligible.  The time required to compute the full pixel map increases more or less linearly with the number of nearest neighbors.  
  We find that repeated iterations become too computationally expensive when more than 50 nearest neighbors are considered. 
  The computational time required per iteration is a multiple of the number of nearest neighbors and the 1.2 million data points 
  in each of the 3.6 and 4.5 ?m data sets.}\label{num_nbr}
\end{figure}

We calculate the Gaussian smoothing kernel $K$ for the $i$th data point with 
respect to its $j$th nearest neighbor, $K_i(j)$, according to
\begin{eqnarray}\label{wfunc}
K_i(j)& = & \exp\left(- \frac{(x_j-x_i)^2}{2\sigma_{x,i}^2} - \frac{(y_j-y_i)^2}{2\sigma_{y,i}^2}\right. \nonumber \\ 
&& \left. - \frac{(\tilde{\beta_j}^{1/2}-\tilde{\beta_i}^{1/2})^2}{2\sigma_{\tilde{\beta}^{1/2},j}^2}\right), 
\end{eqnarray}
where $\sigma_{x,i}$, $\sigma_{y,i}$, and $\sigma_{\tilde{\beta}^{1/2},i}$, are determined by the standard deviation of 
the $x$, $y$, and $\tilde{\beta}^{1/2}$ values for the $n$ nearest neighbors.  This formulation for $\sigma_{x,i}$, $\sigma_{y,i}$, and $\sigma_{\tilde{\beta}^{1/2},i}$ 
gives a wider filter in poorly sampled regions and narrower filter in regions with higher spatial resolution.  Since the $x$, $y$, and $\tilde{\beta}$ vectors
are independent of the fit parameters, we need only determine the $n$ nearest neighbors and calculate $K_i(n)$ once for use in our iterative fitting routines.  

When selecting the optimal number of nearest neighbors to use in this calculation, there is a direct tradeoff between increased precision and 
increased computational time.  Figure \ref{num_nbr} shows the change in the standard deviation of the residuals and 
the time required to compute $F_{0,j}$ for all $j$ as a function of the number of nearest neighbors used.  We find that using more 
than 50 nearest neighbors reduces the standard deviation of the residuals by less than $1\%$.  We also find a significant increase in 
the computational time required to use more than 20-50 nearest neighbors.  We therefore elect to limit the number of nearest 
neighbors considered in our calculation of Equation \ref{wfunc} to 50.


\begin{thebibliography}{}

\bibitem[{{Agol} {et~al.}(2010){Agol}, {Cowan}, {Knutson}, {Deming}, {Steffen},
  {Henry}, \& {Charbonneau}}]{agol10}
{Agol}, E., {Cowan}, N.~B., {Knutson}, H.~A., {et~al.} 2010, \apj, 721, 1861

\bibitem[{{Albrecht} {et~al.}(2012){Albrecht}, {Winn}, {Johnson}, {Howard},
  {Marcy}, {Butler}, {Arriagada}, {Crane}, {Shectman}, {Thompson}, {Hirano},
  {Bakos}, \& {Hartman}}]{alb12}
{Albrecht}, S., {Winn}, J.~N., {Johnson}, J.~A., {et~al.} 2012, \apj, 757, 18

\bibitem[{{Bakos} {et~al.}(2004){Bakos}, {Noyes}, {Kov{\'a}cs}, {Stanek},
  {Sasselov}, \& {Domsa}}]{bak04}
{Bakos}, G., {Noyes}, R.~W., {Kov{\'a}cs}, G., {et~al.} 2004, \pasp, 116, 266

\bibitem[{{Bakos} {et~al.}(2002){Bakos}, {L{\'a}z{\'a}r}, {Papp}, {S{\'a}ri},
  \& {Green}}]{bak02}
{Bakos}, G.~{\'A}., {L{\'a}z{\'a}r}, J., {Papp}, I., {S{\'a}ri}, P., \&
  {Green}, E.~M. 2002, \pasp, 114, 974

\bibitem[{{Bakos} {et~al.}(2007{\natexlab{a}}){Bakos}, {Shporer}, {P{\'a}l},
  {Torres}, {Kov{\'a}cs}, {Latham}, {Mazeh}, {Ofir}, {Noyes}, {Sasselov},
  {Bouchy}, {Pont}, {Queloz}, {Udry}, {Esquerdo}, {Sip{\H o}cz}, {Kov{\'a}cs},
  {Stefanik}, {L{\'a}z{\'a}r}, {Papp}, \& {S{\'a}ri}}]{bak07b}
{Bakos}, G.~{\'A}., {Shporer}, A., {P{\'a}l}, A., {et~al.} 2007{\natexlab{a}},
  \apjl, 671, L173

\bibitem[{{Bakos} {et~al.}(2007{\natexlab{b}}){Bakos}, {Kov{\'a}cs}, {Torres},
  {Fischer}, {Latham}, {Noyes}, {Sasselov}, {Mazeh}, {Shporer}, {Butler},
  {Stefanik}, {Fern{\'a}ndez}, {Sozzetti}, {P{\'a}l}, {Johnson}, {Marcy},
  {Winn}, {Sip{\H o}cz}, {L{\'a}z{\'a}r}, {Papp}, \& {S{\'a}ri}}]{bak07a}
{Bakos}, G.~{\'A}., {Kov{\'a}cs}, G., {Torres}, G., {et~al.}
  2007{\natexlab{b}}, \apj, 670, 826

\bibitem[{{Bakos} {et~al.}(2010){Bakos}, {Torres}, {P{\'a}l}, {Hartman},
  {Kov{\'a}cs}, {Noyes}, {Latham}, {Sasselov}, {Sip{\H o}cz}, {Esquerdo},
  {Fischer}, {Johnson}, {Marcy}, {Butler}, {Isaacson}, {Howard}, {Vogt},
  {Kov{\'a}cs}, {Fernandez}, {Mo{\'o}r}, {Stefanik}, {L{\'a}z{\'a}r}, {Papp},
  \& {S{\'a}ri}}]{bak10}
{Bakos}, G.~{\'A}., {Torres}, G., {P{\'a}l}, A., {et~al.} 2010, \apj, 710, 1724

\bibitem[{{Ballard} {et~al.}(2010){Ballard}, {Charbonneau}, {Deming},
  {Knutson}, {Christiansen}, {Holman}, {Fabrycky}, {Seager}, \&
  {A'Hearn}}]{bal10}
{Ballard}, S., {Charbonneau}, D., {Deming}, D., {et~al.} 2010, \pasp, 122, 1341

\bibitem[{{Baraffe} {et~al.}(2008){Baraffe}, {Chabrier}, \& {Barman}}]{bar08}
{Baraffe}, I., {Chabrier}, G., \& {Barman}, T. 2008, \aap, 482, 315

\bibitem[{{Barnes} \& {O'Brien}(2002)}]{bob02}
{Barnes}, J.~W., \& {O'Brien}, D.~P. 2002, \apj, 575, 1087

\bibitem[{{Batygin} {et~al.}(2009){Batygin}, {Bodenheimer}, \&
  {Laughlin}}]{bbl09}
{Batygin}, K., {Bodenheimer}, P., \& {Laughlin}, G. 2009, \apjl, 704, L49

\bibitem[{{Bean} {et~al.}(2008){Bean}, {Benedict}, {Charbonneau}, {Homeier},
  {Taylor}, {McArthur}, {Seifahrt}, {Dreizler}, \& {Reiners}}]{bea08}
{Bean}, J.~L., {Benedict}, G.~F., {Charbonneau}, D., {et~al.} 2008, \aap, 486,
  1039

\bibitem[{{Beerer} {et~al.}(2011){Beerer}, {Knutson}, {Burrows}, {Fortney},
  {Agol}, {Charbonneau}, {Cowan}, {Deming}, {Desert}, {Langton}, {Laughlin},
  {Lewis}, \& {Showman}}]{bee11}
{Beerer}, I.~M., {Knutson}, H.~A., {Burrows}, A., {et~al.} 2011, \apj, 727, 23

\bibitem[{{Bodenheimer} {et~al.}(2001){Bodenheimer}, {Lin}, \&
  {Mardling}}]{blm01}
{Bodenheimer}, P., {Lin}, D.~N.~C., \& {Mardling}, R.~A. 2001, \apj, 548, 466

\bibitem[{{Burrows} {et~al.}(2008){Burrows}, {Budaj}, \& {Hubeny}}]{bur08}
{Burrows}, A., {Budaj}, J., \& {Hubeny}, I. 2008, \apj, 678, 1436

\bibitem[{{Carter} \& {Winn}(2009)}]{car09}
{Carter}, J.~A., \& {Winn}, J.~N. 2009, \apj, 704, 51

\bibitem[{{Charbonneau} {et~al.}(2008){Charbonneau}, {Knutson}, {Barman},
  {Allen}, {Mayor}, {Megeath}, {Queloz}, \& {Udry}}]{cha08}
{Charbonneau}, D., {Knutson}, H.~A., {Barman}, T., {et~al.} 2008, \apj, 686,
  1341

\bibitem[{{Charbonneau} {et~al.}(2005){Charbonneau}, {Allen}, {Megeath},
  {Torres}, {Alonso}, {Brown}, {Gilliland}, {Latham}, {Mandushev}, {O'Donovan},
  \& {Sozzetti}}]{cha05}
{Charbonneau}, D., {Allen}, L.~E., {Megeath}, S.~T., {et~al.} 2005, \apj, 626,
  523

\bibitem[{{Cowan} \& {Agol}(2008)}]{cow08}
{Cowan}, N.~B., \& {Agol}, E. 2008, \apjl, 678, L129

\bibitem[{{Cowan} \& {Agol}(2011)}]{cow11}
---. 2011, \apj, 726, 82

\bibitem[{{Cowan} {et~al.}(2012{\natexlab{a}}){Cowan}, {Machalek}, {Croll},
  {Shekhtman}, {Burrows}, {Deming}, {Greene}, \& {Hora}}]{cow12a}
{Cowan}, N.~B., {Machalek}, P., {Croll}, B., {et~al.} 2012{\natexlab{a}}, \apj,
  747, 82

\bibitem[{{Cowan} {et~al.}(2012{\natexlab{b}}){Cowan}, {Voigt}, \&
  {Abbot}}]{cow12b}
{Cowan}, N.~B., {Voigt}, A., \& {Abbot}, D.~S. 2012{\natexlab{b}}, \apj, 757,
  80

\bibitem[{{Demory} {et~al.}(2011){Demory}, {Seager}, {Madhusudhan}, {Kjeldsen},
  {Christensen-Dalsgaard}, {Gillon}, {Rowe}, {Welsh}, {Adams}, {Dupree},
  {McCarthy}, {Kulesa}, {Borucki}, \& {Koch}}]{dem11}
{Demory}, B.-O., {Seager}, S., {Madhusudhan}, N., {et~al.} 2011, \apjl, 735,
  L12

\bibitem[{{Eastman} {et~al.}(2010){Eastman}, {Siverd}, \& {Gaudi}}]{eas10}
{Eastman}, J., {Siverd}, R., \& {Gaudi}, B.~S. 2010, \pasp, 122, 935

\bibitem[{{Eggleton} {et~al.}(1998){Eggleton}, {Kiseleva}, \& {Hut}}]{egg98}
{Eggleton}, P.~P., {Kiseleva}, L.~G., \& {Hut}, P. 1998, \apj, 499, 853

\bibitem[{{Fabrycky}(2008)}]{fab08}
{Fabrycky}, D. 2008, \apjl, 677, L117

\bibitem[{{Fabrycky} \& {Tremaine}(2007)}]{fab07}
{Fabrycky}, D., \& {Tremaine}, S. 2007, \apj, 669, 1298

\bibitem[{{Fazio} {et~al.}(2004){Fazio}, {Hora}, {Allen}, {Ashby}, {Barmby},
  {Deutsch}, {Huang}, {Kleiner}, {Marengo}, {Megeath}, {Melnick}, {Pahre},
  {Patten}, {Polizotti}, {Smith}, {Taylor}, {Wang}, {Willner}, {Hoffmann},
  {Pipher}, {Forrest}, {McMurty}, {McCreight}, {McKelvey}, {McMurray}, {Koch},
  {Moseley}, {Arendt}, {Mentzell}, {Marx}, {Losch}, {Mayman}, {Eichhorn},
  {Krebs}, {Jhabvala}, {Gezari}, {Fixsen}, {Flores}, {Shakoorzadeh}, {Jungo},
  {Hakun}, {Workman}, {Karpati}, {Kichak}, {Whitley}, {Mann}, {Tollestrup},
  {Eisenhardt}, {Stern}, {Gorjian}, {Bhattacharya}, {Carey}, {Nelson},
  {Glaccum}, {Lacy}, {Lowrance}, {Laine}, {Reach}, {Stauffer}, {Surace},
  {Wilson}, {Wright}, {Hoffman}, {Domingo}, \& {Cohen}}]{faz04}
{Fazio}, G.~G., {Hora}, J.~L., {Allen}, L.~E., {et~al.} 2004, \apjs, 154, 10

\bibitem[{{Ford}(2005)}]{for05}
{Ford}, E.~B. 2005, \aj, 129, 1706

\bibitem[{{Fortney} {et~al.}(2007){Fortney}, {Marley}, \& {Barnes}}]{for07}
{Fortney}, J.~J., {Marley}, M.~S., \& {Barnes}, J.~W. 2007, \apj, 659, 1661

\bibitem[{{Gehrz} {et~al.}(2007){Gehrz}, {Roellig}, {Werner}, {Fazio}, {Houck},
  {Low}, {Rieke}, {Soifer}, {Levine}, \& {Romana}}]{geh07}
{Gehrz}, R.~D., {Roellig}, T.~L., {Werner}, M.~W., {et~al.} 2007, Review of
  Scientific Instruments, 78, 011302

\bibitem[{{Harris}(1990)}]{har90}
{Harris}, W.~E. 1990, \pasp, 102, 949

\bibitem[{{Hauschildt} {et~al.}(1999){Hauschildt}, {Allard}, \&
  {Baron}}]{hau99}
{Hauschildt}, P.~H., {Allard}, F., \& {Baron}, E. 1999, \apj, 512, 377

\bibitem[{{Holman} \& {Murray}(2005)}]{hm05}
{Holman}, M.~J., \& {Murray}, N.~W. 2005, Science, 307, 1288

\bibitem[{{Hut}(1981)}]{hut81}
{Hut}, P. 1981, \aap, 99, 126

\bibitem[{{Iro} \& {Deming}(2010)}]{iro10}
{Iro}, N., \& {Deming}, L.~D. 2010, \apj, 712, 218

\bibitem[{{Ivanov} \& {Papaloizou}(2007)}]{iva07}
{Ivanov}, P.~B., \& {Papaloizou}, J.~C.~B. 2007, \mnras, 376, 682

\bibitem[{{Jackson} {et~al.}(2008){Jackson}, {Greenberg}, \& {Barnes}}]{jac08}
{Jackson}, B., {Greenberg}, R., \& {Barnes}, R. 2008, \apj, 678, 1396

\bibitem[{{Jenkins} {et~al.}(2002){Jenkins}, {Caldwell}, \& {Borucki}}]{jen02}
{Jenkins}, J.~M., {Caldwell}, D.~A., \& {Borucki}, W.~J. 2002, \apj, 564, 495

\bibitem[{{Kane} \& {Gelino}(2010)}]{kan10}
{Kane}, S.~R., \& {Gelino}, D.~M. 2010, \apj, 724, 818

\bibitem[{{Kataria} {et~al.}(2012){Kataria}, {Showman}, {Lewis}, {Fortney},
  {Marley}, \& {Freedman}}]{kat12}
{Kataria}, T., {Showman}, A.~P., {Lewis}, N.~K., {et~al.} 2012, ArXiv e-prints

\bibitem[{{Kipping}(2009)}]{kip09}
{Kipping}, D.~M. 2009, \mnras, 392, 181

\bibitem[{{Kipping} \& {Spiegel}(2011)}]{kip11}
{Kipping}, D.~M., \& {Spiegel}, D.~S. 2011, \mnras, 417, L88

\bibitem[{{Knutson} {et~al.}(2008){Knutson}, {Charbonneau}, {Allen}, {Burrows},
  \& {Megeath}}]{knu08}
{Knutson}, H.~A., {Charbonneau}, D., {Allen}, L.~E., {Burrows}, A., \&
  {Megeath}, S.~T. 2008, \apj, 673, 526

\bibitem[{{Knutson} {et~al.}(2009{\natexlab{a}}){Knutson}, {Charbonneau},
  {Cowan}, {Fortney}, {Showman}, {Agol}, \& {Henry}}]{knu09a}
{Knutson}, H.~A., {Charbonneau}, D., {Cowan}, N.~B., {et~al.}
  2009{\natexlab{a}}, \apj, 703, 769

\bibitem[{{Knutson} {et~al.}(2010){Knutson}, {Howard}, \& {Isaacson}}]{knu10}
{Knutson}, H.~A., {Howard}, A.~W., \& {Isaacson}, H. 2010, \apj, 720, 1569

\bibitem[{{Knutson} {et~al.}(2007){Knutson}, {Charbonneau}, {Allen}, {Fortney},
  {Agol}, {Cowan}, {Showman}, {Cooper}, \& {Megeath}}]{knu07}
{Knutson}, H.~A., {Charbonneau}, D., {Allen}, L.~E., {et~al.} 2007, \nat, 447,
  183

\bibitem[{{Knutson} {et~al.}(2009{\natexlab{b}}){Knutson}, {Charbonneau},
  {Cowan}, {Fortney}, {Showman}, {Agol}, {Henry}, {Everett}, \&
  {Allen}}]{knu09b}
{Knutson}, H.~A., {Charbonneau}, D., {Cowan}, N.~B., {et~al.}
  2009{\natexlab{b}}, \apj, 690, 822

\bibitem[{{Knutson} {et~al.}(2012){Knutson}, {Lewis}, {Fortney}, {Burrows},
  {Showman}, {Cowan}, {Agol}, {Aigrain}, {Charbonneau}, {Deming}, {D{\'e}sert},
  {Henry}, {Langton}, \& {Laughlin}}]{knu12}
{Knutson}, H.~A., {Lewis}, N., {Fortney}, J.~J., {et~al.} 2012, \apj, 754, 22

\bibitem[{{Kozai}(1962)}]{koz62}
{Kozai}, Y. 1962, \aj, 67, 591

\bibitem[{{Langton} \& {Laughlin}(2008)}]{lan08}
{Langton}, J., \& {Laughlin}, G. 2008, \apj, 674, 1106

\bibitem[{{Lewis} {et~al.}(2010){Lewis}, {Showman}, {Fortney}, {Marley},
  {Freedman}, \& {Lodders}}]{lew10}
{Lewis}, N.~K., {Showman}, A.~P., {Fortney}, J.~J., {et~al.} 2010, \apj, 720,
  344

\bibitem[{{Liddle}(2007)}]{lid07}
{Liddle}, A.~R. 2007, \mnras, 377, L74

\bibitem[{{Lin} {et~al.}(1996){Lin}, {Bodenheimer}, \& {Richardson}}]{lin96}
{Lin}, D.~N.~C., {Bodenheimer}, P., \& {Richardson}, D.~C. 1996, \nat, 380, 606

\bibitem[{{Loeillet} {et~al.}(2008){Loeillet}, {Shporer}, {Bouchy}, {Pont},
  {Mazeh}, {Beuzit}, {Boisse}, {Bonfils}, {da Silva}, {Delfosse}, {Desort},
  {Ecuvillon}, {Forveille}, {Galland}, {Gallenne}, {H{\'e}brard}, {Lagrange},
  {Lovis}, {Mayor}, {Moutou}, {Pepe}, {Perrier}, {Queloz}, {S{\'e}gransan},
  {Sivan}, {Santos}, {Tsodikovich}, {Udry}, \& {Vidal-Madjar}}]{loe08}
{Loeillet}, B., {Shporer}, A., {Bouchy}, F., {et~al.} 2008, \aap, 481, 529

\bibitem[{{Mandel} \& {Agol}(2002)}]{man02}
{Mandel}, K., \& {Agol}, E. 2002, \apjl, 580, L171

\bibitem[{{Markwardt}(2009)}]{mar09}
{Markwardt}, C.~B. 2009, in Astronomical Society of the Pacific Conference
  Series, Vol. 411, Astronomical Data Analysis Software and Systems XVIII, ed.
  {D.~A.~Bohlender, D.~Durand, \& P.~Dowler}, 251

\bibitem[{{Matsumura} {et~al.}(2008){Matsumura}, {Takeda}, \& {Rasio}}]{mat08}
{Matsumura}, S., {Takeda}, G., \& {Rasio}, F.~A. 2008, \apjl, 686, L29

\bibitem[{{Mighell}(2005)}]{mig05}
{Mighell}, K.~J. 2005, \mnras, 361, 861

\bibitem[{{Morales-Calder{\'o}n} {et~al.}(2006){Morales-Calder{\'o}n},
  {Stauffer}, {Kirkpatrick}, {Carey}, {Gelino}, {Barrado y Navascu{\'e}s},
  {Rebull}, {Lowrance}, {Marley}, {Charbonneau}, {Patten}, {Megeath}, \&
  {Buzasi}}]{mor06}
{Morales-Calder{\'o}n}, M., {Stauffer}, J.~R., {Kirkpatrick}, J.~D., {et~al.}
  2006, \apj, 653, 1454

\bibitem[{{Muller} \& {Buffington}(1974)}]{mul74}
{Muller}, R.~A., \& {Buffington}, A. 1974, Journal of the Optical Society of
  America (1917-1983), 64, 1200

\bibitem[{{Murray} \& {Dermott}(1999)}]{mur99}
{Murray}, C.~D., \& {Dermott}, S.~F. 1999, {Solar system dynamics}, ed.
  {Murray, C.~D.~\& Dermott, S.~F.}

\bibitem[{{Naoz} {et~al.}(2011){Naoz}, {Farr}, {Lithwick}, {Rasio}, \&
  {Teyssandier}}]{nao11}
{Naoz}, S., {Farr}, W.~M., {Lithwick}, Y., {Rasio}, F.~A., \& {Teyssandier}, J.
  2011, ArXiv e-prints

\bibitem[{{P{\'a}l}(2008)}]{pal08}
{P{\'a}l}, A. 2008, \mnras, 390, 281

\bibitem[{{P{\'a}l} {et~al.}(2010){P{\'a}l}, {Bakos}, {Torres}, {Noyes},
  {Fischer}, {Johnson}, {Henry}, {Butler}, {Marcy}, {Howard}, {Sip{\H o}cz},
  {Latham}, \& {Esquerdo}}]{pal10}
{P{\'a}l}, A., {Bakos}, G.~{\'A}., {Torres}, G., {et~al.} 2010, \mnras, 401,
  2665

\bibitem[{{Pritchet} \& {Kline}(1981)}]{pri81}
{Pritchet}, C., \& {Kline}, M.~I. 1981, \aj, 86, 1859

\bibitem[{{Rasio} \& {Ford}(1996)}]{ras96}
{Rasio}, F.~A., \& {Ford}, E.~B. 1996, Science, 274, 954

\bibitem[{{Reach} {et~al.}(2005){Reach}, {Megeath}, {Cohen}, {Hora}, {Carey},
  {Surace}, {Willner}, {Barmby}, {Wilson}, {Glaccum}, {Lowrance}, {Marengo}, \&
  {Fazio}}]{rea05}
{Reach}, W.~T., {Megeath}, S.~T., {Cohen}, M., {et~al.} 2005, \pasp, 117, 978

\bibitem[{{Sasaki} {et~al.}(2012){Sasaki}, {Barnes}, \& {O'Brien}}]{sbo12}
{Sasaki}, T., {Barnes}, J.~W., \& {O'Brien}, D.~P. 2012, \apj, 754, 51

\bibitem[{{Showman} {et~al.}(2010){Showman}, {Cho}, \& {Menou}}]{sho10}
{Showman}, A.~P., {Cho}, J.~Y.-K., \& {Menou}, K. 2010, {Atmospheric
  Circulation of Exoplanets}, ed. {Seager, S.}, 471--516

\bibitem[{{Showman} {et~al.}(2009){Showman}, {Fortney}, {Lian}, {Marley},
  {Freedman}, {Knutson}, \& {Charbonneau}}]{sho09}
{Showman}, A.~P., {Fortney}, J.~J., {Lian}, Y., {et~al.} 2009, \apj, 699, 564

\bibitem[{{Showman} \& {Guillot}(2002)}]{sho02}
{Showman}, A.~P., \& {Guillot}, T. 2002, \aap, 385, 166

\bibitem[{{Sing}(2010)}]{sin10}
{Sing}, D.~K. 2010, \aap, 510, A21

\bibitem[{{Southworth}(2008)}]{sou08}
{Southworth}, J. 2008, \mnras, 386, 1644

\bibitem[{{Stevenson} {et~al.}(2010){Stevenson}, {Harrington}, {Nymeyer},
  {Madhusudhan}, {Seager}, {Bowman}, {Hardy}, {Deming}, {Rauscher}, \&
  {Lust}}]{ste10}
{Stevenson}, K.~B., {Harrington}, J., {Nymeyer}, S., {et~al.} 2010, \nat, 464,
  1161

\bibitem[{{Stevenson} {et~al.}(2011){Stevenson}, {Harrington}, {Fortney},
  {Loredo}, {Hardy}, {Nymeyer}, {Bowman}, {Cubillos}, {Bowman}, \&
  {Hardin}}]{ste11}
{Stevenson}, K.~B., {Harrington}, J., {Fortney}, J., {et~al.} 2011, ArXiv
  e-prints

\bibitem[{{Todorov} {et~al.}(2012){Todorov}, {Deming}, {Knutson}, {Burrows},
  {Sada}, {Cowan}, {Agol}, {Desert}, {Fortney}, {Charbonneau}, {Laughlin},
  {Langton}, {Showman}, \& {Lewis}}]{tod12}
{Todorov}, K.~O., {Deming}, D., {Knutson}, H.~A., {et~al.} 2012, \apj, 746, 111

\bibitem[{{Vogt} {et~al.}(1994){Vogt}, {Allen}, {Bigelow}, {Bresee}, {Brown},
  {Cantrall}, {Conrad}, {Couture}, {Delaney}, {Epps}, {Hilyard}, {Hilyard},
  {Horn}, {Jern}, {Kanto}, {Keane}, {Kibrick}, {Lewis}, {Osborne},
  {Pardeilhan}, {Pfister}, {Ricketts}, {Robinson}, {Stover}, {Tucker}, {Ward},
  \& {Wei}}]{vogt94}
{Vogt}, S.~S., {Allen}, S.~L., {Bigelow}, B.~C., {et~al.} 1994, in Society of
  Photo-Optical Instrumentation Engineers (SPIE) Conference Series, Vol. 2198,
  Society of Photo-Optical Instrumentation Engineers (SPIE) Conference Series,
  ed. D.~L. {Crawford} \& E.~R. {Craine}, 362

\bibitem[{{Weidenschilling} \& {Marzari}(1996)}]{wei96}
{Weidenschilling}, S.~J., \& {Marzari}, F. 1996, \nat, 384, 619

\bibitem[{{Werner} {et~al.}(2004){Werner}, {Roellig}, {Low}, {Rieke}, {Rieke},
  {Hoffmann}, {Young}, {Houck}, {Brandl}, {Fazio}, {Hora}, {Gehrz}, {Helou},
  {Soifer}, {Stauffer}, {Keene}, {Eisenhardt}, {Gallagher}, {Gautier}, {Irace},
  {Lawrence}, {Simmons}, {Van Cleve}, {Jura}, {Wright}, \&
  {Cruikshank}}]{wer04}
{Werner}, M.~W., {Roellig}, T.~L., {Low}, F.~J., {et~al.} 2004, \apjs, 154, 1

\bibitem[{{Winn}(2010)}]{win10}
{Winn}, J.~N. 2010, {Exoplanet Transits and Occultations}, ed. S.~{Seager},
  55--77

\bibitem[{{Winn} {et~al.}(2009){Winn}, {Johnson}, {Albrecht}, {Howard},
  {Marcy}, {Crossfield}, \& {Holman}}]{win09}
{Winn}, J.~N., {Johnson}, J.~A., {Albrecht}, S., {et~al.} 2009, \apjl, 703, L99

\bibitem[{{Winn} {et~al.}(2007{\natexlab{a}}){Winn}, {Johnson}, {Peek},
  {Marcy}, {Bakos}, {Enya}, {Narita}, {Suto}, {Turner}, \& {Vogt}}]{win07}
{Winn}, J.~N., {Johnson}, J.~A., {Peek}, K.~M.~G., {et~al.} 2007{\natexlab{a}},
  \apjl, 665, L167

\bibitem[{{Winn} {et~al.}(2007{\natexlab{b}}){Winn}, {Holman}, {Bakos},
  {P{\'a}l}, {Johnson}, {Williams}, {Shporer}, {Mazeh}, {Fernandez}, {Latham},
  \& {Gillon}}]{win07a}
{Winn}, J.~N., {Holman}, M.~J., {Bakos}, G.~{\'A}., {et~al.}
  2007{\natexlab{b}}, \aj, 134, 1707

\bibitem[{{Winn} {et~al.}(2008){Winn}, {Holman}, {Torres}, {McCullough},
  {Johns-Krull}, {Latham}, {Shporer}, {Mazeh}, {Garcia-Melendo}, {Foote},
  {Esquerdo}, \& {Everett}}]{win08}
{Winn}, J.~N., {Holman}, M.~J., {Torres}, G., {et~al.} 2008, \apj, 683, 1076

\bibitem[{{Wizinowich} {et~al.}(2000){Wizinowich}, {Acton}, {Shelton},
  {Stomski}, {Gathright}, {Ho}, {Lupton}, {Tsubota}, {Lai}, {Max}, {Brase},
  {An}, {Avicola}, {Olivier}, {Gavel}, {Macintosh}, {Ghez}, \&
  {Larkin}}]{wiz00}
{Wizinowich}, P., {Acton}, D.~S., {Shelton}, C., {et~al.} 2000, \pasp, 112, 315

\bibitem[{{Wu} \& {Lithwick}(2011)}]{wu11}
{Wu}, Y., \& {Lithwick}, Y. 2011, \apj, 735, 109

\bibitem[{{Wu} \& {Murray}(2003)}]{wu03}
{Wu}, Y., \& {Murray}, N. 2003, \apj, 589, 605

\end{thebibliography}
\end{document}